\documentclass[10pt]{article}
\usepackage{graphicx}
\usepackage{epstopdf}
\usepackage{amsmath,amssymb,amsthm,amsfonts}
\usepackage{color}
\usepackage{hyperref}
\usepackage{cite}
\usepackage{verbatim}
\usepackage{appendix}

\newtheorem{thm}{Theorem}[section]

\newtheorem{lem}[thm]{Lemma}

\theoremstyle{definition}

\theoremstyle{remark}
\newtheorem{rem}{Remark}[section]

\numberwithin{equation}{section}

\DeclareMathSymbol{\C}{\mathalpha}{AMSb}{"43}

\newcommand{\eps}{\varepsilon}
\newcommand{\Om}{\Omega}

\newcommand{\lam}{\lambda}
\newcommand{\gam}{\gamma}
\newcommand{\B}{{\beta}}
\newcommand{\R}{{\mathbb{R}}}
\newcommand{\h}{{\mathcal{H}}}
\newcommand{\m}{{\mathcal{M}}}

\newcommand{\A}{{\mathcal{A}}}
\newcommand{\inte}{\int_{\mathbb{R}^2}}

\newcommand{\f}{a_1,a_2,\beta}

\def\ds{\displaystyle}

\def\R{{\mathbb R}}
\def\C{{\mathbb C}}

\newcommand{\bsub}{\begin{subequations}}
\newcommand{\esub}{\end{subequations}$\!$}
\title {Nonexistence of Vortices for Rotating Two-Component Focusing Bose Gases}
\author{ Yongshuai Gao\thanks{Email: ysgao@mails.ccnu.edu.cn. Y. Gao is partially supported by the Graduate Education Innovation Funds $\#$2022CXZZ088 at Central China Normal University in P. R. China.}
, Yujin Guo\thanks{Email: yguo@ccnu.edu.cn. Y.  Guo is partially supported by NSFC under Grants 12225106 and 11931012, as well as the Fundamental Research Funds $\#$CCNU22LJ002 for the Central Universities in P. R. China.}, Yan Li\thanks{Email: yanlimath@mails.ccnu.edu.cn.},
and Yong Luo\thanks{Email: yluo@ccnu.edu.cn. Y. Luo is partially supported by NSFC under Grant 12201231.}
\\
\small \it	School of Mathematics and Statistics,\\
\small \it  Hubei key Laboratory of Mathematical Sciences,\\
\small \it Central China Normal University, Wuhan 430079, P. R. China\\}

\begin{document}
\baselineskip= 15pt
\maketitle
\begin{abstract}
This paper is concerned with ground states of two-component Bose gases confined in a harmonic trap $V(x)=x_1^2+\Lambda^2 x_2^2$ rotating at the velocity $\Om >0$, where $\Lambda\ge 1$ and $(x_1, x_2)\in\R^2$. We focus on the case where the intraspecies interaction $(-a_1,-a_2)$ and the interspecies interaction $-\B$ are both attractive, i.e, $a_1, a_2$ and $\B $ are all positive. It is shown that for any $0<\Om <\Om ^*:=2$, ground states exist if and only if $0<a_1,\, a_2<a^*:=\|w\|^2_2$ and $0<\B<\B^*:=a^*+\sqrt{(a^*-a_1)(a^*-a_2)}$, where $w>0$ is the unique positive solution of $-\Delta w+ w-w^3=0$ in $\R^2$. By developing the argument of refined expansions, we further prove the nonexistence of vortices for ground states as $\B\nearrow\B^*$, where $0<\Om<\Om^*$ and $0<a_1,\, a_2<a^*$ are fixed.




\end{abstract}

\tableofcontents

\bigskip

\section{Introduction}
In this paper, we study ground states of the following  coupled rotational Gross-Pitaevskii equations
\begin{equation}\label{1.1equation system}
\left\{\begin{array}{lll}
-\Delta u_1+V(x)u_1+i\, \Om \big(x^{\perp}\cdot \nabla u_1\big)=\mu u_1+a_1|u_1|^2u_1+\B |u_2|^2u_1\  \ \hbox{in}\  \ \R^2,\\[3mm]
-\Delta u_2+V(x)u_2+i\, \Om \big(x^{\perp}\cdot \nabla u_2\big)=\mu u_2+a_2|u_2|^2u_2+\B |u_1|^2u_2\  \ \hbox{in}\  \ \R^2,
\end{array}\right.
\end{equation}
where $x^\perp=(-x_2,x_1)$ holds for $x=(x_1,x_2)\in\R^2$, and $\mu= \mu (a_1, a_2, \B)\in\R$ is a chemical potential. The system \eqref{1.1equation system} is used to model the rotating two-component Bose gases (e.g., \cite{GRPG,HMEWC,HS,KTU1,KTU2,LS,LS1,MH,ZBL}), where $V(x)\geq0$ is a trapping potential rotating at the velocity $\Om>0$, $|a_j|$ denotes the intraspecies interaction strength of cold atoms inside the $j^{th}$ component ($j=1,2$), and $|\B|$ describes the interspecies interaction strength of cold atoms between two components. Here $a_j>0$ (resp. $<0$) represents that the intraspecies interaction is attractive (resp. repulsive), and while $\B>0$ (resp. $<0$) denotes that the interspecies interaction is attractive (resp. repulsive).

Our main interest of the present paper is to study ground states of
\eqref{1.1equation system} for the case where the intraspecies interaction and interspecies interaction are both attractive, i.e., $a_1, a_2$ and $\B $ are all positive.
Similar to \cite [Proposition A.1] {GLWZ1}, ground states of
\eqref{1.1equation system} in this case can be described equivalently by minimizers of the following complex-valued constraint variational problem
\begin{equation}\label{1.3miniproble}
e(\Om,\f):=\inf_{(u_1,u_2)\in\m}F_{\Om,\f}(u_1,u_2),\  \ \Om >0,\ \, a_1>0,\  \ a_2>0,\  \ \B>0,
\end{equation}
where the  energy functional $F_{\Om,\f}(u_1,u_2)$ is given by
\begin{equation}\label{1.4energyfunctional}
\begin{split}
 F_{\Om,\f}(u_1,u_2):=
 &\sum_{j=1}^2\inte\Big[|\nabla u_j|^2+V(x)|u_j|^2-\frac{a_j}{2}|u_j|^4
 -\Om\, x^{\perp}\cdot\big(iu_j,\nabla u_j\big)\Big]dx\\
 &-\B \inte |u_1|^2|u_2|^2dx,
\end{split}
\end{equation}
and $(iu_j,\nabla u_j)=i(u_j\nabla \bar u_j-\bar u_j\nabla u_j)/2$ for $j=1, 2$. Here the space $\m$ is defined as
\begin{equation}\label{1.2def:m}
\m:=\Big\{(u_1,u_2)\in \h \times \h: \ \inte (|u_{1}|^2+|u_{2}|^2)dx=1\Big\},
\end{equation}
and
\begin{equation*}
\h:=\Big\{u\in  H^1(\R^2,\C):\ \inte V(x)|u(x)|^2dx<\infty \Big\}
\end{equation*}
is equipped with the norm
$$\|u\|_{_{\h}}=\Big\{\inte \Big[|\nabla u|^2+(V(x)+1)|u(x)|^2\Big] dx\Big\}^{\frac{1}{2}}.$$
In order to study ground states of \eqref{1.1equation system}, in this paper we therefore focus on the constraint variational problem \eqref{1.3miniproble}.

When there is no rotation (i.e., $\Om=0$) for the system,  stimulated by \cite{GLW,GS},  the existence, nonexistence, uniqueness and the semi-trivial limiting behavior of minimizers for $e(0,\f)$ were studied in \cite{GLWZ1,GLWZ2}  and the references therein.
Further, the authors in \cite{GX} analyzed excited states of BEC systems, where they generalized many analytical  results of ground states established in \cite{GLWZ1}.
In addition, when the constraint condition $\m$ of \eqref{1.3miniproble} is replaced by
$$\m _1:=\Big\{(u_1, u_2)\in \h \times \h: \, \inte|u_{1}|^2dx=\inte |u_{2}|^2dx=1\Big\},$$
the existence, nonexistence, mass concentration, and other analytical properties of minimizers for $e(0,\f)$
were also analyzed in \cite{GZZ2} and the references therein.

When there is  rotation (i.e., $\Om>0$) for the system, the following single-component constraint variational problem
\begin{equation}\label{1.3-1miniproble}
e(\Omega, a):=\inf\limits_{\{u\in\h,\, \inte|u|^2dx=1\}} F_a(u),\ \, \Omega >0, \  \ a>0,
\end{equation}
where
\begin{equation*}
 F_a(u):=
 \inte\big(|\nabla u|^2+V(x)|u|^2\big)dx-\frac{a}{2}\inte|u|^4dx-\Om\inte x^{\perp}\cdot(iu,\nabla u)dx,
\end{equation*}
was studied more recently in \cite{ANS1,BC,GLP,GLY,IM-1,LS,LNR} and the references therein.
More precisely, it was proved in \cite{BC,GLY,LNR} that there exists a critical rotational velocity $0<\Om^*:=\Om^*\big(V(x)\big)\leq\infty$ such that for any $\Om\in(0,\Om^*$), the minimizers of $e(\Omega, a)$ exist if and only if $0<a<a^*:=\|w\|^2_2$. Here $w=w(|x|)>0$ denotes (cf. \cite{GNN,K,W}) the unique, up to translation, positive solution of the following nonlinear scalar field equation
\begin{equation}\label{1.8equation:w}
-\Delta u+ u-u^3=0\  \  \mbox{in} \  \  \R^2,\  \  \mbox{where}\, \ u\in H^1(\R^2,\R).
\end{equation}
Especially,
by developing the method of inductive symmetry, the authors in \cite{GLY} proved the nonexistence of vortices of minimizers for $e(\Omega, a)$, provided that $V(x)=|x|^2$ is radially symmetric.
As a continuation of \cite{GLY},  the nonexistence of vortices in a very large region of $\R^2$ was recently obtained in \cite{Guo}  under the non-radially symmetric trap $V(x)=x_1^2+\Lambda^2x_2^2$ $(0<\Lambda<1)$ by analyzing the asymptotic expansions of minimizers.
On the other hand, we also remark that there exist some interesting works on ground states of rotating two-component defocusing Bose gases, see \cite {AMW,ANS,AS,MA} and the references therein. In particular, the authors in \cite{ANS} proved analytically the nonexistence of vortices for two-component defocusing Bose gases under rotation. In spite of these facts, as far as we know, there are rarely analytical results on ground states of two-component focusing Bose gases under rotation, i.e. the variational problem \eqref{1.3miniproble} for the case $\Omega>0$.

By developing a new approach, which is simpler than those of \cite{GLY,Guo}, the main purpose of this paper is to investigate the nonexistence of vortices for complex-valued minimizers of \eqref{1.3miniproble}, in the case where  the trapping potential $V(x)$ is  harmonic, i.e.,
\begin{equation}\label{V(x)}
V(x)=x_1^2+\Lambda^2x_2^2\, \ge 0,
\ \ \hbox{where}\ \ \Lambda\ge1 \ \ \hbox{and} \ \ x=(x_1,x_2)\in\R^2.
\end{equation}
We comment that the analysis of the case $0<\Lambda <1$ is similar to that of the case $ \Lambda>1$.
Throughout the whole paper, without special notations we thus
always consider the harmonic trap (\ref{V(x)}) with the range $ \Lambda \ge 1$.

\subsection{Main results}
In this subsection, we shall introduce briefly the main results of this paper. For convenience, we define
\begin{equation}\label{main 1.11def:beta*}
\B^*:=\B^*(a_1,a_2)=a^*+\sqrt{(a^*-a_1)(a^*-a_2)},\ \  0< a_1,a_2<a^*:=\|w\|_2^2,
\end{equation}
where $w>0$ is the unique positive solution of \eqref{1.8equation:w}.
Our first result is concerned with the following existence and nonexistence of minimizers for $e(\Om,\f)$.

\begin{thm}\label{main thm-1.1}
Suppose $V(x)$ satisfies \eqref{V(x)} for some $\Lambda\ge1$, and define $a^*>0$ and $\beta^*>0$ by (\ref{main 1.11def:beta*}).
Then there exists a critical velocity $\Om^* =2$
such that
\begin{enumerate}
\item If $\Om>\Om^*$, then for any $a_1>0$, $a_2>0$ and $\B>0$, there is no minimizer of $e(\Om,\f)$.
\item If $0<\Om <\Om^*$, $0<a_1,a_2< a^*$ and $0<\B<\B^*$, then there exists at least one minimizer of $e(\Om,\f)$.
\item  If $0<\Om <\Om^*$ and either $a_1> a^*$ or $a_2> a^*$ or $\B>\B^*$,
then there is no minimizer of $e(\Om,\f)$.
\end{enumerate}
\end{thm}

We shall illustrate in Remark \ref{rem2.1} how the above critical velocity $\Om^* =2$ can be obtained.
The existence of Theorem \ref{main thm-1.1} is proved by employing the compactness of Lemma \ref{Lem:compact}, which is different from the usual ones (e.g. \cite{GS,GLY,LNR}).
One can note that the existence of minimizers for $e(\Om,\f)$ at the threshold, where either $a_1=a^*$ or $a_2=a^*$ or $\B=\B^*$, is not addressed in Theorem
\ref{main thm-1.1}. We shall establish the following  nonexistence results for this critical case.

\begin{thm}\label{main thm-1.2}
Suppose $V(x)$ satisfies \eqref{V(x)} for some $\Lambda\ge1$, and define $a^*>0$ and $\beta^*>0$ by (\ref{main 1.11def:beta*}). Then for any $0<\Om<\Om^*:=2$,  we have
\begin{enumerate}
\item If either $a_1=a^*$, $a_2\leq a^*$ and $0<\B\leq \B^*(=a^*)$ or $a_2=a^*$, $a_1\leq a^*$ and $0<\B\leq \B^*(=a^*)$, then there is no minimizer of $e(\Om,\f)$.
\item  If $\B=\B^*$ and $0<a_1,a_2<a^*$, then there is no minimizer of $e(\Om,\f)$.
\end{enumerate}
\end{thm}

\begin{figure}
\centering
\includegraphics[scale=0.3]{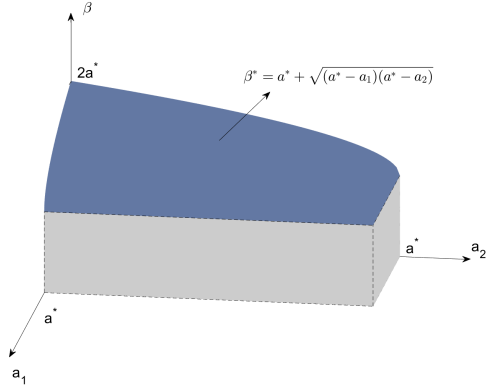}
\caption{When $0<\Om<\Om^*$, $e(\Om,\f)$ has minimizers if and
only if the point $(\f)$ lies within the cuboid.}
\label{1}
\end{figure}

When $0<\Om<\Om^*$, Theorems \ref{main thm-1.1} and \ref{main thm-1.2} give a complete classification that $e(\Om,\f)$ admits the minimizers, if and only if the point $(\f)$ lies within the cuboid as illustrated by Figure 1 below. On the other hand, when $\Om>\Om^*$, there is no minimizer of $e(\Om,\f)$ as soon as the point $(a_1, a_2,\B)$ lies in the first quadrant of Figure 1 below.
In Section 2, we shall establish Theorems \ref{thm-1.1} and \ref{thm-1.2} on the existence and nonexistence of minimizers for $e(\Om,\f)$ under a more general class of  trapping potentials $V(x)$, which then imply above Theorems \ref{main thm-1.1} and  \ref{main thm-1.2} in a direct way.

By the variational theory, if $(u_{1\B},u_{2\B})$ is a minimizer of $e(\Om,\f)$, then $(u_{1\B},u_{2\B})$ solves the following Euler-Lagrange system
\begin{equation}\label{1.1-1equation system}
\left\{\begin{array}{lll}
-\Delta u_{1\B}+V(x)u_{1\B}+i\,\Om \big(x^{\perp}\cdot \nabla u_{1\B}\big)=\mu_\B u_{1\B}+a_1|u_{1\B}|^2u_{1\B}+\B |u_{2\B}|^2u_{1\B}\  \ \hbox{in}\  \ \R^2,\\[3mm]
-\Delta u_{2\B}+V(x)u_{2\B}+i\,\Om \big(x^{\perp}\cdot \nabla u_{2\B}\big)=\mu_\B u_{2\B}+a_2|u_{2\B}|^2u_{2\B}+\B |u_{1\B}|^2u_{2\B}\  \ \hbox{in}\  \ \R^2,
\end{array}\right.
\end{equation}
where $\mu_\B\in\R$ is a suitable Lagrange multiplier and satisfies
\begin{equation}\label{1.5Lmu}
\mu_\B=e(\Om,\f)-\inte \Big(\frac{a_1}{2}|u_{1\B}|^4+\frac{a_2}{2}|u_{2\B}|^4+\B|u_{1\B}|^2|u_{2\B}|^2\Big)dx.
\end{equation}
Under the harmonic trap $V(x)=x_1^2+\Lambda^2x_2^2$ $(\Lambda\ge1)$ in \eqref{V(x)}, the main result of this paper is the  following nonexistence of vortices for minimizers.

\begin{thm}\label{thm-nonvorex}
Suppose $V(x)$ satisfies \eqref{V(x)} for some $\Lambda\ge1$, and define $a^*>0$ and $\beta ^*>0$ by (\ref{main 1.11def:beta*}). Let $(u_{1\B},u_{2\B})$ be a complex-valued minimizer of $e(\Om,\f)$, where $0<\Om<\Om^*:=2$, $0<a_1,a_2<a^*$ and $0<\beta<\beta^*$. Then we have
\begin{enumerate}
  \item If $\Lambda=1$, then up to a constant phase, all minimizers $(u_{1\B},u_{2\B})$ of $e(\Om,\f)$ are real-valued, unique and free of vortices when $\B^*-\B>0$ is small enough.
  \item If $\Lambda>1$, then there exists a constant $C>0$, independent of $0<\B<\B^*$, such that
\begin{equation}\label{thm-nonvorex lambda dayu 1}
|u_{1\B}(x)|,\ \ |u_{2\B}(x)|>0\,\  \mbox{in}\,\ R(\B):=\big\{x\in\R^2:\, |x|\le C(\B^*-\B)^{-\frac{1}{12}} \big\}\ \ \hbox{as}\ \ \B\nearrow\B^*,
\end{equation}
i.e., $(u_{1\B},u_{2\B})$ does not admit any vortex in the region $R(\B)$ when $\B^*-\B>0$ is small enough.
\end{enumerate}
\end{thm}

There are several comments on Theorem \ref{thm-nonvorex}. Firstly, as discussed in Subsection 1.2, we shall develop the so-called argument of refined expansions to address the proof of Theorem \ref{thm-nonvorex}. It deserves to remark that this new approach simplifies the proof procedure of Theorem \ref{thm-nonvorex} (1), comparing with the existing arguments, such as the method of inductive symmetry (cf. \cite{GLY}). Secondly,
as a byproduct, Theorem \ref{thm-nonvorex} yields essentially the refined limiting expansion (\ref{solution u1B u2B expan}) of the minimizer $(u_{1\B},u_{2\B})$, see Section 4 for details. Thirdly, under the assumptions of Theorem \ref{thm-nonvorex}, where $V(x)$ satisfies \eqref{V(x)} for any $\Lambda\ge1$, we conclude from Theorems \ref{thm-1.3} and \ref{thm-equation.uniqueness} that up to a constant phase, $e(\Om,\f)$ admits a unique minimizer when $\B^*-\B>0$ is small enough.



\subsection{Proof strategy of Theorem \ref{thm-nonvorex}}
The main purpose of this subsection is to explain by five steps the general strategy of proving Theorem \ref{thm-nonvorex}, which can be summarized as the argument of refined expansions.
Roughly speaking, the basic idea of this argument is to derive several different types of refined limiting expansions.

In the first step, under the assumptions of Theorem \ref{thm-nonvorex} we define
\begin{equation*}
  \eps_\B:=\sqrt{-\frac{1}{\mu_\B}}>0\ \ \hbox{as}\ \ \B\nearrow\B^*,
\end{equation*}
where the Lagrange multiplier $\mu_{\B}\in\R$ is given in (\ref{1.5Lmu}). We shall derive that $\mu_\B\to -\infty$ and hence $ \eps_\B\to 0$ as $\B\nearrow\B^*$. For the minimizer $(u_{1\B},u_{2\B})$ of $e(\Om,\f)$, we then prove in Lemmas \ref{lem-3.4} and \ref{small}  that
after the transformation
\begin{equation}\label{sec1.2-1}
\left\{\begin{array}{lll}
v_{1\B}(x):=\sqrt{a^*}\eps_\B u_{1\B}\big(\eps_\B x\big)
e^{i\widetilde{\theta}_{1\B}}, \\[3mm]
v_{2\B}(x):=\sqrt{a^*}\eps_\B u_{2\B}\big(\eps_\B x\big)
e^{i\widetilde{\theta}_{2\B}},
\end{array}\right.
\end{equation}
the function $(v_{1\B},v_{2\B})$ satisfies
\begin{equation}\label{sec1.2-2}
\lim_{\B\nearrow \B^*}(v_{1\B},v_{2\B})=(\sqrt{\gamma_1}w,\sqrt{\gamma_2}w)\,\ \hbox{strongly in $H^1(\R^2,\C)\cap L^\infty(\R^2,\C)$,}
\end{equation}
where $\gam_j:=1-\frac{\sqrt{a^*-a_j}}{\sqrt{a^*-a_1}+\sqrt{a^*-a_2}}\in(0,1)$ for $j=1,2$, and $(\widetilde{\theta}_{1\B},\widetilde{\theta}_{2\B})\in[0,2\pi)\times [0,2\pi)$ is a suitable constant phase.

Different from the existing works \cite{Guo,GLP,GLY,GLWZ1} and the references therein, the transform \eqref{sec1.2-1} is defined in terms of $\eps_\B>0$, instead of $\big[\inte \big(|\nabla u_{1\B}|^2+|\nabla u_{2\B}|^2\big)dx\big]^{-\frac{1}{2}}$, and does not employ the global maximum point $(x_{1\B},x_{2\B})$ of the minimizer.
The advantages of the transform \eqref{sec1.2-1} lie in the facts that the corresponding elliptic system of the function $(v_{1\B},v_{2\B})$ becomes simpler, and  one does not need to analyze yet the refined estimates of $(x_{1\B},x_{2\B})$. Unfortunately, the transform \eqref{sec1.2-1} however leads to extra difficulties in proving the convergence (\ref{sec1.2-2}). To overcome such difficulties, we shall establish an important
Pohozaev identity involved with the magnetic Laplacian operator, see  Lemma \ref{lem.A.1} for details, which is also used in other steps.

As the second step, we shall follow the first step to establish Theorem \ref{thm-1.3} on the following refined limiting behavior of minimizers $(u_{1\B},u_{2\B})$ for $e(\Om,\f)$:
for $j=1,2$,
\begin{equation}\label{1:MM}
w_{j\B}(x):=\sqrt{a^*}\alpha_\B u_{j\B}(\alpha_\B x)e^{i\theta_{j\B}}
\to\sqrt{\gam_j}w(x)
\ \ \hbox{strongly in}\ \ H^1(\R^2,\C)\cap L^\infty(\R^2,\C)
   \end{equation}
as $\B\nearrow\B^*$, where $\theta_{j\B}\in [0,2\pi)$ is a properly chosen constant,
and $\alpha_\B$ is defined as
\begin{equation}\label{AA:sec-1.3}
 \alpha_\B:=\Big[\frac{2\gam_1\gam_2}{\lam_0}(\B^*-\B)\Big]^\frac{1}{4}>0, \ \
 \lam_0:=\ds\inte V(x)w^2(x)dx>0.
\end{equation}
Since the function $w_{j\B}$ of (\ref{1:MM}) is not scaled around the global maximum point $(x_{1\B},x_{2\B})$  of the minimizer, the proof of the convergence (\ref{1:MM}) is different slightly from the existing approaches, e.g. \cite{Guo,GLY,GS,GZZ2} and the references therein.

To establish (\ref{1:MM}), actually we shall first use \eqref{sec1.2-2} to obtain the refined lower bound of $e(\Om,\f)$, together with its refined upper bound of  Lemma \ref{lem-3.2}, which then yields that
\begin{equation}\label{sec-1.3}
\eps_\B :=\sqrt{-\frac{1}{\mu_\B}}=\alpha_\B+o(\alpha_\B)\ \ \hbox{as}\ \ \B\nearrow\B^*,
\end{equation}
where $\alpha_\B>0$ is as in \eqref{AA:sec-1.3}. Applying \eqref{sec1.2-2} and \eqref{sec-1.3}, we shall further prove that
for $j=1,2$,
\begin{equation}\label{AA:Y12}
     \sqrt{a^*}\alpha_\B u_{j\B}(\alpha_\B x)
     e^{i\widetilde{\theta}_{j\B}}
     \to\sqrt{\gam_j}w(x)
     \ \ \text{strongly in}\  \ H^1(\R^2,\C)\cap L^\infty(\R^2,\C)\  \ \text{as}\  \ \B\nearrow\B^*,
   \end{equation}
where the constant phase $(\widetilde{\theta}_{1\B},\widetilde{\theta}_{2\B})\in[0,2\pi)\times [0,2\pi)$ is as in (\ref{sec1.2-1}). It is proved that
\begin{equation}\label{1.2}
  \lim\limits_{\B\nearrow\B^*}|\theta_{j\B}-\tilde{\theta}_{j\B}|=0,\ \ j=1,\ 2.
\end{equation}
The convergence (\ref{1:MM}) is then proved by making full use of (\ref{AA:Y12}) and \eqref{1.2}.

In the third step, we shall establish the refined expansions,  in terms of $\eps_{\B}:=\sqrt{-\frac{1}{\mu_\B}}>0$, of the normalized concentration solutions $(\tilde u_{1\B}, \tilde u_{2\B})\in\m$ for the elliptic system
(\ref{1.1-1equation system}) satisfying (\ref{1:MM}), where the parameter $\mu_\B$ satisfies $\mu_\B\alpha_\B^2\to -1$ as $\B\nearrow\B^*$.
More precisely, the purpose of this step is to prove that for $j=1,2$,
\begin{equation}\label{1:4}
\begin{split}
\ds v_{j\B}(x):&=\sqrt{a^*}\eps_\B \tilde u_{j\B}(\eps_\B x)e^{i\widetilde\theta_{j\B}}\\
&=\rho_{j\B}\big\{w+\eps_\B^4\psi_0+O(\eps_\B^{8})\big\}
+i\big\{\rho_{j\B}\,\Omega\eps_\B^6\phi_0
+O(\eps_\B^{10})\big\}\ \ \hbox{in}\,\  L^\infty(\R^2,\C)
\end{split}
\end{equation}
as $\B\nearrow\B^*$, where $\psi_0$ and $\phi_0$ are uniquely given by \eqref{psi 0.equation.thm} and \eqref{phi 0.equation.thm}, respectively, and the constant $\rho_{j\B}>0$ varies in $\B$ and satisfies
\[
\ds\rho_{j\B}:=\sqrt{\frac{a^*(\B-a_m)}{\B^2-a_1a_2}}\rightarrow\sqrt{\gamma_{j}}>0\ \ \hbox{as}\ \ \B\nearrow\B^*,
\ \ m\neq j,\ \ m=1,2.
\]

We comment that the right hand side of the expansion (\ref{1:4}) contains the varying parameter $\rho_{j\B}>0$. Therefore, the form (\ref{1:4}) and its proof are slightly different from  the recent works  \cite{Guo,GLW}, which handle the single component case of Bose gases. To prove (\ref{1:4}), actually we need employ the following ``limiting" system
\begin{equation*}
\left\{\begin{array}{lll}
-\ds\Delta Q_{1\B}+Q_{1\B}-\frac{a_1}{a^*}Q_{1\B}^3
-\frac{\B}{a^*}Q_{2\B}^2Q_{1\B}=0
\quad \hbox{in}\  \ \R^2,\\[3mm]
-\ds\Delta Q_{2\B}+Q_{2\B}-\frac{a_2}{a^*}Q_{2\B}^3
-\frac{\B}{a^*}Q_{1\B}^2Q_{2\B}=0
\quad \hbox{in}\  \ \R^2,
\end{array}\right.
\ \  \max\{a_1,a_2\}<\B<\B ^*.
\end{equation*}
This is different from the usually limiting system (4.19) in \cite{GLWZ1}, which contains the fixed coefficient $\B^*$, instead of $\B>0$.



As the fourth step, we shall improve the main result (\ref{1:4}) of the third step to establish Theorem
\ref{thm-expan solution} on the following refined expansions  in terms of $\alpha_\B>0$:  for $j=1,2$,
\begin{equation}\label{solution u1B u2B expan1}
\begin{split}
\ds w_{j\B}(x):&=\sqrt{a^*}\alpha_\B \tilde u_{j\B}(\alpha_\B x)e^{i\theta_{j\B}}\\
&=\rho_{j\B}\big\{w+\alpha_\B^4\big[\psi_0-\frac{1}{2}C(a_1,a_2)
\big(w+x\cdot\nabla w\big)\big]+O(\alpha_\B^{8})\big\}\\[1mm]
&\quad+i\big\{\rho_{j\B}\,\Omega\alpha_\B^6\phi_0
+O(\alpha_\B^{10})\big\}\,\ \hbox{in}\,\  L^\infty(\R^2,\C) \ \ \hbox{as}\ \ \B\nearrow\B^*,
\end{split}
\end{equation}
where the constant $C(a_1,a_2)$ is independent of $\B >0$, and $(\tilde u_{1\B}, \tilde u_{2\B})\in\m$ denotes a normalized concentration solution of the elliptic system
(\ref{1.1-1equation system}) satisfying (\ref{1:MM}) and $\mu_\B\alpha_\B^2\to -1$ as $\B\nearrow\B^*$. For this purpose, since  $\mu_\B\alpha_\B^2\to -1$ as $\B\nearrow\B^*$,
we shall first derive from \eqref{1:4} that
\begin{equation}\label{alphaB2 muB expan1}
\alpha_\B^2\mu_\B=-1+C(a_1,a_2)\alpha_\B^4+O(\alpha_\B^8)\ \ \hbox{as}\ \ \B\nearrow\B^*.
\end{equation}
The refined expansion \eqref{solution u1B u2B expan1} is then proved  by applying \eqref{1:4} and \eqref{alphaB2 muB expan1}. We remark from above four steps that the minimizer $(u_{1\B},u_{2\B})$  of $e(\Om,\f)$ satisfies the refined expansion (\ref{solution u1B u2B expan1}) as $\B\nearrow\B^*$.

Even though a similar expansion of \eqref{alphaB2 muB expan1} was established earlier (e.g. \cite{Guo} and the references therein) for the single component case of Bose gases, there appear new difficulties in the proof of \eqref{alphaB2 muB expan1}, due to the fact that the expansion \eqref{1:4} contains the varying parameter $\rho_{j\B}>0$. For this reason, we first use the expansion \eqref{1:4} to derive the relationship between $\rho_{1\B}^2+\rho_{2\B}^2$ and $\eps_\B$. The relationship between $\rho_{1\B}^2+\rho_{2\B}^2$ and $\alpha_\B$ is then established by analyzing the Taylor's expansion of $\rho_{1\B}^2+\rho_{2\B}^2$ around the point $\B^*$. Applying these results, we finally obtain the expansion \eqref{alphaB2 muB expan1}, see Lemma \ref{lem 4} for details.

In the fifth step, we shall complete the proof of Theorem \ref{thm-nonvorex} by considering separately the following two cases:

{\em Case 1: $\Lambda=1$.} Applying the expansion \eqref{solution u1B u2B expan1}, in this case we shall prove in Theorem \ref{thm-equation.uniqueness} that the normalized concentration solutions, which satisfy (\ref{1:MM}) and $\mu_\B\alpha_\B^2\to -1$ as $\B\nearrow\B^*$, of the elliptic system (\ref{1.1-1equation system}) in $\m$ must be unique as $\B\nearrow\B^*$.
Note from the previous step that the minimizer $(u_{1\B},u_{2\B})$ is such a normalized concentration solution of (\ref{1.1-1equation system}) as $\B\nearrow\B^*$. On the other hand, it follows from  \cite[Theorems 1.4 and 1.5]{GLWZ1} that $(u_{10},u_{20})$ is a unique real-valued positive minimizer of $e(0,\f)$ as $\B\nearrow\B^*$. By the symmetric-decreasing rearrangement, we also have the radial symmetry of $(u_{10},u_{20})$. We shall further derive that $(u_{10},u_{20})$ is also such a normalized concentration solution of the elliptic system (\ref{1.1-1equation system}) as $\B\nearrow\B^*$. By the uniqueness of Theorem \ref{thm-equation.uniqueness}, we thus conclude that up to a constant phase $(\theta_{1\B},\theta_{2\B})\in [0,2\pi)\times[0,2\pi)$, $(u_{1\B}e^{i\theta_{1\B}},u_{2\B}e^{i\theta_{2\B}})\equiv(u_{10},u_{20})$ is real-valued, unique and positive as $\B\nearrow\B^*$, and Theorem \ref{thm-nonvorex} (1) therefore follows.

We expect that Theorem \ref{thm-nonvorex} (1) can be proved by the method of inductive symmetry (cf. \cite[Theorem 1.3]{GLY}), which is however involved with more complicated analysis. Moreover, the nonexistence of vortices for defocusing Bose gases under rotation was studied earlier by jacobian estimates, vortex ball constructions, and some other arguments, see \cite{AJ,ANS,IM-1,A,CR,SSbook,Roug} and the references therein.  Unfortunately, to our best knowledge, the above arguments seem not applicable to our focusing case.

{\em Case 2: $\Lambda>1$.} In this case, we note from
\cite[Proposition 4.1]{GNN} that the unique positive solution $w(x)$ of \eqref{1.8equation:w} satisfies
\begin{equation}\label{exp decay}
w(x),\,|\nabla w(x)|=O(|x|^{-\frac 12}e^{-|x|})\quad\hbox{as $|x|\to\infty$}.
\end{equation}
Applying \eqref{solution u1B u2B expan1} and \eqref{exp decay}, we shall use the comparison principle to obtain the following optimal estimate of the lower order terms:  for $j=1,2$,
\begin{equation*}
\begin{split}
|\tilde{w}_{j\B}(x)|:&=\big|\sqrt{a^*}\alpha_\B u_{j\B}(\alpha_\B x)e^{i\theta_{j\B}}-\rho_{j\B}w(x)\big|\\
&\leq C_{1}\alpha_\B^{4}|x|^{\frac{5}{2}}e^{-\sqrt{1-C_{2}\alpha_\B^{4}}\,|x|}\ \ \hbox{uniformly in}\ \ \R^2 \ \ \mbox{as} \ \ \B\nearrow \B^*,
\end{split}
\end{equation*}
where the positive constants $C_{1}$ and $C_{2}$ are independent of $0<\B<\B^*$, and $\alpha_\B>0$ is defined by \eqref{AA:sec-1.3}.
Following above estimates, we then conclude from (\ref{exp decay}) that  for $j=1,2$,
\[
  |\sqrt{a^*}\alpha_\B u_{j\B}(\alpha_\B x)e^{i\theta_{j\B}}|\geq \rho_{j\B}w(x)-|\tilde{w}_{j\B}(x)|>0,  \ \ \mbox{if} \ \ |x|\le C(\B^*-\B)^{-\frac{1}{12}} ,
\]
which therefore completes the proof of Theorem \ref{thm-nonvorex} (2).

This paper is organized as follows. Section 2 is devoted to the proof of Theorems \ref{main thm-1.1} and \ref{main thm-1.2} on the existence and nonexistence of minimizers for $e(\Om,\f)$. In Section 3, we shall establish Theorem \ref{thm-1.3} on the limiting behavior of minimizers for $e(\Om,\f)$ as $\B\nearrow\B^*$. Section 4 is then devoted to the proof of Theorem \ref{thm-expan solution} on the refined expansions of solutions $(\tilde u_{1\B}, \tilde u_{2\B})\in\m$ for the Euler-Lagrange system (\ref{1.1-1equation system}).
In Section 5, we shall prove Theorem \ref{thm-equation.uniqueness} on the local uniqueness of   $(\tilde u_{1\B}, \tilde u_{2\B})$ as $\B\nearrow\B^*$. Applying Theorems \ref{thm-expan solution} and \ref{thm-equation.uniqueness}, the proof of Theorem \ref{thm-nonvorex} is then complete in Section 6 by the argument of refined expansions.  In Appendix A.1 we shall establish Lemma \ref{lem.A.1} on an important integral identity, and the proofs of \eqref{4u19.bar.ujBvjB.Linfty} and Lemma \ref{lem 5} are given finally in Appendix A.2.

\section{Existence and Nonexistence of Minimizers}
In this section we prove Theorems \ref{main thm-1.1} and \ref{main thm-1.2} on  the existence and nonexistence of minimizers for $e(\Om,\f)$. Towards this purpose, in this section we consider a more general class of trapping potentials $0\leq V(x)\in L^{\infty}_{loc}(\R^2)$ satisfying
\begin{equation}\label{1.6condition1:V(x)}
\varliminf\limits_{|x|\to\infty}\frac{V(x)}{|x|^2}>0.
\end{equation}
Under the general assumption (\ref{1.6condition1:V(x)}), one may define a critical rotational velocity $\Om^*$ by
\begin{equation}\label{main 1.7def:Omgea*}
\Om^*:=\sup\Big\{\Om>0: \ V_\Om(x)=V(x)-\frac{\Om^2}{4}|x|^2\to\infty \ \text{as} \ |x|\to\infty\Big\},
\end{equation}
so that $0<\Omega ^*=\Omega ^*\big(V(x)\big)\le \infty$ holds true.

\begin{rem}\label{rem2.1}
The above definition (\ref{main 1.7def:Omgea*}) yields that
$$ \mbox{if $V(x)$ satisfies \eqref{V(x)} for some $\Lambda\ge1$, then}\,\ \Om^*=2.$$
\end{rem}

Under the general trapping potential \eqref{1.6condition1:V(x)}, we shall establish the following existence results of minimizers for $e(\Om,\f)$, which imply directly Theorem  \ref{main thm-1.1}.

\begin{thm}\label{thm-1.1}
Suppose $V(x)\in L^\infty_{loc}(\R^2)$ satisfies
\eqref{1.6condition1:V(x)} such that $\Om^*\in (0,+\infty]$ exists in \eqref{main 1.7def:Omgea*}, and define $a^*>0$ and $\beta^*>0$ by (\ref{main 1.11def:beta*}). Then we have
\begin{enumerate}
\item If
    $\Om>\Om^*$, then for any $a_1>0$, $a_2>0$ and $\B>0$, there is no minimizer of $e(\Om,\f)$.
\item If $0<\Om <\Om ^*$, $0<a_1,a_2< a^*$ and $0<\B<\B^*$, then there exists at least one minimizer of $e(\Om,\f)$.
\item  If $0<\Om <\Om ^*$ and either $a_1> a^*$ or $a_2> a^*$ or $\B>\B^*$,
then there is no minimizer of $e(\Om,\f)$.
\end{enumerate}
\end{thm}

We also have the following results on the nonexistence  of minimizers for $e(\Om,\f)$ at the threshold,
 where either $a_1=a^*$ or $a_2=a^*$ or $\B=\B^*$.

\begin{thm}\label{thm-1.2}
Suppose that $V(x)\in L^\infty_{loc}(\R^2)$ satisfies
\eqref{1.6condition1:V(x)}, such that $\Om^*\in (0,+\infty]$ exists in \eqref{main 1.7def:Omgea*}, where $0<\Om<\Om^*$ is fixed. Then we have
\begin{enumerate}
\item If either $a_1=a^*$, $a_2\leq a^*$ and $0<\B\leq \B^*(=a^*)$ or $a_2=a^*$, $a_1\leq a^*$ and $0<\B\leq \B^*(=a^*)$, then there is no minimizer of $e(\Om,\f)$. Moreover, we have $\lim\limits_{a_1\nearrow a^*} e(\Om,\f)=e(\Om,a^*,a_2,\B)=\inf\limits_{x\in\R^2}V_{\Om}(x)$ for $a_2<a^*$,
and $\lim\limits_{a_2\nearrow a^*} e(\Om,\f)=e(\Om,a_1,a^*,\B)=\inf\limits_{x\in\R^2}V_{\Om}(x)$ for $a_1<a^*$, where $0<\B\leq \B^*(=a^*)$ and $V_\Om(x)=V(x)-\frac{\Om^2}{4}|x|^2$.
\item  If $\B=\B^*$ and $0<a_1,a_2<a^*$, then there is no minimizer of $e(\Om,\f)$. Furthermore, we have
    $\lim\limits_{\B\nearrow\B^*}e(\Om,\f)=e(\Om,a_1,a_2,\B^*)
    =\inf\limits_{x\in\R^2}V_{\Om}(x)$ for $a_1,a_2<a^*$.
\end{enumerate}
\end{thm}

Specially, Theorem \ref{thm-1.2} yields directly Theorem  \ref{main thm-1.2}. The rest part of this section is to prove above two theorems.

\subsection{Proof of Theorems \ref{thm-1.1} and \ref{thm-1.2}}
In this subsection, we shall complete the proof of Theorems  \ref{thm-1.1} and \ref{thm-1.2} under the general assumption \eqref{1.6condition1:V(x)}.
Recall  the following Gagliardo-Nirenberg inequality
\begin{equation}\label{2.1ineq:GNQ}
\inte|u(x)|^4dx\leq
\frac{2} {\|w\|^2_2}\inte |\nabla u(x) |^2 dx \inte |u(x)|^2 dx,
\  \ u\in H^1(\R^2,\R),
\end{equation}
where the identity is attained at the unique positive  solution $w$ of \eqref{1.8equation:w}, see \cite{W}.
Furthermore, it follows from \cite [Lemma 8.1.2] {C} and \cite[Proposition 4.1]{GNN}  that $w$ satisfies
\begin{equation}\label{1.9identity:w}
\inte|\nabla w|^2dx=\inte w^2dx=\frac{1}{2}\inte w^4dx ,
\end{equation}
and 
\begin{equation} \label{1.10decay:w}
w(x) \, , \ |\nabla w(x)| = O(|x|^{-\frac{1}{2}}e^{-|x|})
\ \ \text{as} \  \ |x|\to \infty.
\end{equation}
As a generalization of (\ref{2.1ineq:GNQ}), we have (cf. \cite [Lemma A.2] {GLWZ1}) the following coupled Gagliardo-Nirenberg inequality : for $(u_1,u_2)\in H^1(\R^2,\R)\times H^1(\R^2,\R)$,
 \begin{equation}\label{2.2Ineq:GN}
\inte\big(|u_1|^2+|u_2|^2  \big)^2 dx
\le\frac{2}{ \|w\|_2^2}
\inte\big(|\nabla u_1|^2+|\nabla u_2|^2\big)dx\inte\big(|u_1|^2+|u_2|^2\big)dx,
\end{equation}
where $\frac{2}{ \|w\|_2^2}>0$ is the best constant of \eqref{2.2Ineq:GN}, and the identity is attained at $(w\cos\theta,w\sin\theta)$ for any $\theta\in[0,2\pi)$.
Recall also from \cite{LL} the following diamagnetic inequality
\begin{equation}\label{2.3ineq:diam}
\big|(\nabla-i\A)u\big|^2\geq\big|\nabla |u|\big|^2
\ \ \hbox{a.e. in}\ \ \R^2, \  \ u\in H^1(\R^2,\C),
\end{equation}
where $\A\in L^2_{loc}(\R^2,\R^2)$ is any given vector function.

To prove Theorem \ref{thm-1.1}, we first establish the following compactness lemma, which is different from the usual ones (e.g. \cite{GS,GLY,LNR}).

\begin{lem}\label{Lem:compact}
Suppose $V(x)\in L^\infty_{loc}(\R^2)$ satisfies
\eqref{1.6condition1:V(x)} such that $\Om^*\in (0,+\infty]$ exists in \eqref{main 1.7def:Omgea*}, and consider $V_\Om(x)=V(x)-\frac{\Om^2}{4}|x|^2$, where $0<\Om<\Om^*$ is fixed. Define the space $H^1_{A,V_\Om}$ as
\begin{equation}\label{com1}
H^1_{A,V_\Om}:=\Big\{u\in H^1_A(\R^2,\C):\  \ \inte |V_\Om(x)||u|^2dx<\infty\Big\}
\end{equation}
equipped with the norm
\begin{equation}\label{com2}
\|u\|_{H^1_{A,V_\Om}}=\Big(\inte\big|(\nabla-i\frac{\Om}{2} x^\perp)u\big|^2dx
                          +\inte \big(|V_\Om(x)|+1\big)|u|^2dx\Big)^{\frac{1}{2}},
\end{equation}
where $H^1_{A}(\R^2,\C)$ is the magnetic Sobolev space defined by $$H^1_{A}(\R^2,\C):=\Big\{\Big(\nabla-i\frac{\Om }{2}x^\perp\Big)u\in L^2(\R^2,\C),
\ \ u\in L^2(\R^2,\C)\Big\}.$$
Then $H^1_{A,V_\Om}\times H^1_{A,V_\Om}=\h\times \h$, and the embedding $H^1_{A,V_\Om}\times H^1_{A,V_\Om}\hookrightarrow
	L^{q}(\R^2,\C)\times L^{q}(\R^2,\C)$ is compact for all $2\leq q<\infty$.
\end{lem}

{\noindent \bf Proof.} Since we do not know whether the proof of Lemma \ref{Lem:compact} exists somewhere, we address it for the reader's convenience. Because it yields from \eqref{1.6condition1:V(x)} that $\lim\limits_{|x|\to\infty}V(x)=\infty$, we can deduce that $\h\hookrightarrow L^{q}(\R^2,\C)$ is compact for all $2\leq q<\infty$ (e.g. \cite [Theorem XIII.67] {RS} ). Therefore, in order to establish Lemma \ref{Lem:compact}, it suffices to show that $H^1_{A,V_\Om}=\h$.
Since $0<\Om<\Om^*$, using Cauchy inequality, we obtain from \eqref{main 1.7def:Omgea*} and (\ref{com2}) that
\begin{equation}\label{com3}
\begin{split}
\|u\|_{H^1_{A,V_\Om}}^2&=\inte\big|(\nabla-i\frac{\Om}{2} x^\perp)u\big|^2dx
                         +\inte \big(|V_\Om(x)|+1\big)|u|^2dx\\
&=\inte|\nabla u|^2dx-\Om\inte x^\perp\cdot(iu,\nabla u)dx
                     +\frac{\Om^2}{4}\inte |x|^2|u|^2dx\\
&\quad+\inte \big(|V_\Om(x)|+1\big)|u|^2dx\\
&\leq\frac{3}{2}\inte|\nabla u|^2dx+\frac{3\Om^2}{4}\inte |x|^2|u|^2dx
                   +\inte \big(V(x)+\frac{\Om^2}{4}|x|^2+1\big)|u|^2dx\\
&\leq\frac{3}{2}\inte|\nabla u|^2dx+C_{0}'(\Om)\inte|u|^2dx
                                   +C_{0}''(\Om)\inte \big(V(x)+1\big)|u|^2dx\\
&\leq C_{0}(\Om)\|u\|^2_\h,
\end{split}
\end{equation}
where the constants $C_{0}'(\Om)>0$, $C_{0}''(\Om)>0$ and $C_{0}(\Om)>0$.

We next address the reverse inequality of (\ref{com3}). Since $0<\Om<\Om^*$, the definition of $\Om^*$ in \eqref{main 1.7def:Omgea*} implies that there exists a constant $C_{1}'\big(\Om,\Om^*\big)>0$ such that
\begin{equation}\label{com5}
|x|^2\leq C_{1}'\big(\Om,\Om^*\big)\big(|V_{\Om}(x)|+1\big)\ \ \mbox{in}\ \ \R^2.
\end{equation}
Using Cauchy inequality again, we deduce from \eqref{com2} and \eqref{com5} that
\begin{equation}\label{com4}
\begin{split}
\|u\|_{H^1_{A,V_\Om}}^2&=\inte|\nabla u|^2dx-\Om\inte x^\perp\cdot(iu,\nabla u)dx
                                            +\frac{\Om^2}{4}\inte |x|^2|u|^2dx\\
&\quad+\inte \big(|V_\Om(x)|+1\big)|u|^2dx\\
&\geq\inte|\nabla u|^2dx-\Om\inte x^\perp\cdot(iu,\nabla u)dx
    +\inte \big(V(x)+1\big)|u|^2dx\\
&\geq\frac{1}{2}\inte|\nabla u|^2dx-\frac{\Om^2}{2}\inte |x|^2|u|^2dx
    +\inte \big(V(x)+1\big)|u|^2dx\\
&\geq\frac{1}{2}\inte|\nabla u|^2dx
    -C_{1}''\big(\Om,\Om^*\big)\inte \big(|V_{\Om}(x)|+1\big)|u|^2dx
    +\inte \big(V(x)+1\big)|u|^2dx,
\end{split}
\end{equation}
where $C_{1}''\big(\Om,\Om^*\big)>0$.
On the other hand, we obtain from  \eqref{com2} that
\begin{equation}\label{com6}
\|u\|_{H^1_{A,V_\Om}}^2=\inte\big|(\nabla-i\frac{\Om}{2} x^\perp)u\big|^2dx
                        +\inte \big(|V_\Om(x)|+1\big)|u|^2dx
                        \geq \inte \big(|V_\Om(x)|+1\big)|u|^2dx.
\end{equation}
Combining \eqref{com4} and \eqref{com6} yields that
\begin{equation}\label{com7}
\|u\|_{H^1_{A,V_\Om}}^2
\geq \frac{1}{1+C_{1}''\big(\Om,\Om^*\big)}\Big[\frac{1}{2}\inte|\nabla u|^2dx
                                           +\inte \big(V(x)+1\big)|u|^2dx\Big]
\geq C_{1}\big(\Om,\Om^*\big)\|u\|^2_\h,
\end{equation}
where $C_{1}\big(\Om,\Om^*\big)>0$.
It then follows from \eqref{com3} and \eqref{com7} that $H^1_{A,V_\Om}=\h$, and Lemma \ref{Lem:compact} is thus proved.
\qed

Applying Lemma \ref{Lem:compact},  we are now ready to prove Theorem  \ref{thm-1.1}
 under the general assumption \eqref{1.6condition1:V(x)}.
\vskip 0.05truein

\noindent {\bf Proof of Theorem \ref{thm-1.1}.}
\noindent{\bf 1.}
Suppose $0\leq \varphi \in C_0^\infty(\R^2)$ is a cutoff function satisfying $\varphi(x)=1$ for $|x|\leq 1$ and $\varphi(x)=0$ for $|x|\geq 2$. For any $\tau>0$, define
\begin{equation}\label{2.13def:trial}
\begin{split}
u_{1\tau}(x)&:=\sqrt{\theta}\frac{A_\tau\tau}{\| w\|_2}\varphi(x-x_0)
                 w(\tau|x-x_0|)e^{i\frac{\Om}{2}x\cdot x_0^\perp}, \\[3mm]
u_{2\tau}(x)&:=\sqrt{1-\theta}\frac{A_\tau\tau}{\| w\|_2}\varphi(x-x_0)
                 w(\tau|x-x_0|)e^{i\frac{\Om}{2}x\cdot x_0^\perp},
\end{split}
\end{equation}
where $x_0\in\R^2$ and $\theta \in [0,1]$ are to be determined later, and $A_{\tau}>0$ is chosen such that $\int_{\R^2}\big(|u_{1\tau}(x)|^2+|u_{2\tau}(x)|^2\big)dx=1$. Applying \eqref{1.9identity:w} and the exponential decay \eqref{1.10decay:w} of $w$, direct calculations give that as $\tau\to\infty$,
\begin{equation}\label{2.14sup:F.w}
\begin{split}
e(\Om,\f)&\le  F_{\Om,\f}(u_{1\tau},u_{2\tau}) \\
&\leq\frac{A_\tau^2 \tau^2}{\|w\|_2^2}\Big[
 \inte|\nabla w|^2dx-\frac{a_1\theta^2 +a_2 (1-\theta)^2+2\beta\theta(1-\theta)}
      {2a^*}\inte w^4 dx\Big]  \\
&\quad+\frac{A_\tau^2}{\|w\|_2^2}
          \inte\Big[V_{\Om}\Big(\frac{x}{\tau}+x_0\Big)
                   +\frac{\Om^2|x|^2}{4\tau^2} \Big]w^2 dx+Ce^{-2\tau}\\
&=\tau^2\Big[1-\frac{a_1\theta^2+a_2 (1-\theta)^2
             +2\beta\theta(1-\theta)}{a^*}\Big]+V_{\Om}(x_0)+o(1).
\end{split}
\end{equation}

If $\Om>\Om ^* $, then the definition of $\Om ^*$ implies that
$\inf\limits_{x\in\R^2}V_\Om(x)=-\infty$. For
any fixed and sufficiently large $\tau >0$, take a point $x_0\in\R^2$ such that
$V_\Om(x_0)\leq-2\tau ^2$ and take $\theta=\frac{1}{2}$. We then derive from \eqref{2.14sup:F.w} that for any $a_1>0$, $a_2>0$ and $\B>0$,
\begin{align*}
F_{\Om,\f}(u_{1\tau},u_{2\tau})
&\le V_\Om(x_0)
+\tau^2\Big(1-\frac{\frac{a_1}{4}+\frac{a_2}{4}+\frac{\B}{2}}{a^*}\Big)+o(1)
\  \ \text{as}\  \ \tau\to\infty.
\end{align*}
Since $\tau >0$ is arbitrary, the above estimate implies that $e(\Om,\f)$ is unbounded from below. Therefore,  there is no minimizer of $e(\Om,\f)$ for the case where $\Om >\Om ^*$, $a_1>0$, $a_2>0$ and $\B>0$.
\vskip 0.05truein

\noindent{\bf 2.} For any given $0<a_1,a_2< a^*$ and $0<\B<\B^*$, there exists a constant $a>0$ such that $$\max\{a_1,a_2\}\leq a<a^*\  \ \text{and}\  \ \B<a+\sqrt{(a-a_1)(a-a_2)}.$$
We thus rewrite \eqref{1.4energyfunctional} as
\begin{equation}\label{2.1trans2:F}
\begin{split}
F_{\Om,\f}(u_1,u_2)&=\sum_{j=1}^2
\inte\Big(\Big|\big(\nabla-i\frac{\Om}{2}x^\perp\big)u_j\Big|^2+V_\Om(x)|u_j|^2\Big)dx\\
&\quad-\frac{a}{2}\inte(|u_1|^2+|u_2|^2)^2dx
      +\frac{1}{2}\inte \big(\sqrt{a-a_1}|u_1|^2-\sqrt{a-a_2}|u_2|^2\big)^2dx\\
&\quad+\big(a+\sqrt {(a-a_1)(a-a_2)}-\B\big) \inte|u_1|^2|u_2|^2 dx.
\end{split}
\end{equation}
Since $0<\Om<\Om^*$, applying the coupled Gagliardo-Nirenberg  inequality \eqref{2.2Ineq:GN} and the diamagnetic inequality \eqref{2.3ineq:diam}, we derive from \eqref{2.1trans2:F} that for any sufficiently large $R>0$,
\begin{equation}\label{2.2trans2:F-1}
\begin{split}
&\quad F_{\Om,\f}(u_{1},u_{2})\\
&\geq\sum_{j=1}^2\Big[
  \Big(1-\frac{a}{a^*}\Big)\inte\Big|\big(\nabla-i\frac{\Om}{2}x^\perp\big)u_{j}\Big|^2dx
 +\inte V_\Om(x)|u_{j}|^2dx\Big]\\
&\geq\Big(1-\frac{a}{a^*}\Big)\sum_{j=1}^2
\Big[\inte\Big|\big(\nabla-i\frac{\Om}{2}x^\perp\big)u_{j}\Big|^2dx
     +\inte \big|V_\Om(x)\big||u_{j}|^2dx\Big]-C(\Om,R),
\end{split}
\end{equation}
which implies that $F_{\Om,\f}(u_{1},u_{2})$ is bounded from below for all $0<a_1,a_2< a^*$ and $0<\B<\B^*$.
Let $\{(u_{1n},u_{2n})\}\in \m$ be a minimizing sequence of $e(\Om,\f)$. From Lemma \ref{Lem:compact}, we obtain that $H^1_{A,V_\Om}\times H^1_{A,V_\Om}=\h\times \h$, which implies that the minimizing sequence $\{(u_{1n},u_{2n})\}$ is also in $H^1_{A,V_\Om}\times H^1_{A,V_\Om}$.

We then derive from \eqref{2.2trans2:F-1} that the sequence $\{(u_{1n},u_{2n})\}$ is bounded uniformly in $H^1_{A,V_\Om}\times H^1_{A,V_\Om}$. Applying the compactness of Lemma \ref{Lem:compact}, there exist a subsequence, still denoted by $\{(u_{1n},u_{2n})\}$, of $\{(u_{1n},u_{2n})\}$ and $(u_{10},u_{20})\in H^1_{A,V_\Om}\times H^1_{A,V_\Om}$ such that
\begin{equation*}
(u_{1n},u_{2n}) \stackrel{n}{\rightharpoonup} (u_{10},u_{20})
\  \ \hbox {weakly in}\  \  H^1_{A,V_\Om}\times H^1_{A,V_\Om},
\end{equation*}
and
\begin{equation*}\label{2.11:lim:uin.Lq}
(u_{1n},u_{2n}) \xrightarrow{n} (u_{10},u_{20})
\  \ \hbox {strongly in}\  \  L^q(\R^2,\mathbb{C})\times L^{q}(\R^2,\mathbb{C}),
\  \ \hbox {where}\  \  2\leq q<\infty.
\end{equation*}
Therefore, by the weak lower semicontinuity, we conclude that $\inte\big(|u_{10}|^2+|u_{20}|^2 \big)dx=1$ and
$F_{\Om,\f}(u_{10},u_{20})= e(\Om,\f)$,
which further implies that $(u_{10},u_{20})$ is a minimizer of $e(\Om,\f)$ for any given $0<\Om <\Om ^*$, $0<a_1,a_2 < a^*$ and $0<\beta<\beta^*$.
\vskip 0.05truein

\noindent{\bf 3.}
For $0<\Om<\Om ^* $ and $a_1>a^*$, we take the test function \eqref{2.13def:trial} with $\theta=1$.
It then folows from \eqref{2.14sup:F.w} that
\begin{equation*}
e(\Om,\f)\leq -C\tau^2 \  \ \text{as} \  \ \tau\to\infty,
\end{equation*}
where $C>0$ and $x_0\in\R^2$ in \eqref{2.13def:trial} can be chosen arbitrarily.
This implies that $e(\Om,\f)$ is unbounded from below and hence there is no minimizer of $e(\Om,\f)$ in the case where $0<\Om<\Om ^* $ and $a_1>a^*$. Similarly, taking the test function \eqref{2.13def:trial} with $\theta=0$ if $0<\Om<\Om ^* $ and $a_2>a^*$, and taking the test function \eqref{2.13def:trial} with $\theta=\frac{\B-a_2}{2\B-a_1-a_2}$ if $0<\Om<\Om ^* $ and $\B>\B^*$,  the same argument as above then gives that there is no minimizer of $e(\Om,\f)$.
This completes the proof of Theorem \ref{thm-1.1}.
\qed

If $0< a_1,a_2\le a^*:=\|w\|_2^2$, then one can rewrite the energy functional $F_{\Om,\f}(\cdot,\cdot)$ as
\begin{equation}\label{1.12trans1:F}
\begin{split}
&F_{\Om,\f}(u_1,u_2)\\
=& \int_{\R ^2} \Big(\big|(\nabla -i\frac{\Om}{2}x^{\perp})u_1\big|^2+\big|(\nabla -i\frac{\Om}{2}x^{\perp})u_2\big|^2\Big) dx-\frac{a^*}{2}\int_{\R^2}\big(|u_1|^2+|u_2|^2\big)^2 dx \\
&+\int_{\R ^2}V_\Om(x)\big(|u_1|^2+|u_2|^2\big) dx
+\frac{1}{2}\int_{\R ^2} \big(\sqrt{a^*-a_1}|u_1|^2-\sqrt{a^*-a_2}|u_2|^2\big)^2 dx\\
&+(\B^*-\B) \int_{\R ^2}|u_1|^2|u_2|^2 dx,
\end{split}
\end{equation}
where $V_\Om(x)=V(x)-\frac{\Om^2}{4}|x|^2$, and $\beta^*\ge 0$ is defined by (\ref{main 1.11def:beta*}).
Applying (\ref{1.12trans1:F}), we next prove Theorem \ref{thm-1.2} on the nonexistence of minimizers under the general assumption \eqref{1.6condition1:V(x)}.
\vskip 0.05truein

\noindent {\bf Proof of Theorem \ref{thm-1.2}.} Let $(u_{1\tau}, u_{2\tau})$ be the same test function as that of \eqref{2.13def:trial}. We then have the same estimate  \eqref{2.14sup:F.w} of $e(\Om,\f)$, where $0<\Om<\Om^*$ is fixed.
\vskip 0.05truein

\noindent{\bf 1.}  We first consider the case where $a_1=a^*$, $a_2\leq a^*$ and $\B\leq \B^*=a^*$.
 Take $\theta=1$ and $x_0\in\R^2$ such that
\begin{equation}\label{2.15con.test.b1}
V_{\Om}(x_0)=\inf\limits_{x\in\R^2}V_{\Om}(x).
\end{equation}
We then derive from \eqref{2.14sup:F.w} that
\begin{equation*}
e(\Om,a^*,a_2,\B)\leq \inf\limits_{x\in\R^2}V_{\Om}(x).
\end{equation*}
On the other hand,  by the coupled Gagliardo-Nirenberg inequality \eqref{2.2Ineq:GN} and the diamagnetic inequality \eqref{2.3ineq:diam}, we obtain from \eqref{1.12trans1:F} that for any $(u_{1},u_{2})\in\m$,
\begin{equation}\label{2.16}
\begin{split}
  F_{\Om,a^*,a_2,\B}(u_1,u_2)
&\geq\inte\big(\big|\nabla |u_1|\big|^2+\big|\nabla |u_2|\big|^2\big)dx
     -\frac{a^*}{2}\inte\big(|u_1|^2+|u_2|^2\big)^2dx\\
&\quad+\inte V_\Om(x)\big(|u_1|^2+|u_2|^2\big)dx\\
&\geq\inf\limits_{x\in\R^2}V_{\Om}(x),
\end{split}
\end{equation}
which gives that $e(\Om,a^*,a_2,\B)\geq\inf\limits_{x\in\R^2}V_{\Om}(x)$. We thus have
\begin{equation}\label{2.17val:e0}
e(\Om,a^*,a_2,\B)=\inf\limits_{x\in\R^2}V_{\Om}(x).
\end{equation}
Suppose now that there exists a minimizer $(u_1^*,u_2^*)$ of $e(\Om,a^*,a_2,\B)$.  We then deduce from \eqref{2.16} that
\begin{equation}\label{2.18equality:e0}
e(\Om,a^*,a_2,\B)=F_{\Om,a^*,a_2,\B}(u^*_1,u^*_2)>\inf_{x\in\R^2}V_\Om(x),
\end{equation}
which however contradicts with \eqref{2.17val:e0}. Therefore, there is no minimizer of $e(\Om,\f)$ in this case. Similarly, one can prove the nonexistence of minimizers in the case where $a_2=a^*$, $a_1\leq a^*$ and $\B\leq \B^*=a^*$.

When $a_2<a^*$, $\B\leq a^*$ and $a_1\nearrow a^*$, take $\theta=1$, $\tau=(a^*-a_1)^{-\frac{1}{4}}$ and $x_0\in\R^2$ satisfying \eqref{2.15con.test.b1}. Applying \eqref{2.14sup:F.w} and \eqref{2.16}, we then obtain from \eqref{2.17val:e0} that
\begin{equation}\label{2.20lim:e.0}
\begin{split}
\inf\limits_{x\in\R^2}V_{\Om}(x)\leq\lim\limits_{a_1\nearrow a^*} e(\Om,\f)
&\leq\lim\limits_{a_1\nearrow a^*}F_{\Om,\f}(u_{1\tau},u_{2\tau})\\
&\leq\inf\limits_{x\in\R^2}V_{\Om}(x)=e(\Om,a^*,a_2,\B).
\end{split}
\end{equation}
Similarly, one can prove that $\lim\limits_{a_2\nearrow a^*} e(\Om,\f)=e(\Om,a_1,a^*,\B)=\inf\limits_{x\in\R^2}V_{\Om}(x)$ for $a_1< a^*$ and $0<\B\leq a^*$.
\vskip 0.05truein

\noindent{\bf 2.} For $\B= \B^*$ and $0<a_1,a_2<a^*$, take $x_0\in\R^2$ satisfying \eqref{2.15con.test.b1} and $\theta=\frac{\sqrt{a^*-a_2}}{\sqrt{a^*-a_1}+\sqrt{a^*-a_2}}$ for \eqref{2.14sup:F.w}. We then deduce from \eqref{2.14sup:F.w} that
\begin{equation}\label{2.21sup:E.V}
e(\Om,a_1,a_2,\B^*)\leq \inf\limits_{x\in\R^2}V_{\Om}(x).
\end{equation}
Hence, we conclude that  $e(\Om, a_1, a_2, \B^*)=\inf\limits_{x\in\R^2}V_{\Om}(x)$ in view of \eqref{1.12trans1:F}.
By the same argument of proving \eqref{2.18equality:e0}, we then further derive that $e(\Om, a_1, a_2, \B^*)$ has no minimizer in this case.

Furthermore, choose $x_0\in\R^2$ as above, and take $\theta=\frac{\sqrt{a^*-a_2}}{\sqrt{a^*-a_1}+\sqrt{a^*-a_2}}$ and $\tau=(\B^*-\B)^{-\frac{1}{4}}$. Similar to \eqref{2.20lim:e.0}, we then derive from \eqref{2.14sup:F.w} that $\lim\limits_{\B\nearrow \B^*} e(\Om,\f)=e(\Om,a_1,a_2,\B^*)=\inf\limits_{x\in\R^2}V_{\Om}(x)$.
This therefore completes the proof of Theorem \ref{thm-1.2}.
\qed

\section{Limiting Behavior of Minimizers as $\B\nearrow \B^*$}

Starting from this section, we always consider the harmonic trap $V(x)$ satisfying \eqref{V(x)} for some $\Lambda\ge1$. By employing the energy method and blow-up analysis, this section is mainly devoted to establishing Theorem \ref{thm-1.3} below on the limiting behavior of minimizers for $e(\Om,\f)$ as $\B\nearrow\B^*$.

Under the harmonic trap \eqref{V(x)}, where $\Lambda\ge1$, one can note from \eqref{main 1.7def:Omgea*} that $e(\Om,\f)$ admits a critical velocity $\Om^*:=2$. For any fixed $0<\Omega<\Omega ^*:=2$, it gives that
\begin{equation}\label{Vomx ge 0}
 V_\Omega (x):=V(x)-\frac{\Omega^2}{4} |x|^2\ge 0\ \ \hbox{in}\ \ \R^2,
\end{equation}
and
\begin{equation}\label{1.14nondege}
\det\Big(\inte\frac{\partial V_\Om(x)}{\partial x_j}
        \frac{\partial w^2(x)}{\partial x_l}dx\Big)_{j,l=1,2}\neq0.
\end{equation}
Recall from \eqref{1.1-1equation system} and \eqref{1.5Lmu} that any minimizer $(u_{1\B},u_{2\B})$ of $e(\Om,\f)$ satisfies the following Euler-Lagrange system
\begin{equation}\label{Sec 3 1.1-1equation system}
\left\{\begin{array}{lll}
-\Delta u_{1\B}+V(x)u_{1\B}+i\, \Om \big(x^{\perp}\cdot \nabla u_{1\B}\big)
=\mu_\B u_{1\B}+a_1|u_{1\B}|^2u_{1\B}+\B |u_{2\B}|^2u_{1\B}
\  \ \hbox{in}\  \ \R^2,\\[3mm]
-\Delta u_{2\B}+V(x)u_{2\B}+i\, \Om \big(x^{\perp}\cdot \nabla u_{2\B}\big)
=\mu_\B u_{2\B}+a_2|u_{2\B}|^2u_{2\B}+\B |u_{1\B}|^2u_{2\B}
\  \ \hbox{in}\  \ \R^2,
\end{array}\right.
\end{equation}
where $\mu_\B\in\R$ is a suitable Lagrange multiplier satisfying
\begin{equation}\label{Sec 3 1.5Lmu}
\mu_\B=e(\Om,\f)-\inte
\Big(\frac{a_1}{2}|u_{1\B}|^4+\frac{a_2}{2}|u_{2\B}|^4+\B|u_{1\B}|^2|u_{2\B}|^2\Big)dx.
\end{equation}
For convenience, we denote
\begin{equation}\label{1.16def:beta.V.eps}
\alpha_\B:=\Big[\frac{2\gam_1\gam_2}{\lam_0}(\B^*-\B)\Big]^\frac{1}{4}>0,\ \ 0<\B<\B^*,
\end{equation}
where $\B^*>0$ is as in \eqref{main 1.11def:beta*}, $\gam_j>0$ is given by
\begin{equation}\label{1.13gamma}
\gam_j:=1-\frac{\sqrt{a^*-a_j}}{\sqrt{a^*-a_1}+\sqrt{a^*-a_2}}\in(0,1),
 \ \ j=1,2,
\end{equation}
and $\lam_0>0$ is defined by
\begin{equation}\label{1.14lam0}
\lam_0:=\ds\inte V(x)w^2(x)dx.
\end{equation}
Using above notations, the limiting behavior of minimizers for $e(\Om,\f)$ as $\B\nearrow\B^*$ can be stated as follows.

\begin{thm}\label{thm-1.3}
Suppose $V(x)$ satisfies \eqref{V(x)} for some $\Lambda\ge1$, and assume $0<\Om<\Om^*:=2$. Let $(u_{1\B},u_{2\B})$ be a minimizer of $e(\Om,\f)$, where $0< a_1,a_2<a^*$ and $0<\B<\B^*:=a^*+\sqrt{(a^*-a_1)(a^*-a_2)}$. Then we have
\begin{equation}\label{eps2mu}
\alpha_\B^2\mu_\B=-1+o(1)\  \ \text{as}\  \ \B\nearrow\B^*,
\end{equation}
and for $j=1,2$,
\begin{equation}\label{1.15lim:beta.V.u.exp}
w_{j\B}(x):=\sqrt{a^*}\alpha_\B u_{j\B}(\alpha_\B x)e^{i\theta_{j\B}}
\to\sqrt{\gam_j}w(x)
\ \ \hbox{strongly in}\ \ H^1(\R^2,\C)\cap L^\infty(\R^2,\C)
   \end{equation}
as $\B\nearrow\B^*$, where $\mu_\B\in\R$ satisfies \eqref{Sec 3 1.5Lmu}, $\alpha_\B>0$ is given by \eqref{1.16def:beta.V.eps}, $0<\gamma_j<1$ is as in \eqref{1.13gamma}, and $\theta_{j\B}\in [0,2\pi)$ is a properly chosen constant.
\end{thm}

In order to prove Theorem \ref{thm-1.3}, we first establish the following energy bounds of $e(\Om,\f)$ as $\B\nearrow\B^*$:

\begin{lem}\label{lem-3.2}
Suppose $V(x)$ satisfies \eqref{V(x)} for some $\Lambda\ge1$, and assume  $0<\Om<\Om^*:=2$. Then for any given $0<a_1,a_2<a^*$, we have
\begin{equation}\label{3.5}
0\leq e(\Om,\f)
 \leq\frac{2}{a^*}\Big[2\gam_1\gam_2\lam_0(\B^*-\B)\Big]^{\frac{1}{2}}
\big(1+o(1)\big)
 \  \ \hbox{as}\  \ \B\nearrow\B^*,
\end{equation}
where $0<\gam_1,\gam_2<1$ are as in \eqref{1.13gamma}, and $\lam_0>0$ is defined by \eqref{1.14lam0}.
\end{lem}

{\noindent \bf Proof.} By the coupled Gagliardo-Nirenberg inequality \eqref{2.2Ineq:GN} and the diamagnetic inequality \eqref{2.3ineq:diam}, we derive from \eqref{1.12trans1:F} and \eqref{Vomx ge 0} that for any $(u_{1},u_{2})\in\m$,
\begin{align*}
F_{\Om,\f}(u_1,u_2)
&\geq\inte\Big(\big|\nabla |u_1|\big|^2+\big|\nabla |u_2|\big|^2\Big) dx
       -\frac{a^*}{2}\inte\big(|u_1|^2+|u_2|^2\big)^2 dx \\
&\quad+\inte V_\Om(x)\big(|u_1|^2+|u_2|^2\big) dx
      +\frac{1}{2}\inte \big(\sqrt{a^*-a_1}|u_1|^2-\sqrt{a^*-a_2}|u_2|^2\big)^2 dx\\
&\quad+(\B^*-\B)\inte|u_1|^2|u_2|^2 dx
\geq 0,
\end{align*}
which gives the lower bound of \eqref{3.5}.

We next  establish the upper bound of \eqref{3.5}. Choose a test function  $(u_{1\tau},u_{2\tau})$ of the form \eqref{2.13def:trial} with  $x_0=0
$ and $\theta=\gam_1 $, where $\gam_1 \in (0, 1)$ is as in \eqref{1.13gamma}. 
Direct calculations then give that
\begin{equation}\label{3.6}
\begin{split}
e(\Om,\f)&\leq F_{\Om,\f}(u_{1\tau},u_{2\tau})\\
&= \inte\Big(|\nabla u_{1\tau}|^2+|\nabla u_{2\tau}|^2\Big)dx
+\inte V(x)\big(|u_{1\tau}|^2+|u_{2\tau}|^2\big) dx \\
&\quad-\inte\Big(\frac{a_1}{2}|u_{1\tau}|^4+\frac{a_2}{2}|u_{2\tau}|^4
                 +\B|u_{1\tau}|^2|u_{2\tau}|^2\Big)dx\\
&\quad -\Om\inte\Big[x^\perp\cdot\big(iu_{1\tau},\nabla u_{1\tau}\big)
+x^\perp\cdot\big(iu_{2\tau},\nabla u_{2\tau}\big)\Big]dx \\
&\leq \frac{1}{a^*\tau^2}\inte V(x)w^2(x)dx
+\frac{2\gam_1\gam_2}{a^*}\tau^2(\B^*-\B)+C\tau^2e^{-2\tau}\\
&= \frac{\lam_0\big(1+o(1)\big)}{a^*\tau^{2}}+\frac{2\gam_1\gam_2}{a^*}\tau^2(\B^*-\B)
\ \ \text{as} \ \ \tau\to\infty,
\end{split}
\end{equation}
where $\lambda_0>0$ is as in  \eqref{1.14lam0}.
Substituting $\tau=\ds\Big[\frac{\lam_0}{2\gam_1\gam_2(\B^*-\B)}\Big]^{\frac{1}{4}}$ into \eqref{3.6} then yields the upper bound of \eqref{3.5}, and this completes the proof of Lemma \ref{lem-3.2}. \qed

\subsection{Blow-up analysis}
To prove Theorem \ref{thm-1.3}, the main purpose of this subsection is to investigate the limiting
behavior of minimizers for $e(\Om,\f)$ as $\B\nearrow\B^*$.
We start with the following lemma.

\begin{lem}\label{lem-3.3}
Suppose $V(x)$ satisfies \eqref{V(x)} for some $\Lambda\ge1$, and assume  $0<\Om<\Om^*:=2$. Let $(u_{1\B},u_{2\B})$ be a minimizer of $e(\Om,\f)$, where $0< a_1,a_2<a^*$ and $0<\B<\B^*:=a^*+\sqrt{(a^*-a_1)(a^*-a_2)}$. Then we have
\begin{enumerate}
\item $(u_{1\B},u_{2\B})$ blows up  as $\B\nearrow\B^*$, in the sense that
\begin{equation}\label{3.8}
\lim_{\B \nearrow\B^*}\inte |\nabla u_{1\B}|^2dx
=+\infty,\  \ \lim_{\B \nearrow\B^*}\inte |\nabla u_{2\B}|^2dx=+\infty,
\end{equation}
and
\begin{equation}\label{3.9}
\lim_{\B \nearrow\B^*}\inte | u_{1\B}|^4dx=+\infty,\  \ \lim_{\B \nearrow\B^*}\inte |u_{2\B}|^4dx=+\infty.
\end{equation}
\item $(u_{1\B},u_{2\B})$ also satisfies
\begin{equation}\label{3.10}
\lim_{\B \nearrow\B^*}\inte V_{\Om}(x)|u_{1\B}|^2dx=\lim_{\B \nearrow\B^*}\inte V_{\Om}(x)| u_{2\B}|^2dx=0,
\end{equation}
\begin{equation}\label{3.11}
\lim\limits_{\B\nearrow\B^*}\int_{\R^2}\Big(\sqrt{a^*-a_1}|u_{1\B}|^2
-\sqrt{a^*-a_2}|u_{2\B}|^2\Big)^2dx=0,
\end{equation}
\begin{equation}\label{3.12}
\lim\limits_{\B\nearrow\B^*}
\frac{\inte\big(|\nabla u_{1\B}|^2+|\nabla u_{2\B}|^2\big)dx}
{\int_{\R^2}\big(| u_{1\B}|^2+|u_{2\B}|^2\big)^2dx}=\frac{a^*}{2},
\end{equation}
and
\begin{equation}\label{3.13}
\lim\limits_{\B\nearrow\B^*}
\frac{\int_{\R^2}|u_{1\B}|^4dx}{\int_{\R^2}|u_{2\B}|^4dx}
= \frac{{a^*-a_2}}{{a^*-a_1}}.
\end{equation}
\end{enumerate}
\end{lem}

\noindent{\bf Proof.} It first follows from Lemma \ref{lem-3.2} that
\begin{equation}\label{3.5-1}
\lim\limits_{\B\nearrow\B^*} e(\Om,\f)=0.
\end{equation}
Using the coupled Gagliardo-Nirenberg inequality \eqref{2.2Ineq:GN} and the diamagnetic inequality \eqref{2.3ineq:diam}, we derive from \eqref{1.12trans1:F}, \eqref{Vomx ge 0} and \eqref{3.5-1} that
\begin{equation*}
\begin{split}
0=\lim\limits_{\B\nearrow\B^*} e(\Om,\f)
&=\lim\limits_{\B\nearrow\B^*}F_{\Om,\f}(u_{1\B},u_{2\B})\\
&\geq\lim\limits_{\B\nearrow\B^*}
\Big[\inte V_\Om(x)\big(|u_{1\B}|^2+|u_{2\B}|^2\big)dx\\
&\qquad\qquad+\frac{1}{2}\inte \big(\sqrt{a^*-a_1}|u_{1\B}|^2-\sqrt{a^*-a_2}|u_{2\B}|^2\big)^2 dx\Big]\geq0,
\end{split}
\end{equation*}
which implies that both \eqref{3.10} and \eqref{3.11} hold true.

Moreover, using Lemmas \ref{Lem:compact} and \ref{lem-3.2}, the same argument of proving \cite[Lemma 3.2 (1)]{GLY} gives that
\begin{equation}\label{3.8-1}
\lim_{\B \nearrow\B^*}\inte \big(|\nabla u_{1\B}|^2+|\nabla u_{2\B}|^2\big)dx=+\infty.
\end{equation}
By the definition of $\Om^*$ in \eqref{main 1.7def:Omgea*}, we obtain that there exists a constant $C(\Om)>0$ such that
\begin{equation}\label{3.15-1}
|x|^2\leq C(\Om)\big[1+V_{\Om}(x)\big]\ \ \mbox{in}\ \ \R^2.
\end{equation}
By Cauchy inequality and \eqref{3.10}, we deduce from  \eqref{3.8-1} and \eqref{3.15-1} that
    \begin{equation}\label{3.12-2}
     \lim\limits_{\B\nearrow\B^*}
     \frac{\inte\big[|(\nabla -i\frac{\Om}{2}x^{\perp})u_{1\B}|^2+|(\nabla -i\frac{\Om}{2}x^{\perp})u_{2\B}|^2\big]dx}
     {\inte\big(|\nabla u_{1\B}|^2+|\nabla u_{2\B}|^2\big)dx}=1.
    \end{equation}
We obtain from \eqref{3.8-1} and \eqref{3.12-2} that
\begin{equation}\label{3.8-2}
\lim_{\B \nearrow\B^*}
\inte\Big(\Big|(\nabla -i\frac{\Om}{2}x^{\perp})u_{1\B}\Big|^2
         +\Big|(\nabla -i\frac{\Om}{2}x^{\perp})u_{2\B}\Big|^2\Big)dx=+\infty.
\end{equation}
Using \eqref{3.10}, \eqref{3.11}, \eqref{3.5-1} and \eqref{3.8-2}, we then derive from \eqref{1.12trans1:F} that
    \begin{equation}\label{3.12-1}
     \lim\limits_{\B\nearrow\B^*}
     \frac{\inte\big[|(\nabla -i\frac{\Om}{2}x^{\perp})u_{1\B}|^2+|(\nabla -i\frac{\Om}{2}x^{\perp})u_{2\B}|^2\big]dx}
     {\inte\big(| u_{1\B}|^2+|u_{2\B}|^2\big)^2dx}=\frac{a^*}{2}.
    \end{equation}
Combining \eqref{3.12-2} and \eqref{3.12-1} then yields that \eqref{3.12} holds true.

We next prove \eqref{3.9}. We derive from \eqref{3.12} and \eqref{3.8-1} that
\begin{equation*}
\lim_{\B \nearrow\B^*}\inte \big(|u_{1\B}|^4+|u_{2\B}|^4\big)dx=+\infty.
\end{equation*}
Suppose now that \eqref{3.9} is false. Without loss of generality, we may assume that
\begin{equation}\label{3.9-2}
\inte|u_{1\B}|^4dx\leq C\ \ \hbox{uniformly in}\ \ \B \ \ \hbox{and}\ \
 \lim_{\B \nearrow\B^*}\inte|u_{2\B}|^4dx=+\infty.
\end{equation}
By H\"{o}lder inequality, we derive from \eqref{3.11} that
\begin{equation}\label{3.9-3}
\lim\limits_{\B\nearrow\B^*}
\Big(\sqrt{a^*-a_1}\|u_{1\B}\|_4^2-\sqrt{a^*-a_2}\|u_{2\B}\|_4^2\Big)^2
=0.
\end{equation}
Using \eqref{3.9-2}, we then deduce from \eqref{3.9-3} that
\begin{align*}
0&=a^*-a_2
+(a^*-a_1)\lim_{\B \nearrow\B^*}\frac{\|u_{1\B}\|^4_4}{\|u_{2\B}\|^4_4}
-2\sqrt{(a^*-a_1)(a^*-a_2)}\lim_{\B \nearrow\B^*}\frac{\|u_{1\B}\|^2_4}{\|u_{2\B}\|^2_4}\\
&=a^*-a_2>0,
\end{align*}
which is a contradiction, and \eqref{3.9} is thus proved.

Furthermore, using the Gagliardo-Nirenberg inequality \eqref{2.1ineq:GNQ},
it then follows from \eqref{3.9} that \eqref{3.8} holds true. Finally, one can deduce from \eqref{3.9} and \eqref{3.9-3} that \eqref{3.13} holds true, and we are done.
\qed

 Following Lemma \ref{lem-3.2} and \eqref{3.9}, one can derive that the Lagrange multiplier $\mu_\B\in\R$ in \eqref{Sec 3 1.5Lmu} satisfies $\mu_\B\to-\infty$ as $\B\nearrow\B^*$.
 Define
\begin{equation}\label{3.2}
\eps_\B:=\sqrt{\frac{1}{-\mu_\B}}>0\ \ \hbox{as}\ \ \B\nearrow\B^*,
\end{equation}
so that
\begin{equation}\label{3.19}
\eps_{\B}\to 0\ \ \hbox{as}\  \ \B \nearrow \B^*.
\end{equation}
By applying Lemma \ref{lem-3.3}, \eqref{3.2} and \eqref{3.19}, we now study the following refined estimates of minimizers $(u_{1\B},u_{2\B})$ for $e(\Om,\f)$ as $\B\nearrow\B^*$.


\begin{lem}\label{lem-3.4}
Under the assumptions of Theorem \ref{thm-1.3}, let $(u_{1\B},u_{2\B})$ be a minimizer of $e(\Om,\f)$. Define
\begin{equation}\label{Sec3 v1B v2B}
v_{j\B}(x):=\sqrt{a^*}\eps_\B u_{j\B}(\eps_\B x)
e^{i\widetilde{\theta}_{j\B}}
=R_{j\B}(x)+iI_{j\B}(x),\ \ j=1,2,
\end{equation}
where $\eps_\B>0$ is as in \eqref{3.2}, $R_{j\B}(x)$ and
$I_{j\B}(x)$ denote the real and imaginary parts of
$v_{j\B}(x)$, respectively, and the constant $\widetilde{\theta}_{j\B}\in [0,2\pi)$ is properly chosen such that
\begin{equation}\label{3.49.tildew}
\inte w(x)I_{j\B}(x)dx=0, \  \  j=1,2.
\end{equation}
Then we have
\begin{enumerate}
\item There exist constants $R_0>0$ and $\delta>0$ such that
\begin{equation}\label{3.21}
 \liminf_{\B\nearrow\B^*}\int_{B_{2R_0}(0)}|v_{j\B}(x)|^2dx\geq \delta>0,\  \ j=1,2.
\end{equation}
\item $(v_{1\B},v_{2\B})$ satisfies
\begin{equation}\label{3.23}
     v_{j\B}(x)\to\ds\sqrt{\gam_j}w(x)
     \ \ \text{strongly in}\  \ H^1(\R^2,\C)\  \ \text{as}\  \ \B\nearrow\B^*,\ \ j=1,2,
\end{equation}
where $0<\gam_j<1$ is given by \eqref{1.13gamma}.
\end{enumerate}
\end{lem}

\noindent{\bf Proof.}
We note that the condition \eqref{3.49.tildew} can be satisfied by the similar argument of \cite [Lemma 3.2] {GLY}.

1. We first claim that
\begin{equation}\label{lem3.4 2-1}
\lim\limits_{\B\nearrow\B^*}\inte\big(\sqrt{a^*-a_1}|v_{1\B}|^2
-\sqrt{a^*-a_2}|v_{2\B}|^2\big)^2dx=0,
\end{equation}
\begin{equation}\label{lem3.4 2-2}
\lim\limits_{\B\nearrow\B^*}
\inte\big(| v_{1\B}|^2+|v_{2\B}|^2\big)^2dx=2a^*,
\end{equation}
and
\begin{equation}\label{lem3.4 2-3}
\lim\limits_{\B\nearrow\B^*}
\inte\big(|\nabla v_{1\B}|^2+|\nabla v_{2\B}|^2\big)dx
=a^*.
\end{equation}
Indeed, \eqref{lem3.4 2-1} follows directly from \eqref{3.11}, \eqref{3.19} and \eqref{Sec3 v1B v2B}.
Using \eqref{Sec 3 1.5Lmu}, we obtain from \eqref{3.2} and \eqref{Sec3 v1B v2B} that
\begin{equation}\label{lem3.4 2-4}
\begin{split}
-1&=\eps_\B^2 e(\Om,\f)-\eps_\B^2\inte\Big(\frac{a_1}{2}|u_{1\B}|^4
+\frac{a_2}{2}|u_{2\B}|^4+\B|u_{1\B}|^2|u_{2\B}|^2\Big)dx\\
&=\eps_\B^2 e(\Om,\f)-\frac{1}{(a^*)^2}\inte\Big(\frac{a_1}{2}|v_{1\B}|^4
+\frac{a_2}{2}|v_{2\B}|^4+\B|v_{1\B}|^2|v_{2\B}|^2\Big)dx.
\end{split}
\end{equation}
By Lemma \ref{lem-3.2}, we derive from \eqref{3.19} and \eqref{lem3.4 2-4} that as $\B\nearrow\B^*$,
\begin{equation}\label{lem3.4 2-5}
\inte\frac{a^*}{2}\big(|v_{1\B}|^2+|v_{2\B}|^2\big)^2dx
-\frac{1}{2}\inte\big(\sqrt{a^*-a_1}|v_{1\B}|^2
-\sqrt{a^*-a_2}|v_{2\B}|^2\big)^2dx\to(a^*)^2.
\end{equation}
Combining \eqref{lem3.4 2-1} and \eqref{lem3.4 2-5} then yields that
\eqref{lem3.4 2-2} is true. Furthermore,  we obtain from \eqref{3.12} and \eqref{Sec3 v1B v2B} that
\begin{equation}\label{lem3.4 2-6}
\lim\limits_{\B\nearrow\B^*}
\frac{\inte\big(|\nabla v_{1\B}|^2+|\nabla v_{2\B}|^2\big)dx}
{\inte\big(| v_{1\B}|^2+|v_{2\B}|^2\big)^2dx}
=\lim\limits_{\B\nearrow\B^*}\frac{1}{a^*}
\frac{\inte\big(|\nabla u_{1\B}|^2+|\nabla u_{2\B}|^2\big)dx}
{\inte\big(| u_{1\B}|^2+|u_{2\B}|^2\big)^2dx}
=\frac{1}{2}.
\end{equation}
It then follows from \eqref{lem3.4 2-2} and \eqref{lem3.4 2-6} that \eqref{lem3.4 2-3} holds true.

Following above claims \eqref{lem3.4 2-1}--\eqref{lem3.4 2-3}, similar to \cite [Lemma 5.3]{GZZ2}, one can deduce that there exist a sequence $\{y_\B\}\subseteq\R^2$ and constants $R_0>0$ and $\delta>0$ such that
\begin{equation}\label{lem3.4 2-7}
     \liminf_{\B\nearrow\B^*}\int_{B_{R_0}(y_\B)}|v_{j\B}|^2dx\geq \delta>0, \,\ j=1,2.
\end{equation}
We next claim that the sequence $\{y_\B\}$ in \eqref{lem3.4 2-7} satisfies
\begin{equation}\label{lem3.4 2-8}
|y_\B|\le C\ \ \hbox{as} \ \ \B\nearrow\B^*,
\end{equation}
where the constant $C>0$ is independent of $\B$. Actually, using \eqref{2.2Ineq:GN}, \eqref{2.3ineq:diam} and \eqref{lem3.4 2-7}, we derive from \eqref{1.12trans1:F} that
\begin{equation}\label{lem3.4 2-9}
\begin{split}
e(\Om,\f)&\ge \inte V_\Om(x)\big(|u_{1\B}|^2+|u_{2\B}|^2\big)dx\\
&=\frac{1}{a^*}\inte V_\Om\big(\eps_\B x+\eps_\B y_\B\big)
\big(|v_{1\B}(x+y_\B)|^2+|v_{2\B}(x+y_\B)|^2\big)dx\\
&\ge C\int_{B_{R_0}(0)}|\eps_\B x+\eps_\B y_\B|^2
\big(|v_{1\B}(x+y_\B)|^2+|v_{2\B}(x+y_\B)|^2\big)dx\\
&\ge C\int_{B_{R_0}(0)}\big(\eps_\B^2 |y_\B|^2-2\eps_\B^2|y_\B|R_0\big)
\big(|v_{1\B}(x+y_\B)|^2+|v_{2\B}(x+y_\B)|^2\big)dx\\
&\ge C\int_{B_{R_0}(0)}\Big(\frac{1}{2}\eps_\B^2 |y_\B|^2-2\eps_\B^2R_0^2\Big)
\big(|v_{1\B}(x+y_\B)|^2+|v_{2\B}(x+y_\B)|^2\big)dx\\
&=C\int_{B_{R_0}(y_\B)}\Big(\frac{1}{2}\eps_\B^2 |y_\B|^2-2\eps_\B^2R_0^2\Big)
\big(|v_{1\B}(x)|^2+|v_{2\B}(x)|^2\big)dx\\
&\ge C\delta\eps_\B^2 |y_\B|^2-C\eps_\B^2R_0^2
\quad \hbox{as}\ \ \B\nearrow\B^*.
\end{split}
\end{equation}
On the other hand, using \eqref{2.2Ineq:GN}, \eqref{2.3ineq:diam}
and Lemma \ref{lem-3.2}, we deduce from \eqref{1.12trans1:F} that
\begin{equation}\label{lem3.4 2-10}
(\B^*-\B)\inte |u_{1\B}|^2|u_{2\B}|^2dx
\le e(\Om,\f)\le C(\B^*-\B)^{\frac{1}{2}}\ \ \hbox{as}\ \ \B\nearrow\B^*,
\end{equation}
together with \eqref{3.11}, which yields that there exists a constant $C>0$, independent of $\B$, such that
\begin{equation}\label{lem3.4 2-11}
\inte\big(|u_{1\B}|^4+|u_{2\B}|^4\big)dx\le C(\B^*-\B)^{-\frac{1}{2}}
\ \ \hbox{as}\ \ \B\nearrow\B^*.
\end{equation}
By \eqref{lem3.4 2-10} and \eqref{lem3.4 2-11}, we deduce from \eqref{Sec 3 1.5Lmu} and \eqref{3.2} that
\begin{equation}\label{lem3.4 2-12}
\begin{split}
\eps_\B^{-2}=-\mu_\B&=-e(\Om,\f)
+\inte\Big(\frac{a_1}{2}|u_{1\B}|^4+\frac{a_2}{2}|u_{2\B}|^4
+\B|u_{1\B}|^2|u_{2\B}|^2\Big)dx\\
&\le C(\B^*-\B)^{-\frac{1}{2}}
\ \ \hbox{as}\ \ \B\nearrow\B^*.
\end{split}
\end{equation}
Combining \eqref{lem3.4 2-9} and \eqref{lem3.4 2-12}, we thus conclude from \eqref{lem3.4 2-10} that
\begin{equation*}
C\delta\eps_\B^2 |y_\B|^2\le e(\Om,\f)+C\eps_\B^2R_0^2
\le C(\B^*-\B)^{\frac{1}{2}}+C\eps_\B^2R_0^2\le C\eps_\B^2
\ \ \hbox{as}\ \ \B\nearrow\B^*,
\end{equation*}
which implies that \eqref{lem3.4 2-8} holds true.
One can deduce from \eqref{lem3.4 2-7} and \eqref{lem3.4 2-8} that there exist constants $R_0>>1$ and $\delta>0$ such that
\begin{equation*}
 \liminf_{\B\nearrow\B^*}\int_{B_{2R_0}(0)}|v_{j\B}(x)|^2dx\geq \delta>0,\  \ j=1,2,
\end{equation*}
i.e., \eqref{3.21} holds true, and Lemma \ref{lem-3.4} (1) is therefore proved.
\vskip 0.05truein

2. By \eqref{2.3ineq:diam}, we obtain from \eqref{lem3.4 2-3} that $|v_{1\B}|$ and $|v_{2\B}|$ are bounded uniformly in $H^1(\R^2,\R)$. Defining
\begin{equation*}
W_\B(x):=\sqrt{|v_{1\B}(x)|^2+|v_{2\B}(x)|^2},
\end{equation*}
we then derive from \cite [Theorem 6.17] {LL} that
\begin{equation}\label{3.33}
\big|\nabla W_\B\big|^2\leq \big|\nabla |v_{1\B}|\big|^2+\big|\nabla |v_{2\B}|\big|^2\  \ \hbox{a.e.}\  \ \hbox{in}\ \ \R^2.
\end{equation}
Since it follows from \eqref{3.33} that $\big\{W_\B\big\}$ is bounded uniformly in $H^1(\R^2,\R)$, we may assume that up to a subsequence if necessary, $W_{\B}\rightharpoonup W_0\geq0$ weakly in $H^1(\R^2,\R)$ as $\B\nearrow\B^*$ for some $W_0\in H^1(\R^2,\R)$. Furthermore, we get from \eqref{3.21} that $W_0\not\equiv0$. By the weak convergence, we have $W_\B\to W_0\ge 0$ a.e. in $\R^2$ as $\B\nearrow\B^*$. 
Similar to the proof of \cite [Lemma 3.2] {GLY}, one can deduce that $W_0$ is  an optimizer of the Gagliardo-Nirenberg inequality \eqref{2.1ineq:GNQ}, $\|W_0\|^2_2=a^*$ and $\nabla W_\B\to\nabla W_0$ strongly in $L^2(\R^2,\R)$ as $\B\nearrow\B^*$.
Therefore, we conclude that there exists a point $x_0\in\R^2$ such that up to a subsequence if necessary,
\begin{equation}\label{3.34}
W_\B(x)\to W_0(x)=w(x-x_0)\  \ \hbox{strongly in} \  \ H^1(\R^2,\R)\  \ \mbox{as} \  \ \B\nearrow\B^*.
\end{equation}

We next follow \eqref{3.34} to prove \eqref{3.23}. Since $\big\{|v_{1\B}|\big\}$ and $\big\{|v_{2\B}|\big\}$ are bounded uniformly in $H^1(\R^2,\R)$, we may assume that up to a subsequence if necessary, $\{|v_{1\B}|\}$ and $\{|v_{2\B}|\}$ converge to $v_{10}\geq0$ and $v_{20}\geq0$ weakly in $H^1(\R^2,\R)$, respectively, where $v_{10}, v_{20}\in H^1(\R^2,\R)$. Note from \eqref{3.21} that $v_{10}\not\equiv 0$ and $v_{20}\not\equiv 0$ in $\R^2$. Because the embedding $H^1(\R^2,\R)\hookrightarrow L^2_{loc}(\R^2,\R)$ is compact, we derive from \eqref{3.34} that
\begin{equation}\label{v10v20 a^*}
 \inte \big(v_{10}^2+v_{20}^2\big)dx=a^*.
\end{equation}
By the weak lower semicontinuity, we then deduce from above that
$$\lim_{\B\nearrow\B^*}\inte|v_{1\B}|^2dx=\inte v_{10}^2dx\quad \hbox{and}\quad \lim_{\B\nearrow\B^*}\inte|v_{2\B}|^2dx=\inte v_{20}^2dx,$$
which further implies that $|v_{1\B}|$ and $|v_{2\B}|$ converge to $v_{10}$ and $v_{20}$ strongly in $L^q(\R^2,\R)$ $(2\leq q<\infty)$ as $\B\nearrow\B^*$, respectively.
Following the above convergence, by \eqref{2.2Ineq:GN}, \eqref{2.3ineq:diam} and \eqref{v10v20 a^*}, we derive from \eqref{lem3.4 2-2} and \eqref{lem3.4 2-3} that
\begin{equation*}
\begin{split}
a^*&=\lim_{\B\nearrow\B^*}
\inte \big(|\nabla v_{1\B}|^2+|\nabla v_{2\B}|^2\big)dx\\
&\geq\lim_{\B\nearrow\B^*}
\inte \big(\big|\nabla |v_{1\B}|\big|^2+\big|\nabla|v_{2\B}|\big|^2\big)dx\\
&\geq \inte\big(|\nabla v_{10}|^2+|\nabla v_{20}|^2\big)dx\\
&\geq \frac{1}{2}\inte\big(|v_{10}|^2+|v_{20}|^2\big)^2dx\\
&=\frac{1}{2}\lim_{\B\nearrow\B^*}\inte\big(|v_{1\B}|^2+|v_{2\B}|^2\big)^2dx
=a^*,
\end{split}
\end{equation*}
which further implies that $\nabla|v_{1\B}|$ and $\nabla|v_{2\B}|$ converge to $\nabla v_{10}$ and $\nabla v_{20}$ strongly in $L^2(\R^2,\R)$ as $\B\nearrow\B^*$, respectively. Thus, up to a subsequence if necessary,
\begin{equation}\label{3.40}
|v_{1\B}|\to v_{10}\  \ \hbox{and}\  \ |v_{2\B}|\to v_{20}\ \ \hbox{strongly in} \ \ H^1(\R^2,\R) \  \ \text{as} \  \ \B\nearrow\B^*.
\end{equation}
We further derive from \eqref{lem3.4 2-1} that
\begin{equation}\label{3.41}
(a^*-a_1)^{\frac 14}v_{10}(x)=(a^*-a_2)^{\frac 14}v_{20}(x)\  \ \hbox{a.\,e.}\  \ \text{in}\  \ \R^2.
\end{equation}
We conclude from \eqref{3.34}, \eqref{3.40} and \eqref{3.41} that up to a subsequence if necessary,
\begin{equation}\label{3.23-1}
     \lim\limits_{\B\nearrow\B^*}|v_{j\B}(x)|=\sqrt{\gam_j}w(x-x_0)
     \ \ \hbox{strongly in}\ \  H^1(\R^2,\R),\  \ j=1,2,
\end{equation}
where $0<\gam_j<1$ is given by \eqref{1.13gamma}. Similar to the argument of proving \eqref{lem3.4 2-8}, one can deduce that $|x_0|\le C$ for some constant $C>0$.

Following \eqref{Sec 3 1.1-1equation system}, we obtain from \eqref{Sec3 v1B v2B} that $(v_{1\B},v_{2\B})$ satisfies the following system
\begin{equation}\label{3.61}
\left\{\begin{array}{lll}
-\Delta v_{1\B}+i\,\eps_\B^2\,\Omega \big(x^{\perp}\cdot \nabla v_{1\B}\big)
+\Big[
\eps_\B^4V(x)
+1-\ds \frac{a_1}{a^*}|v_{1\B}|^2-\frac{\B}{a^*}|v_{2\B}|^2
\Big]v_{1\B}=0\quad \hbox{in}\  \ \R^2,\\[3mm]
-\Delta v_{2\B}+i\,\eps_\B^2\,\Omega \big(x^{\perp}\cdot \nabla v_{2\B}\big)
+\Big[
\eps_\B^4V(x)
+1-\ds\frac{a_2}{a^*}|v_{2\B}|^2-\frac{\B}{a^*}|v_{1\B}|^2
\Big]v_{2\B}=0\quad \hbox{in}\  \ \R^2.
\end{array}\right.
\end{equation}
The same argument of
\eqref{3.63-1} below yields from \eqref{3.23-1} and \eqref{3.61} that
\begin{equation*}
  \lim_{|x|\to\infty}|v_{j\B}(x)|=0 \  \ \text{as}\  \ \B\nearrow\B^*, \  \ j=1,2.
\end{equation*}
We then obtain from Lemma \ref{lem.A.1} in the Appendix that
\begin{equation}\label{pohozaev v1B plus v2B-1}
\inte \frac{\partial V_\Om(x)}{\partial x_j}
\big(|v_{1\B}|^2+|v_{2\B}|^2\big)
=0\  \ \text{as}\  \ \B\nearrow\B^*, \ \ j=1,2.
\end{equation}
On the other hand, using H\"{o}lder inequality, we deduce from \eqref{3.5}, \eqref{lem3.4 2-12} and \eqref{3.23-1} that
\begin{equation*}
\inte \frac{\partial V_\Om(x)}{\partial x_j}
\big(|v_{1\B}|^2+|v_{2\B}|^2\big)
=\inte \frac{\partial V_\Om(x)}{\partial x_j}w^2(x-x_0)+o(1)
\ \ \hbox{as}\ \ \B\nearrow\B^*,\ \ j=1,2,
\end{equation*}
together with \eqref{pohozaev v1B plus v2B-1}, which yields that $x_0=0$ in view of the form of $V_{\Om}(x)$.
Since the convergence \eqref{3.23-1} is independent of the subsequence $\{|v_{j\B}|\}$,
we deduce that \eqref{3.23-1} holds for the whole sequence.
Then it is not difficult to derive from \eqref{3.23-1} that up to a subsequence if necessary,
\begin{equation}\label{3.58}
\lim\limits_{\B\nearrow\B^*}v_{j\B}=\sqrt{\gam_j}w(x) e^{i\sigma_j}
     \ \ \hbox{strongly in}\  \ H^1(\R^2,\C)\ \ \hbox{for some}\ \ \sigma_j\in \R, \ \ j=1,2.
\end{equation}
In view of \eqref{3.49.tildew}, we further have $\sigma_j=0$. Since the convergence \eqref{3.58} with $\sigma_j=0$ is independent of the subsequence $\{v_{j\B}\}$, we conclude that  \eqref{3.58} holds true for the whole sequence, and hence \eqref{3.23} holds true.
Lemma \ref{lem-3.4} is therefore proved.
\qed

Based on Lemma \ref{lem-3.4}, we next derive the following $L^\infty-$uniform convergence of
$\{v_{j\B}\}$ as $\B\nearrow\B^*$ for $j=1,2$.

\begin{lem}\label{small}
Under the assumptions of Theorem \ref{thm-1.3}, let $(u_{1\B},u_{2\B})$ be a minimizer of $e(\Om,\f)$, and consider the sequence $\{v_{j\B}\}$ defined in Lemma \ref{lem-3.4}. Then we have
\begin{enumerate}
\item There exists a constant $C>0$, independent of $\B$, such that
\begin{equation}\label{3.59}
      |v_{j\B}(x)|\leq Ce^{-\frac{2}{3}|x|},
      \ \ |\nabla v_{j\B}(x)|\leq Ce^{-\frac{1}{2}|x|}
      \ \ \text{in}\ \ \R^2\ \ \hbox{as}\ \ \B\nearrow\B^*,\ \ j=1,2.
     \end{equation}
\item $v_{j\B}(x)$ satisfies
    \begin{equation}\label{3.1-1}
     v_{j\B}(x)\to\ds\sqrt{\gam_j}w(x)
     \ \ \text{strongly in}\  \ L^\infty(\R^2,\C)\  \ \text{as}\  \ \B\nearrow\B^*,\ \ j=1,2,
   \end{equation}
   where $0<\gam_j<1$ is given by \eqref{1.13gamma}.
\item  The following estimate holds:
\begin{equation}\label{G12-1}
\Om\inte \Big[x^\perp\cdot(iv_{1\B},\nabla v_{1\B})
+x^\perp\cdot(iv_{2\B},\nabla v_{2\B})\Big]dx\\
 =o(\eps^2_\B)\  \ \text{as}\  \ \B\nearrow\B^*,
\end{equation}
where $\eps_\B>0$ is as in \eqref{3.2}.
\end{enumerate}
\end{lem}

\noindent{\bf Proof.}
1. Multiplying the first equation of \eqref{3.61} by $\bar v_{1\B}$ and the second equation of \eqref{3.61} by $\bar v_{2\B}$, respectively, we deduce from \eqref{2.3ineq:diam} that
\begin{equation}\label{3.63}
\left\{\begin{array}{lll}
-\ds\frac 12\Delta |v_{1\B}|^2
+\Big[1-\frac{a_1}{a^*}|v_{1\B}|^2-\frac{\B}{a^*}|v_{2\B}|^2\Big]
|v_{1\B}|^2\leq0 \quad \text{in} \  \ \R^2,\\[3mm]
-\ds\frac 12\Delta |v_{2\B}|^2
+\Big[1-\frac{a_2}{a^*}|v_{2\B}|^2-\frac{\B}{a^*}|v_{1\B}|^2\Big]
|v_{2\B}|^2\leq0 \quad \text{in} \  \ \R^2.
\end{array}\right.
\end{equation}
Recall from \eqref{3.23} that
\begin{equation}\label{3.53-1}
\lim\limits_{\B\nearrow\B^*}v_{j\B}(x)
     =\sqrt{\gam_j}w(x)
     \ \ \hbox{strongly in}\  \ H^1(\R^2,\C),\ \ j=1,2,
\end{equation}
where $0<\gam_j<1$ is given by \eqref{1.13gamma}.
Applying De Giorgi-Nash-Moser theory (cf. \cite [Theorem 4.1] {HL}), we derive from \eqref{3.63} and \eqref{3.53-1} that for any $\xi\in\R^2$,
\begin{equation}\label{3.53-2}
\max_{x\in B_1(\xi)}|v_{j\B}(x)|^2\le C\int_{B_2(\xi)}|v_{j\B}(x)|^2dx,\ \ j=1,2.
\end{equation}
Therefore, we derive from \eqref{3.53-1} and \eqref{3.53-2} that
\begin{equation}\label{3.63-1}
|v_{j\B}(x)|^2\le C\ \ \hbox{and}\ \ \lim_{|x|\to\infty}|v_{j\B}(x)|^2=0 \  \ \text{as}\  \ \B\nearrow\B^*, \  \ j=1,2.
\end{equation}
We then deduce from \eqref{3.63} and \eqref{3.63-1} that for sufficiently large $R>0$,
\begin{equation}\label{3.63-2}
-\Delta|v_{j\B}|^2+\frac{16}{9}|v_{j\B}|^2\leq0 \  \ \text{in}\  \ \R^2/B_R(0)
\ \ \hbox{as}\ \ \B\nearrow\B^*,\ \ j=1,2.
\end{equation}
Applying the comparison principle to \eqref{3.63-2}, we deduce that
\begin{equation*}
|v_{j\B}(x)|\leq Ce^{-\frac{2}{3}|x|}
\ \ \text{in}\ \ \R^2/B_R(0)\ \ \hbox{as}\ \ \B\nearrow\B^*,\ \ j=1,2,
\end{equation*}
together with \eqref{3.63-1}, which yields that
\begin{equation}\label{3.63-4-1}
|v_{j\B}(x)|\leq Ce^{-\frac{2}{3}|x|}
\ \ \text{in}\ \ \R^2\ \ \hbox{as}\ \ \B\nearrow\B^*,\ \ j=1,2.
\end{equation}
Similar to the argument of proving \cite [Proposition 2.2]{GLP}, one can deduce from \eqref{3.53-1} and \eqref{3.63-4-1} that there exists a constant $C>0$, independent of $\B$, such that
\begin{equation*}
      |\nabla v_{j\B}(x)|\leq Ce^{-\frac{1}{2}|x|}
      \ \ \text{in}\ \ \R^2\ \ \hbox{as}\ \ \B\nearrow\B^*,\ \ j=1,2,
     \end{equation*}
together with \eqref{3.63-4-1}, which implies that the exponential decay \eqref{3.59} holds true.

2. Similar to \cite [Proposition 3.3 (ii)]{GLY},
by the exponential decay of \eqref{1.10decay:w} and  \eqref{3.59}, and the standard elliptic regularity theory (cf. \cite {GT}), one can further derive from \eqref{3.23} that \eqref{3.1-1} holds.

3. Denote the operator $N_{j\B}$ by
\begin{equation}\label{Sec3 lem3.5-1}
N_{j\B}
:=-\Delta
+\Big(
\eps_\B^4V(x)
+1-\ds\frac{a_j}{a^*}|v_{j\B}|^2-\frac{\B}{a^*}|v_{m\B}|^2
\Big),
\ \ \hbox{where} \ \ j,\ m=1,2\ \ \hbox{and}\ \ j\neq m.
\end{equation}
We then obtain from \eqref{Sec3 v1B v2B}, \eqref{3.49.tildew} and \eqref{3.61} that $I_{j\B}$ satisfies
\begin{equation}\label{Sec3 lem3.5-2}
N_{j\B}I_{j\B}
=-\eps_\B^2\Omega \, \big(x^{\perp}\cdot \nabla R_{j\B}\big)
\ \  \hbox{in}\  \ \R^2,\ \ \inte w(x)I_{j\B}(x)dx=0,\ \ j=1,2.
\end{equation}
Multiplying \eqref{Sec3 lem3.5-2} by $I_{j\B}$ and integrating over $\R^2$, we derive from \eqref{3.59} that
\begin{equation}\label{Sec3 lem3.5-3}
\inte \big(N_{j\B}I_{j\B}\big)I_{j\B}
=-\eps_\B^2\Omega \inte\big(x^{\perp}\cdot \nabla R_{j\B}\big)I_{j\B}
 \le C\eps_\B^2\|I_{j\B}\|_{L^2(\R^2)}
\ \ \hbox{as} \ \ \B\nearrow\B^*, \ \ j=1,2.
\end{equation}
Denote
$$\mathcal{L}:=-\Delta+1-w^2\  \ \text{in} \  \ L^2(\R^2).$$
By a standard argument (e.g. (3.45)
in \cite {GLY}), there exists $\rho>0$ such that
\begin{equation}\label{G6}
\langle \mathcal{L}u,u\rangle\geq\rho\|u\|^2_{H^1(\R^2)}\  \ \text{for all}\  \ u\in \mathcal{S},
\end{equation}
where the space $\mathcal{S}$ is defined as
$$\mathcal{S}:=\Big\{u\in H^1(\R^2,\R): \ \inte w(x)u(x)dx=0\Big\}.$$

Following \eqref{3.1-1} and \eqref{Sec3 lem3.5-1}, we obtain from \eqref{G6} that
\begin{equation}\label{Sec3 lem3.5-4}
\begin{split}
\inte \big(N_{j\B}I_{j\B}\big)I_{j\B}
&\ge\inte \Big[\big(\mathcal{L}I_{j\B}\big)I_{j\B}
-\Big(\ds\frac{a_j}{a^*}|v_{j\B}|^2+\frac{\B}{a^*}|v_{m\B}|^2-w^2\Big)
I_{j\B}^2
\Big]\\
&=\inte \big(\mathcal{L}I_{j\B}\big)I_{j\B}
+o(1)\inte I_{j\B}^2\\
&\ge \frac{\rho}{2}\|I_{j\B}\|_{H^1(\R^2)}^2
\ \ \hbox{as}\ \ \B\nearrow\B^*,\ \ j=1,2,
\end{split}
\end{equation}
where the constant $\rho>0$, independent of $0<\B<\B^*$, is given by \eqref{G6}.
Thus, we obtain from \eqref{Sec3 lem3.5-3} and \eqref{Sec3 lem3.5-4} that
for $j=1,2$,
\begin{equation}\label{G11}
\|I_{j\B}\|_{H^1(\R^2)}
\leq C\eps^2_\B\  \ \text{as}\  \ \B\nearrow\B^*.
\end{equation}
Using \eqref{3.59} and \eqref{3.1-1}, we deduce from \eqref{Sec3 v1B v2B} and \eqref{G11} that
\begin{equation}\label{G12}
\begin{split}
 &\quad\Om\inte \Big[x^\perp\cdot\big(iv_{1\B},\nabla v_{1\B}\big)
 +x^\perp\cdot\big(iv_{2\B},\nabla v_{2\B}\big)\Big]dx\\
 &=2\Om\inte \Big[x^\perp\cdot\big(R_{1\B}\nabla I_{1\B}\big)
 +x^\perp\cdot\big(R_{2\B}\nabla I_{2\B}\big)\Big]dx\\
 &=2\Om\inte \Big[x^\perp\cdot\big(\sqrt{\gamma_1}w\nabla I_{1\B}\big)
 +x^\perp\cdot\big(\sqrt{\gamma_2}w\nabla I_{2\B}\big)\Big]dx+o(\eps^2_\B)\\
 &=-2\Om\inte \Big[\big(x^\perp\cdot\nabla (\sqrt{\gamma_1}w) \big)I_{1\B}
  +\big(x^\perp\cdot\nabla(\sqrt{\gamma_2}w)\big)I_{2\B}\Big]dx
  +o(\eps^2_\B)\\
&=o(\eps^2_\B)\  \ \text{as}\  \ \B\nearrow\B^*,
\end{split}
\end{equation}
where $0<\gam_1,\gam_2<1$ are as in \eqref{1.13gamma}.
It then follows from \eqref{G12} that \eqref{G12-1} holds true, and Lemma \ref{small} is thus proved.
\qed

\subsection{Proof of Theorem \ref{thm-1.3}}
In this subsection, we shall complete the proof of Theorem \ref{thm-1.3} on the limiting behavior of minimizers for $e(\Om,\f)$ as $\B\nearrow\B^*$, where $0<a_1,a_2<a^*$ are fixed as above.
\vskip 0.05truein

\noindent {\bf Proof of Theorem \ref{thm-1.3}.} \  We first claim that
\begin{equation}\label{Y1}
\eps_\B=\alpha_\B+o(\alpha_\B)
\ \ \hbox{as} \ \ \B\nearrow\B^*,
\end{equation}
where $\alpha_\B>0$ and $\eps_\B>0$ are given by \eqref{1.16def:beta.V.eps} and \eqref{3.2}, respectively.

Indeed, by  the definition of $e(\Om,\f)$, we deduce from \eqref{2.2Ineq:GN}, \eqref{2.3ineq:diam} and \eqref{Sec3 v1B v2B} that
\begin{equation}\label{Y2}
\begin{split}
&\quad e(\Om,\f)\\
&=F_{\Om,\f}(u_{1\B},u_{2\B})\\
&=\frac{\eps^{-2}_\B}{a^*}\Big[\inte \big(|\nabla v_{1\B}|^2+|\nabla v_{2\B}|^2\big)dx
-\frac{1}{2}\inte\big(|v_{1\B}|^2+|v_{2\B}|^2\big)^2dx\Big]\\
&\quad+ \frac{\eps^2_\B}{a^*}\inte V(x)\big(|v_{1\B}|^2
+|v_{2\B}|^2\big)dx
+\frac{(\B^*-\B)\eps^{-2}_\B}{(a^*)^2}\inte|v_{1\B}|^2|v_{2\B}|^2 dx\\
&\quad+\frac{\eps^{-2}_\B}{2(a^*)^2}
\inte \big(\sqrt{a^*-a_1}|v_{1\B}|^2-\sqrt{a^*-a_2}|v_{2\B}|^2\big)^2 dx\\
&\quad-\frac{\Om}{a^*}\inte \Big[x^\perp\cdot\big(iv_{1\B},\nabla v_{1\B}\big)
+x^\perp\cdot\big(iv_{2\B},\nabla v_{2\B}\big)\Big]dx\\
&\geq \frac{\eps^2_\B}{a^*}\inte V(x)\big(|v_{1\B}|^2
+|v_{2\B}|^2\big)dx
+\frac{(\B^*-\B)\eps^{-2}_\B}{(a^*)^2} \inte|v_{1\B}|^2|v_{2\B}|^2 dx\\
&\quad-\frac{\Om}{a^*}\inte\Big[x^\perp\cdot\big(iv_{1\B},\nabla v_{1\B}\big)
+x^\perp\cdot\big(iv_{2\B},\nabla v_{2\B}\big)\Big]dx.
\end{split}
\end{equation}
Note from \eqref{G12-1} that
\begin{equation}\label{Y3}
\Om\inte \Big[x^\perp\cdot\big(iv_{1\B},\nabla v_{1\B}\big)
+x^\perp\cdot\big(iv_{2\B},\nabla v_{2\B}\big)\Big]dx\\
=o(\eps^2_\B)\  \ \text{as}\  \ \B\nearrow\B^*.
\end{equation}
Applying \eqref{1.9identity:w}, we deduce from  \eqref{3.59} and \eqref{3.1-1} that as $\B\nearrow\B^*$,
\begin{equation*}
\frac{(\B^*-\B)\eps^{-2}_\B}{(a^*)^2} \inte|v_{1\B}|^2|v_{2\B}|^2dx
=\big[1+o(1)\big](\B^*-\B)\frac{2\gamma_1\gamma_2}{a^*\eps^2_\B},
\end{equation*}
and
\begin{equation*}
 \frac{\eps^2_\B}{a^*}\inte V(x)\big(|v_{1\B}|^2
+|v_{2\B}|^2\big)dx=[1+o(1)]\frac{\eps^2_\B}{a^*}\inte V(x)w^2dx.
\end{equation*}
Note that
\begin{equation}\label{Y7}
\begin{split}
 \frac{\eps^2_\B}{a^*}\inte V(x)w^2dx
+(\B^*-\B)\frac{2\gamma_1\gamma_2}{a^*\eps^2_\B}
\geq\frac{2}{a^*}
\Big[2\gamma_1\gamma_2\lam_0(\B^*-\B)\Big]^{\frac{1}{2}}  \  \ \text{as}\  \ \B\nearrow\B^*,
\end{split}
\end{equation}
where the identity holds if and only if \eqref{Y1} holds true.
By the upper bound of \eqref{3.5}, we conclude from \eqref{Y2}--\eqref{Y7} that
\begin{equation*}
e(\Om,\f)\approx\frac{2}{a^*}
\Big[2\gamma_1\gamma_2\lam_0(\B^*-\B)\Big]^{\frac{1}{2}}\  \ \text{as}\  \ \B\nearrow\B^*,
\end{equation*}
and \eqref{Y1} holds true.
It then follows from \eqref{3.2} and \eqref{Y1} that \eqref{eps2mu} holds true.

We finally prove \eqref{1.15lim:beta.V.u.exp}. By \eqref{Y1}, we deduce from \eqref{3.23} and \eqref{3.1-1} that for $j=1,2$,
\begin{equation}\label{Y12}
     \sqrt{a^*}\alpha_\B u_{j\B}(\alpha_\B x)
     e^{i\widetilde{\theta}_{j\B}}
     \to\sqrt{\gam_j}w(x)
     \ \ \text{strongly in}\  \ H^1(\R^2,\C)\cap L^\infty(\R^2,\C)\  \ \text{as}\  \ \B\nearrow\B^*,
   \end{equation}
where $\alpha_\B>0$ is given by \eqref{1.16def:beta.V.eps}, $0<\gam_j<1$ is as in \eqref{1.13gamma}, and $\widetilde{\theta}_{j\B}\in[0,2\pi)$ 
is chosen such that \eqref{3.49.tildew} holds true.
Define
\begin{equation}\label{hatw.define}
w_{j\B}(x):=\sqrt{a^*}\alpha_\B u_{j\B}(\alpha_\B x)
     e^{i\theta_{j\B}},\  \ j=1,2,
\end{equation}
where $\alpha_\B>0$ is as in \eqref{1.16def:beta.V.eps},
and similar to \cite[Lemma 3.2]{GLY}, the constant phase $\theta_{j\B}\in[0,2\pi)$ is chosen properly such that
\begin{equation}\label{Sec 3 tilde theta-2-1}
\inte wIm\big(w_{j\B}(x)\big)=0,\ \ j=1,2.
\end{equation}
Following \eqref{Sec3 v1B v2B}, \eqref{3.49.tildew} and \eqref{Sec 3 tilde theta-2-1}, we have
\begin{equation}\label{Sec 3 tilde theta-2}
\inte wIm\Big(w_{j\B}(x)-v_{j\B}(x)\Big)=0,\ \ j=1,2.
\end{equation}
On the other hand, we obtain from \eqref{Sec3 v1B v2B}, \eqref{3.1-1}, \eqref{Y12} and \eqref{hatw.define} that for $j=1,2$,
\begin{equation}\label{Sec 3 tilde theta-3}
\begin{split}
&\quad Im\big(w_{j\B}(x)-v_{j\B}(x)\big)\\
&=Im\Big[\sqrt{a^*}\alpha_\B u_{j\B}(\alpha_\B x)e^{i\theta_{j\B}}
-\sqrt{a^*}\eps_\B u_{j\B}(\eps_\B x)e^{i\widetilde{\theta}_{j\B}}\Big]\\
&=Im\Big[\sqrt{a^*}\alpha_\B u_{j\B}(\alpha_\B x)e^{i\widetilde{\theta}_{j\B}}
e^{i(\theta_{j\B}-\widetilde{\theta}_{j\B})}
-\sqrt{a^*}\eps_\B u_{j\B}(\eps_\B x)e^{i\widetilde{\theta}_{j\B}}\Big]\\
&=Im\Big[\sqrt{\gamma_{j}}w e^{i(\theta_{j\B}-\widetilde{\theta}_{j\B})}+o(1)
-\sqrt{\gamma_{j}}w +o(1)\Big]\\
&=\sqrt{\gamma_{j}}w\sin(\theta_{j\B}-\widetilde{\theta}_{j\B})+o(1)
\ \ \hbox{as}\ \ \B\nearrow\B^*.
\end{split}
\end{equation}
Substituting \eqref{Sec 3 tilde theta-3} into \eqref{Sec 3 tilde theta-2} then yields that
\begin{equation*}
\lim_{\B\nearrow\B^*}|\theta_{j\B}-\widetilde{\theta}_{j\B}|=0,\  \ j=1,2,
\end{equation*}
and \eqref{1.15lim:beta.V.u.exp} is therefore proved in view of \eqref{Y12} and \eqref{hatw.define}. The proof of Theorem \ref{thm-1.3} is thus complete.\qed

\section{Refined Expansions of Solutions}
The main purpose of this section is to establish Theorem \ref{thm-expan solution} below on the refined expansions of solutions for an elliptic system.

Suppose $V(x)$ satisfies \eqref{V(x)} for some $\Lambda\ge1$, and assume $0<\Om<\Om^*:=2$ and $0<a_1,a_2<a^*$ are fixed.
Let $(u_{1\B},u_{2\B})\in\m$  be a normalized concentration solution of the following system
\begin{equation}\label{1.1equation system-1}
\left\{\begin{array}{lll}
-\Delta u_{1\B}+V(x)u_{1\B}+i\, \Om \big(x^{\perp}\cdot \nabla u_{1\B}\big)
=\mu_\B u_{1\B}+a_1|u_{1\B}|^2u_{1\B}+\B |u_{2\B}|^2u_{1\B}
\  \ \hbox{in}\  \ \R^2,\\[3mm]
-\Delta u_{2\B}+V(x)u_{2\B}+i\, \Om \big(x^{\perp}\cdot \nabla u_{2\B}\big)
=\mu_\B u_{2\B}+a_2|u_{2\B}|^2u_{2\B}+\B |u_{1\B}|^2u_{2\B}\  \ \hbox{in}\  \ \R^2,
\end{array}\right.
\end{equation}
satisfying
\begin{equation}\label{eps2mu-1}
\alpha_\B^2\mu_\B\to-1\  \ \text{as}\  \ \B\nearrow\B^*,
\end{equation}
and for $j=1,2$,
\begin{equation}\label{1.15lim:beta.V.u.exp-1}
w_{j\B}(x):=\sqrt{a^*}\alpha_\B u_{j\B}(\alpha_\B x)e^{i\theta_{j\B}}
\to\sqrt{\gam_j}w(x)
   \ \ \hbox{strongly in} \ \ H^1(\R^2,\C)\cap L^\infty(\R^2,\C)
   \end{equation}
as $\B\nearrow\B^*$, where $\m$ and $\alpha_\B:=\Big[\frac{2\gamma_1\gamma_2}{\lambda_0}(\B^*-\B)\Big]^{\frac{1}{4}}
>0$ are as in \eqref{1.2def:m} and \eqref{1.16def:beta.V.eps}, respectively,
 $0<\gam_j<1$ is as in \eqref{1.13gamma}, and
$\B^*:=a^*+\sqrt{(a^*-a_1)(a^*-a_2)}>0$. Here the constant phase $\theta_{j\B}\in [0,2\pi)$ is chosen properly such that
\begin{equation}\label{Sec 3 tilde theta-2-2}
\inte wIm\big(w_{j\B}(x)\big)=0,\ \ j=1,2.
\end{equation}

For convenience, we define the following operator
\begin{equation}\label{G4}
\mathcal{L}:=-\Delta+1-w^2\  \ \text{in} \  \ L^2(\R^2).
\end{equation}
Applying \cite [Corollary 11.9 and Theorem 11.8] {LL}, we obtain that
\begin{equation}\label{G5}
ker \mathcal{L}=\{w\}\  \ \text{and}\  \ \langle \mathcal{L}u,u\rangle\geq0\  \ \text{for all}\  \ u\in L^2(\R^2).
\end{equation}
We also define the linearized operator $\widetilde{\mathcal{L}}$ of \eqref{1.8equation:w} by
\begin{equation}\label{line oper L}
 \widetilde{\mathcal{L}}:=-\Delta+1-3w^2\  \ \text{in} \  \ L^2(\R^2).
\end{equation}
It then follows from \cite{K,Wei96} that
\begin{equation}\label{ker line oper L}
ker \widetilde{\mathcal{L}}
=\Big\{\frac{\partial w}{\partial x_1},\frac{\partial w}{\partial x_2}\Big\}.
\end{equation}
For simplicity, we denote $\psi_{0}(x)\in C^2(\R^2)\cap L^\infty(\R^2)$ to be the unique solution of
\begin{equation}\label{psi 0.equation.thm}
\nabla\psi_{0}(0)=0,\quad
\widetilde{\mathcal{L}}\psi_{0}(x)=-V(x)w\quad \hbox{in} \ \ \R^2,
\end{equation}
where $\widetilde{\mathcal{L}}$ is defined by \eqref{line oper L}. Here the uniqueness of $\psi_0(x)$ follows from $\nabla \psi_{0}(0)=0$ and the property
\eqref{ker line oper L} (see also \cite[Lemma 4.1]{Wei96}).
We also denote $\phi_{0}(x)\in C^2(\R^2)\cap L^\infty(\R^2)$ to be the unique solution of
\begin{equation}\label{phi 0.equation.thm}
\inte \phi_{0}wdx=0,\quad
\mathcal{L}\phi_{0}(x)=-\big(x^\perp \cdot \nabla \psi_{0}\big)
\quad \hbox{in}\  \ \R^2,
\end{equation}
where $\mathcal{L}$ is defined by \eqref{G4} and $\psi_{0}(x)$ is as in \eqref{psi 0.equation.thm}. Here the uniqueness of $\phi_{0}(x)$ follows from the constriction $\inte \phi_{0}wdx=0$ and the property \eqref{G5} (cf. \cite[Lemma 4.1]{Wei96}).

In this section, we shall establish mainly the following refined expansions of solutions $(u_{1\B},u_{2\B})\in\m$ of \eqref{1.1equation system-1} satisfying \eqref{eps2mu-1} and \eqref{1.15lim:beta.V.u.exp-1}.

\begin{thm}\label{thm-expan solution}
Suppose $V(x)$ satisfies \eqref{V(x)} for some $\Lambda\ge1$, and assume $0<\Om<\Om^*:=2$ and $0<a_1,a_2<a^*$ are fixed. Let $(u_{1\B},u_{2\B})\in\m$ be a normalized concentration solution of \eqref{1.1equation system-1} satisfying \eqref{eps2mu-1} and \eqref{1.15lim:beta.V.u.exp-1}, where $0<\B<\B ^*$. Then we have
\begin{equation}\label{alphaB2 muB expan}
\alpha_\B^2\mu_\B=-1+C(\lambda_0,a_1,a_2,\B^*)\alpha_\B^4+O(\alpha_\B^8)
\ \ \hbox{as}\ \ \B\nearrow\B^*,
\end{equation}
and
\begin{equation}\label{solution u1B u2B expan}
\begin{split}
w_{j\B}(x):&=\sqrt{a^*}\alpha_\B u_{j\B}(\alpha_\B x)e^{i\theta_{j\B}}\\
           &=\rho_{j\B}w+\rho_{j\B}\alpha_\B^4\Big[
           \psi_0-\frac{1}{2}C(\lambda_0,a_1,a_2,\B^*)\big(w+x\cdot\nabla w\big)\Big]
           +O(\alpha_\B^{8})\\[3mm]
&\quad+i\Big[\rho_{j\B}\Omega\,\alpha_\B^6\phi_0
      +O(\alpha_\B^{10})\Big]\ \ \hbox{in}\ \ L^\infty(\R^2, \mathbb{C})
\ \ \hbox{as}\ \ \B\nearrow\B^*,\ \ j=1,2,
\end{split}
\end{equation}
where $\alpha_\B:=\Big[\frac{2\gamma_1\gamma_2}{\lambda_0}(\B^*-\B)\Big]^{\frac{1}{4}}
>0$ is as in \eqref{1.16def:beta.V.eps}, $\theta_{j\B}\in[0,2\pi)$ is a suitable constant phase such that \eqref{Sec 3 tilde theta-2-2} holds true. Here $\psi_0$ and $\phi_0$ are uniquely given by \eqref{psi 0.equation.thm} and \eqref{phi 0.equation.thm}, respectively, and the constant
$\rho_{j\B}$ is defined as
\begin{equation}\label{rho jB define}
\ds\rho_{j\B}:=\sqrt{\frac{a^*(\B-a_m)}{\B^2-a_1a_2}}\rightarrow\sqrt{\gamma_{j}}>0\ \ \hbox{as}\ \ \B\nearrow\B^*,
\ \ j,m=1,2,\ \ \hbox{and} \ \ j\neq m.
\end{equation}
Moreover, the above constant $C(\lambda_0,a_1,a_2,\B^*)$ is independent of $\Om$ and satisfies
\begin{equation*}
C(\lambda_0,a_1,a_2,\B^*):=\frac{3\langle\widetilde{\mathcal{L}}\psi_0,\psi_0\rangle}{2\lambda_0}
                          +\frac{\lambda_0\big(4\gamma_1\gamma_2-1\big)}
                                {\big(2\B^*-a_1-a_2\big)8\gamma_1^2\gamma_2^2},
\end{equation*}
where $0<\gamma_1,\gamma_2<1$ are as in \eqref{1.13gamma}, $\lambda_0>0$ is given by \eqref{1.14lam0}, and the operator $\widetilde{\mathcal{L}}$ is defined by \eqref{line oper L}.
\end{thm}

To prove Theorem \ref{thm-expan solution}, we define
\begin{equation}\label{ep}
\eps_\B:=\sqrt{\frac{1}{-\mu_\B}}>0 \ \ \hbox{as}\ \ \B\nearrow\B^*,
\end{equation}
where $\mu_\B$ is a parameter of \eqref{1.1equation system-1} and satisfies \eqref{eps2mu-1}.
Denote
\begin{equation}\label{v1B v2B-1}
v_{j\B}(x):=\sqrt{a^*}\eps_\B u_{j\B}(\eps_\B x)e^{i\widetilde{\theta}_{j\B}}
=R_{j\B}(x)+iI_{j\B}(x),\ \ j=1,2,
\end{equation}
where 
$R_{j\B}(x)$ and
$I_{j\B}(x)$ denote the real and imaginary parts of
$v_{j\B}(x)$, respectively, and the constant phase $\widetilde{\theta}_{j\B}\in [0,2\pi)$ is chosen such that
\begin{equation}\label{wIjB}
\inte w(x)I_{j\B}(x)dx=0, \ \ j=1,2.
\end{equation}
Following \eqref{eps2mu-1}, we obtain from \eqref{ep} that
\begin{equation}\label{ep alpha B}
\eps_\B=\alpha_\B\big[1+o(1)\big]\ \ \hbox{as}\ \ \B\nearrow \B^*,
\end{equation}
where $\alpha_\B>0$ is defined by \eqref{1.16def:beta.V.eps}.
Similar to the proof of \eqref{1.15lim:beta.V.u.exp}, one can derive from \eqref{1.15lim:beta.V.u.exp-1},
\eqref{Sec 3 tilde theta-2-2} and \eqref{v1B v2B-1}--\eqref{ep alpha B} that
\begin{equation}\label{v1B gama w v2B 1gama w}
v_{j\B}(x)
\to\sqrt{\gamma_j}w(x)
\ \ \hbox{strongly in} \ \ H^1(\R^2,\C)\cap L^\infty(\R^2,\C)\ \ \hbox{as} \ \ \B\nearrow\B^*,\ \ j=1,2,
\end{equation}
where $0<\gamma_j<1$ is as in \eqref{1.13gamma}.

In the coming subsection we shall derive the refined estimates of $(v_{1\B},v_{2\B})$ defined by \eqref{v1B v2B-1} as $\B\nearrow\B^*$ , based on which we shall complete the proof of Theorem \ref{thm-expan solution} in Subsection 4.2.

\subsection{Refined estimates of $(v_{1\B},v_{2\B})$ as $\B\nearrow\B^*$}
The purpose of this subsection is to derive the refined estimates of $(v_{1\B},v_{2\B})$ defined in \eqref{v1B v2B-1} as $\B\nearrow\B^*$. Towards this purpose, we first introduce the following system
\begin{equation}\label{Q1BQ2B.equation}
 \left\{\begin{array}{lll}
-\ds\Delta Q_{1\B}+Q_{1\B}-\frac{a_1}{a^*}Q_{1\B}^3-\frac{\B}{a^*}Q_{2\B}^2Q_{1\B}=0
\quad \hbox{in}\  \ \R^2,\\[3mm]
-\ds\Delta Q_{2\B}+Q_{2\B}-\frac{a_2}{a^*}Q_{2\B}^3-\frac{\B}{a^*}Q_{1\B}^2Q_{2\B}=0
\quad \hbox{in}\  \ \R^2,
\end{array}\right.
\ \ \hbox{where} \ \ \max\{a_1,a_2\}<\B<\B ^*.
\end{equation}
It then follows from \cite[Lemma 2.2 and Theorem 3.1]{DW} and
\cite [Theorem 1.3] {WY} that \eqref{Q1BQ2B.equation} admits a unique real-valued positive solution $(Q_{1\B},Q_{2\B})$ (i.e., $Q_{1\B}>0$ and $Q_{2\B}>0$) satisfying
\begin{equation}\label{Q1BQ2B}
(Q_{1\B},Q_{2\B})=(\rho_{1\B}w,\rho_{2\B}w),
\end{equation}
where $\rho_{j\B}>0$ is defined by \eqref{rho jB define} for $j=1,2$,
and $w>0$ is the unique positive solution of \eqref{1.8equation:w}.
Here the unique positive solution $(Q_{1\B},Q_{2\B})$ is non-degenerate, in the sense that the solution of the following linearized system
\begin{equation}\label{linea opera}
 \left\{\begin{array}{lll}
\widetilde{\mathcal{L}}_{1\B}(\phi_1,\phi_2)
:=-\ds\Delta \phi_1+\phi_1-\frac{3a_1}{a^*}Q_{1\B}^2\phi_1
  -\frac{\B}{a^*}Q_{2\B}^2\phi_1-\frac{2\B}{a^*}Q_{1\B}Q_{2\B}\phi_2=0
\  \ \hbox{in}\  \ \R^2,\\[5mm]
\widetilde{\mathcal{L}}_{2\B}(\phi_2,\phi_1)
:=-\ds\Delta \phi_2+\phi_2-\frac{3a_2}{a^*}Q_{2\B}^2\phi_2
  -\frac{\B}{a^*}Q_{1\B}^2\phi_2-\frac{2\B}{a^*}Q_{1\B}Q_{2\B}\phi_1=0
\  \ \hbox{in}\  \ \R^2,
\end{array}\right.
\end{equation}
is given by
\begin{equation}\label{ker linea opera}
\begin{pmatrix}
  \phi_1\\[5mm]
  \phi_2
\end{pmatrix}
=\sum^2_{l=1}b_l
\begin{pmatrix}
\ds\frac{\partial Q_{1\B}}{\partial x_l}\\[5mm]
\ds\frac{\partial Q_{2\B}}{\partial x_l}
\end{pmatrix}
\end{equation}
for some constants $b_l\in\R$.

Following \eqref{v1B v2B-1}, we now rewrite $v_{j\B}(x)$ as
\begin{equation}\label{v1B v2B}
v_{j\B}(x):=R_{j\B}(x)+iI_{j\B}(x)=Q_{j\B}(x)+T_{j\B}(x)+iI_{j\B}(x),
\ \ j=1,2,
\end{equation}
where $Q_{j\B}(x)$ is given by \eqref{Q1BQ2B}.
Since $Q_{j\B}(x)\to\sqrt{\gamma_j}w(x)$ uniformly in $\R^2$ as $\B\nearrow \B^*$ for $j=1,2$, where $0<\gamma_j<1$ is as in \eqref{1.13gamma}, we obtain from \eqref{v1B gama w v2B 1gama w} and \eqref{v1B v2B} that
\begin{equation}\label{TB0 IB0}
T_{j\B}\to0,\ \ I_{j\B}\to0
\ \ \hbox{uniformly in}\ \ \R^2 \ \ \hbox{as}\ \ \B\nearrow \B^*,
\ \ j=1,2.
\end{equation}
Note from \eqref{1.1equation system-1} and \eqref{ep} that $\big(v_{1\B}(x),v_{2\B}(x)\big)$ defined in \eqref{v1B v2B-1} satisfies the following system
\begin{equation}\label{v1Bv2B.equation}
\left\{\begin{array}{lll}
\begin{split}
  &\quad -\Delta v_{1\B}(x)+i\,\eps_\B^2\,\Om\big[x^\perp\cdot\nabla v_{1\B}(x)\big]
         +\eps_\B^4V(x)v_{1\B}(x)\\
&=-v_{1\B}(x)+\ds\frac{a_1}{a^*}|v_{1\B}|^2v_{1\B}(x)+\frac{\B}{a^*}|v_{2\B}|^2v_{1\B}(x)
\quad \hbox{in}\  \ \R^2,\\[2mm]
  &\quad -\Delta v_{2\B}(x)+i\,\eps_\B^2\,\Om\big[x^\perp\cdot\nabla v_{2\B}(x)\big]
         +\eps_\B^4V(x)v_{2\B}(x)\\
&=-v_{2\B}(x)+\ds\frac{a_2}{a^*}|v_{2\B}|^2v_{2\B}(x)+\frac{\B}{a^*}|v_{1\B}|^2v_{2\B}(x)
\quad\hbox{in}\  \ \R^2.
\end{split}
\end{array}\right.
\end{equation}
For convenience, we denote the operator $L_{j\B}$ by
\begin{equation}\label{LjB}
L_{j\B}:=-\Delta+\Big(\eps_\B^4 V(x)+1-\frac{a_j}{a^*}|v_{j\B}|^2
         -\frac{\B}{a^*}|v_{m\B}|^2\Big)
\quad\hbox{in}\ \ \R^2,\ \ \hbox{where}\ \ j,m=1,2\ \ \hbox{and}\ \ j\neq m.
\end{equation}
It then follows from \eqref{wIjB},
\eqref{v1B v2B}, \eqref{v1Bv2B.equation} and \eqref{LjB} that $(I_{1\B},I_{2\B})$ satisfies
\begin{equation}\label{I1B I2B.equation}
\left\{\begin{array}{lll}
L_{1\B}I_{1\B}
=-\eps_\B^2\Om\, \big(x^\bot\cdot\nabla T_{1\B}\big)
\ \ \mbox{in}\,\  \R^2,\quad \inte wI_{1\B}dx\equiv 0,\\[5mm]
L_{2\B}I_{2\B}
=-\eps_\B^2\Om\, \big(x^\bot\cdot\nabla T_{2\B}\big)
\ \ \mbox{in}\,\  \R^2,\quad \inte wI_{2\B}dx\equiv 0,
\end{array}\right.
\end{equation}
and $(T_{1\B},T_{2\B})$ satisfies
\begin{equation}\label{T1B T2B equation}
\left\{\begin{array}{lll}
\begin{split}
\widetilde{\mathcal{L}}_{1\B}(T_{1\B},T_{2\B})&=F_{1\B}(x)
\quad \mbox{in}\,\  \R^2,\\[3mm]
\widetilde{\mathcal{L}}_{2\B}(T_{2\B},T_{1\B})&=F_{2\B}(x)
\quad \mbox{in}\,\  \R^2,
\end{split}
\end{array}\right.
\end{equation}
where the operators $\widetilde{\mathcal{L}}_{1\B}$ and $\widetilde{\mathcal{L}}_{2\B}$ are as in \eqref{linea opera}, $F_{1\B}(x)$ and $F_{2\B}(x)$ are defined by
\begin{equation}\label{F1B F2B equation}
\left\{\begin{array}{lll}
\begin{split}
F_{1\B}(x)
:=&-\ds\eps_\B^4V(x)R_{1\B}
   +\frac{a_1}{a^*}I_{1\B}^2R_{1\B}+\frac{\B}{a^*}I_{2\B}^2R_{1\B}
   +\eps_\B^2\Om\,\big(x^\perp\cdot\nabla I_{1\B}\big)\\[3mm]
  &+\Big(\ds\frac{3a_1}{a^*}Q_{1\B}T_{1\B}^2+\frac{2\B}{a^*}Q_{2\B}T_{1\B}T_{2\B}
   +\frac{\B}{a^*}Q_{1\B}T_{2\B}^2\\[3mm]
&\qquad+\ds\frac{a_1}{a^*}T_{1\B}^3+\frac{\B}{a^*}T_{2\B}^2T_{1\B}\Big)
\quad \hbox{in} \ \ \R^2,\\[3mm]
F_{2\B}(x)
:=&-\ds\eps_\B^4V(x)R_{2\B}
   +\frac{a_2}{a^*}I_{2\B}^2R_{2\B}+\frac{\B}{a^*}I_{1\B}^2R_{2\B}
   +\eps_\B^2\Om\,\big(x^\perp\cdot\nabla I_{2\B}\big)\\[3mm]
  &+\Big(\ds\frac{3a_2}{a^*}Q_{2\B}T_{2\B}^2+\frac{2\B}{a^*}Q_{1\B}T_{2\B}T_{1\B}
   +\frac{\B}{a^*}Q_{2\B}T_{1\B}^2\\[3mm]
&\qquad+\ds\frac{a_2}{a^*}T_{2\B}^3+\frac{\B}{a^*}T_{1\B}^2T_{2\B}\Big)
\quad \hbox{in} \ \ \R^2,
\end{split}
\end{array}\right.
\end{equation}
where the parts $(\cdot)$ are  lower orders of $T_{j\B}$ as $\B\nearrow\B^*$ for $j=1,2$.

Note from \eqref{3.59} that there exists a constant $C>0$, independent of $\B$, such that for $j=1,2$,
\begin{equation}\label{v1Bv2B.decay}
      |v_{j\B}(x)|\leq Ce^{-\frac{2}{3}|x|}, \quad
      |\nabla v_{j\B}(x)|\leq Ce^{-\frac{1}{2}|x|} \  \ \text{in} \  \  \R^2
      \ \ \hbox{as} \ \ \B\nearrow\B^*.
     \end{equation}
By \eqref{1.10decay:w}, \eqref{G6}, \eqref{v1B gama w v2B 1gama w} and \eqref{v1Bv2B.decay}, the similar argument of \cite [Lemma 2.3] {GLP} yields from \eqref{I1B I2B.equation} that for $j=1,2$,
\begin{equation}\label{I1B I2B tidu decay}
      |I_{j\B}(x)|\leq C_{j}(\eps_\B)e^{-\frac{1}{4}|x|}, \quad
      |\nabla I_{j\B}(x)|\leq C_{j}(\eps_\B)e^{-\frac{1}{8}|x|} \ \
      \text{in} \  \  \R^2\ \ \hbox{as} \ \ \B\nearrow\B^*,
     \end{equation}
where the constant $C_{j}(\eps_\B)>0$ satisfies
$C_{j}(\eps_\B)=o(\eps_\B^2)$ as $\B\nearrow\B^*$.
Applying above estimates, we now establish the following ``rough'' limiting profiles in terms of $\eps_\B$.

\begin{lem}\label{lem 1}
Under the assumptions of Theorem \ref{thm-expan solution}, we have
\begin{enumerate}
\item  The real part $(T_{1\B},T_{2\B})$ of \eqref{v1B v2B} satisfies
\begin{equation}\label{T1B T2B expan}
T_{j\B}(x)=\eps_\B^4\rho_{j\B}\psi_{0}(x)+o(\eps_\B^4)
\quad \hbox{in}\, \ \R^2 \,\ \mbox{as} \,\ \B\nearrow \B^*,
\ \ j=1,2.
\end{equation}
Here $\rho_{j\B}>0$ is defined by \eqref{rho jB define}, and $\psi_{0}(x)\in C^2(\R^2)\cap L^\infty(\R^2)$ is the unique solution of
\begin{equation}\label{psi 0.equation}
\nabla\psi_{0}(0)=0,\quad
\widetilde{\mathcal{L}} \psi_0(x)=-V(x)w\quad \hbox{in} \ \ \R^2,
\end{equation}
where the operator $\widetilde{\mathcal{L}}$ is defined by \eqref{line oper L}.

\item There exists a  constant $C>0$, independent of $0<\B<\B^*$, such that the imaginary part $(I_{1\B},I_{2\B})$ of \eqref{v1B v2B} satisfies
\begin{equation}\label{I1B I2B decay}
\ |I_{j\B} (x)|,\ \ |\nabla I_{j\B}(x)|\leq C \eps_\B^6e^{-\frac{1}{14} |x|}
\quad \hbox{uniformly in}\, \ \R^2\,\ \mbox{as} \,\ \B\nearrow \B^*,
\ \ j=1,2.
\end{equation}
\end{enumerate}
\end{lem}

{\noindent \bf Proof.}
1. We first claim that there exists a constant $C>0$, independent of $0<\B<\B^*$, such that
\begin{equation}\label{TjB bounded}
\big| T_{j\B}(x)\big|\leq C\eps_\B^4
\ \ \hbox{as}\ \ \B\nearrow \B^*,\ \ j=1,2.
\end{equation}
Indeed, using \eqref{linea opera} and \eqref{ker linea opera}, we derive from \eqref{T1B T2B equation} that there exist some constants $c_{1\B}$ and $c_{2\B}$ such that
\begin{equation}\label{TjB decompose}
T_{j\B}(x)=\widetilde{T}_{j\B}(x)
           +\sum^2_{l=1}c_{l\B}\frac{\partial Q_{j\B}}{\partial x_l},
\ \ \inte\widetilde{T}_{j\B}\frac{\partial Q_{j\B}}{\partial x_l}=0
\ \ \hbox{for} \ \ j,l=1,2,
\end{equation}
where $\widetilde{T}_{j\B}$ satisfies
\begin{equation}\label{tilde mathcal L1BL2B tilde TjB}
\left\{\begin{array}{lll}
\begin{split}
\widetilde{\mathcal{L}}_{1\B}\big(\widetilde{T}_{1\B},\widetilde{T}_{2\B}\big)
&=F_{1\B}(x)
\quad \mbox{in}\,\  \R^2,\\[3mm]
\widetilde{\mathcal{L}}_{2\B}\big(\widetilde{T}_{2\B},\widetilde{T}_{1\B}\big)
&=F_{2\B}(x)
\quad \mbox{in}\,\  \R^2,
\end{split}
\end{array}\right.
\end{equation}
where $F_{1\B}(x)$ and $F_{2\B}(x)$ are as in \eqref{F1B F2B equation}.
We now prove that $c_{l\B}$ in \eqref{TjB decompose} satisfies
\begin{equation}\label{cijB bounded}
|c_{l\B}|\le C\eps_\B^4
\ \ \hbox{as}\, \ \B\nearrow \B^*, \ \ l=1,2.
\end{equation}
Without loss of generality, we may assume $|c_{1\B}|\le|c_{2\B}|$.
By \eqref{TB0 IB0}, \eqref{v1Bv2B.decay}, \eqref{I1B I2B tidu decay} and \cite [Proposition 5.1] {DW}, we deduce from \eqref{F1B F2B equation}, \eqref{TjB decompose} and \eqref{tilde mathcal L1BL2B tilde TjB} that
\begin{equation}\label{TjB perp bounded}
\begin{split}
&\quad\big\|\widetilde{T}_{1\B}\big\|_{L^\infty(\R^2)}
     +\big\|\widetilde{T}_{2\B}\big\|_{L^\infty(\R^2)}\\
&\le C\Big(\big\| \widetilde{T}_{1\B}\big\|_{H^2(\R^2)}
     +\big\| \widetilde{T}_{2\B}\big\|_{H^2(\R^2)}\Big)\\
&\le C\Big(\big\| F_{1\B}\big\|_{L^2(\R^2)}+\big\| F_{2\B}\big\|_{L^2(\R^2)}\Big)\\
&\le C\eps_\B^4+\delta_{\B}
\ \ \hbox{uniformly in} \ \ \R^2\ \ \hbox{as} \ \ \B\nearrow \B^*,
\end{split}
\end{equation}
where $\delta_{\B}>0$ satisfies $\delta_{\B}=o(|c_{2\B}|)$ as $\B\nearrow\B^*$.

On the other hand, using \eqref{v1Bv2B.decay}, \eqref{I1B I2B tidu decay},
\eqref{TjB perp bounded} and Lemma \ref{lem.A.1}, we obtain from \eqref{v1B v2B}, \eqref{v1Bv2B.equation} and \eqref{TjB decompose} that
\begin{equation}\label{cijB bounded-1}
\begin{split}
0&=\inte \frac{\partial V_\Om(x)}{\partial x_2}
    \big(|v_{1\B}|^2+|v_{2\B}|^2\big)\\
&=\sum^2_{j=1}\inte \frac{\partial V_\Om(x)}{\partial x_2}
    \big(Q_{j\B}^2+T_{j\B}^2+2Q_{j\B}T_{j\B}+I_{j\B}^2\big)\\
&=\Big[\sum^2_{l,j=1}\inte \frac{\partial V_\Om(x)}{\partial x_2}
                             c_{l\B}\frac{\partial Q_{j\B}^2}{\partial x_l}
+O(\eps_\B^4)+o(|c_{2\B}|)\Big]\\
&=c_{2\B}\big(C+o(1)\big)+O(\eps_\B^4)\ \ \hbox{as} \ \ \B\nearrow\B^*,
\end{split}
\end{equation}
where $C>0$ is independent of $\B$.  It then follows from \eqref{cijB bounded-1} that
\begin{equation*}
c_{2\B}=O(\eps_\B^4)
\ \ \hbox{as}\ \ \B\nearrow \B^*,
\end{equation*}
which implies that \eqref{cijB bounded} holds true.
Combining \eqref{cijB bounded} and \eqref{TjB perp bounded}, we obtain from
\eqref{TjB decompose} that the claim \eqref{TjB bounded} holds true.

Denote
\begin{equation}\label{mathcal TjB}
\mathcal{T}_{j\B}(x):=T_{j\B}(x)-\eps_\B^4\rho_{j\B}\psi_{0}(x),\ \ j=1,2,
\end{equation}
where $\rho_{j\B}>0$ is defined by \eqref{rho jB define}, and
$\psi_{0}(x)\in C^2(\R^2)\cap L^\infty(\R^2)$ is a solution of
\eqref{psi 0.equation}.
One can derive from $\nabla \psi_{0}(0)=0$ and the property
\eqref{ker line oper L} (see also \cite[Lemma 4.1]{Wei96}) that $\psi_{0}(x)$ is unique.
Furthermore, using the comparison principle, we deduce from \eqref{1.10decay:w} and \eqref{psi 0.equation} that
\begin{equation}\label{psi 0. deacy}
\big|\psi_{0}(x)\big|,\ \big|\nabla\psi_{0}(x)\big|\le Ce^{-\delta|x|}
\ \ \mbox{in}\ \ \R^2, \ \ \mbox{where}\ \ \frac{4}{5}<\delta <1.
\end{equation}
Note from \eqref{T1B T2B equation}, \eqref{psi 0.equation} and \eqref{mathcal TjB} that
\begin{equation}\label{mathcal T1B T2B.equation}
\left\{\begin{array}{lll}
\widetilde{\mathcal{L}}_{1\B}(\mathcal{T}_{1\B},\mathcal{T}_{2\B})
=F_{1\B}(x)+\eps_\B^4V(x)Q_{1\B}:=\mathcal{F}_{1\B}(x)
\quad \mbox{in}\,\  \R^2,\\[3mm]
\widetilde{\mathcal{L}}_{2\B}(\mathcal{T}_{2\B},\mathcal{T}_{1\B})
=F_{2\B}(x)+\eps_\B^4V(x)Q_{2\B}:=\mathcal{F}_{2\B}(x)
\quad \mbox{in}\,\  \R^2.
\end{array}\right.
\end{equation}
By \eqref{v1Bv2B.decay}, \eqref{I1B I2B tidu decay} and \eqref{TjB bounded}, we deduce from \eqref{F1B F2B equation} that the terms $\mathcal{F}_{1\B}(x)$ and $\mathcal{F}_{2\B}(x)$ satisfy
\begin{equation}\label{FJB DECAY}
\left\{\begin{array}{lll}
\begin{split}
 \ds|\mathcal{F}_{1\B}(x)|:&=\Big|F_{1\B}(x)+\eps_\B^4V(x)Q_{1\B}\Big|
 \le C_{1\B}\eps_\B^4e^{-\frac{1}{10}|x|}\\[3mm]
 \ds|\mathcal{F}_{2\B}(x)|:&=\Big|F_{2\B}(x)+\eps_\B^4V(x)Q_{2\B}\Big|
 \le C_{2\B}\eps_\B^4e^{-\frac{1}{10}|x|}
\end{split}
\end{array}\right.
\ \ \hbox{uniformly in}\ \ \R^2\ \ \hbox{as}\ \  \B\nearrow \B^*,
\end{equation}
where $C_{j\B}>0$ satisfies $C_{j\B}=o(1)$ as $\B\nearrow\B^*$ for $j=1,2$.
Similar to \eqref{TjB bounded}, one can derive from \eqref{mathcal T1B T2B.equation} and \eqref{FJB DECAY} that
\begin{equation}\label{mathcal TjB bounded}
\big|\mathcal{T}_{j\B}(x)\big|\leq C_{j\B}\eps_\B^4
\quad \hbox{uniformly in}\ \  \R^2\ \ \hbox{as}\ \ \B\nearrow \B^*,
\ \ j=1,2,
\end{equation}
where $C_{j\B}>0$ satisfies $C_{j\B}=o(1)$ as $\B\nearrow\B^*$.
In view of \eqref{mathcal TjB}, we obtain from \eqref{mathcal TjB bounded} that \eqref{T1B T2B expan} holds true.

2. By the argument of proving \eqref{3.59}, one can deduce from
\eqref{mathcal T1B T2B.equation}--\eqref{mathcal TjB bounded} that
\begin{equation}\label{mathcal TjB decay}
\big| \mathcal{T}_{j\B}(x)\big|,\ \ \big|\nabla \mathcal{T}_{j\B}(x)\big|
\leq C_{j\B} \eps_\B^4e^{-\frac{1}{11} |x|}\quad \hbox{uniformly in}\ \ \R^2
\ \ \hbox{as}\, \ \B\nearrow \B^*,\ \  j=1,2,
\end{equation}
where $C_{j\B}>0$ satisfies $C_{j\B}=o(1)$ as $\B\nearrow\B^*$. Note from \eqref{mathcal TjB} that
\[
x^\bot\cdot\nabla T_{j\B}
=x^\bot\cdot\nabla \big[\mathcal{T}_{j\B}+\eps_\B^4\rho_{j\B}\psi_{0} \big]
\quad\hbox{in}\ \ \R^2,\ \ j=1,2.\]
We then obtain from \eqref{psi 0. deacy} and \eqref{mathcal TjB decay}  that
\begin{equation}\label{x bot cdot tidu TjB}
|x^\bot\cdot\nabla T_{j\B}|\leq C\eps_\B^4e^{-\frac {1}{12}|x|}
\quad \hbox{uniformly in}\, \  \R^2\,\ \hbox{as}\, \ \B\nearrow \B^*,
\ \ j=1,2.
\end{equation}
By \eqref{G6}, \eqref{v1B gama w v2B 1gama w}, \eqref{v1Bv2B.decay} and
\eqref{x bot cdot tidu TjB}, the similar argument of proving \cite [Lemma 2.3] {GLP} yields from \eqref{I1B I2B.equation}  that
\begin{equation}\label{IjB decay}
\big| I_{j\B}(x)\big|,\ \ \big| \nabla I_{j\B}(x)\big|
\leq C\eps_\B^6e^{-\frac {1}{14}|x|}
\quad \hbox{uniformly in}\, \ \R^2\,\ \hbox{as}\, \ \B\nearrow\B^*,\ \ j=1,2.
\end{equation}
It then follows from \eqref{IjB decay} that
\eqref{I1B I2B decay} holds true, and Lemma \ref{lem 1} is therefore proved.\qed

We now give the following refined estimate of $(I_{1\B},I_{2\B})$ as $\B\nearrow\B^*$.
\begin{lem}\label{lem 2}
Under the assumptions of Theorem \ref{thm-expan solution}, the imaginary part $(I_{1\B},I_{2\B})$ of \eqref{v1B v2B} satisfies
\begin{equation}\label{IjB expan}
I_{j\B}(x):=\eps_\B^6\,\Omega\,\rho_{j\B}\phi_0(x) +o(\eps_\B^6)
\quad \hbox{in}\, \ \R^2 \ \ \mbox{as} \ \ \B\nearrow \B^*,\ \ j=1,2,
\end{equation}
where $\rho_{j\B}>0$ is as in \eqref{rho jB define}, and $\phi_{0}(x)\in C^2(\R^2)\cap L^\infty(\R^2)$ solves uniquely
\begin{equation}\label{phi 0.equation}
\inte \phi_{0}wdx=0,\quad
\mathcal{L}\phi_{0}=-\big(x^\perp \cdot \nabla \psi_{0}\big)
\quad \hbox{in}\  \ \R^2,
\end{equation}
where the operator $\mathcal{L}$ is defined by \eqref{G4}, and
$\psi_{0}(x)\in C^2(\R^2)\cap L^\infty(\R^2)$ is given by \eqref{psi 0.equation}.
\end{lem}
Since the proof of Lemma \ref{lem 2} is similar to \cite[Lemma 2.3]{Guo}, we omit it for simplicity.
Applying Lemmas \ref{lem 1} and \ref{lem 2}, we next address the refined estimate of $(T_{1\B},T_{2\B})$ in \eqref{v1B v2B} as $\B\nearrow\B^*$.
\begin{lem}\label{lem 3}
Under the assumptions of Theorem \ref{thm-expan solution}, we have
\begin{equation}\label{TjB expan}
T_{j\B}(x):=\rho_{j\B}\big[\eps_\B^4\psi_0(x)+\eps_\B^8\psi_1(x)\big]+O(\eps_\B^{12})
\ \ \hbox{as}\ \ \B\nearrow\B^*,\ \ j=1,2,
\end{equation}
where $\rho_{j\B}>0$ is as in \eqref{rho jB define}, $\psi_{0}(x)\in C^2(\R^2)\cap L^\infty(\R^2)$ is given by
\eqref{psi 0.equation}, and $\psi_{1}(x)\in C^2(\R^2)\cap L^\infty(\R^2)$ solves uniquely
\begin{equation}\label{psi l.equation}
\nabla\psi_1(0)=0,\quad
\widetilde{\mathcal{L}} \psi_1(x)
=-V(x)\psi_0+3\psi_0^2w+\Om^2\big(x^\perp\cdot\nabla\phi_0\big)
 \ \ \mbox{in}\ \ \R^2.
\end{equation}
Here the operator $\widetilde{\mathcal{L}}$ is defined by \eqref{line oper L}, and $\phi_{0}(x)\in C^2(\R^2)\cap L^\infty(\R^2)$ is as in
\eqref{phi 0.equation}.
\end{lem}

\noindent{\bf Proof.}
Set
\begin{equation}\label{hat TjB}
\hat T_{j\B}(x):=T_{j\B}(x)-\rho_{j\B}\eps_\B^4\psi_{0}(x),
\ \ j=1,2,
\end{equation}
where $\rho_{j\B}>0$ is given by \eqref{rho jB define}, $T_{j\B}(x)$ and $\psi_0(x)$ are as in \eqref{v1B v2B} and \eqref{psi 0.equation}, respectively. We then derive from \eqref{v1B v2B}, \eqref{T1B T2B equation} and \eqref{psi 0.equation} that $\big(\hat T_{1\B},\hat T_{2\B}\big)$ satisfies
\begin{equation}\label{hat T1B T2B equation}
\left\{\begin{array}{lll}
\begin{split}
\widetilde{\mathcal{L}}_{1\B}\big(\hat T_{1\B},\hat T_{2\B}\big)
&=\ds F_{1\B}(x)+\eps_\B^4V(x)Q_{1\B}\\
&=\ds-\eps_\B^4V(x)T_{1\B}
     +\frac{a_1}{a^*}I_{1\B}^2R_{1\B}+\frac{\B}{a^*}I_{2\B}^2R_{1\B}
     +\eps_\B^2\Om\,\big(x^\perp\cdot\nabla I_{1\B}\big)\\[3mm]
&\quad+\ds\Big(\frac{3a_1}{a^*}Q_{1\B}T_{1\B}^2+\frac{2\B}{a^*}Q_{2\B}T_{1\B}T_{2\B}
      +\frac{\B}{a^*}Q_{1\B}T_{2\B}^2\\[3mm]
&\qquad\quad+\ds\frac{a_1}{a^*}T_{1\B}^3+\frac{\B}{a^*}T_{2\B}^2T_{1\B}\Big)
\quad \hbox{in} \ \ \R^2,\\[5mm]
\widetilde{\mathcal{L}}_{2\B}\big(\hat T_{2\B},\hat T_{1\B}\big)
&=\ds F_{2\B}(x)+\eps_\B^4V(x)Q_{2\B}\\
&=\ds-\eps_\B^4V(x)T_{2\B}
     +\frac{a_2}{a^*}I_{2\B}^2R_{2\B}+\frac{\B}{a^*}I_{1\B}^2R_{2\B}
     +\eps_\B^2\Om\,\big(x^\perp\cdot\nabla I_{2\B}\big)\\[3mm]
&\quad+\ds\Big(\frac{3a_2}{a^*}Q_{2\B}T_{2\B}^2+\frac{2\B}{a^*}Q_{1\B}T_{2\B}T_{1\B}
      +\frac{\B}{a^*}Q_{2\B}T_{1\B}^2\\[3mm]
&\qquad\quad+\ds\frac{a_2}{a^*}T_{2\B}^3+\frac{\B}{a^*}T_{1\B}^2T_{2\B}\Big)
\quad \hbox{in} \ \ \R^2,
\end{split}
\end{array}\right.
\end{equation}
where the parts $(\cdot)$ are lower orders of $T_{j\B}$ as $\B\nearrow\B^*$ for $j=1,2$.
Using Lemmas \ref{lem 1} and \ref{lem 2}, we obtain from \eqref{hat T1B T2B equation} that as $\B\nearrow\B^*$,
\begin{equation}\label{hat T1B T2B equation-1}
\left\{\begin{array}{lll}
\begin{split}
\widetilde{\mathcal{L}}_{1\B}\big(\hat T_{1\B},\hat T_{2\B}\big)
&=\eps_\B^8\rho_{1\B}\big[-V(x)\psi_0+3\psi_0^2w
                          +\Omega^2\,\big(x^\bot\cdot \nabla \phi_0\big)\big]
  +o(\eps_\B^8),\\[3mm]
\widetilde{\mathcal{L}}_{2\B}\big(\hat T_{2\B},\hat T_{1\B}\big)
&=\eps_\B^8\rho_{2\B}\big[-V(x)\psi_0+3\psi_0^2w
                          +\Omega^2\,\big(x^\bot\cdot \nabla \phi_0\big)\big]
  +o(\eps_\B^8),
\end{split}
\end{array}\right.
\end{equation}
where $\psi_{0}(x)$ and $\phi_{0}(x)$ are  as in \eqref{psi 0.equation} and
\eqref{phi 0.equation}, respectively.
Following \eqref{hat T1B T2B equation-1}, the same argument of Lemma \ref{lem 1} then yields that
\begin{equation}\label{hat T1B T2B expan}
\hat T_{j\B}=\rho_{j\B}\eps_\B^8\psi_1+o(\eps_\B^8)
\quad \hbox{in}\, \ \R^2 \ \ \hbox{as} \ \ \B\nearrow\B^*,\ \ j=1,2,
\end{equation}
where $\psi_{1}(x)\in C^2(\R^2)\cap L^\infty(\R^2)$ is given by
\eqref{psi l.equation}.

Combining \eqref{hat TjB} and \eqref{hat T1B T2B expan}, we obtain that
\begin{equation}\label{TjB expan-1}
T_{j\B}=\rho_{j\B}\big(\eps_\B^4\psi_0+\eps_\B^8\psi_1\big)+o(\eps_\B^{8})
\quad \hbox{in}\, \ \R^2 \ \ \hbox{as}\ \ \B\nearrow\B^*,\ \ j=1,2.
\end{equation}
Applying \eqref{TjB expan-1}, the similar argument of Lemma \ref{lem 2} then yields that
\begin{equation}\label{IjB expan-1}
I_{j\B}(x):=\eps_\B^6\,\Omega\,\rho_{j\B}\phi_0(x) +O(\eps_\B^{10})
\quad \hbox{in}\, \ \R^2 \ \ \mbox{as} \ \ \B\nearrow \B^*,\ \ j=1,2.
\end{equation}
By \eqref{TjB expan-1} and \eqref{IjB expan-1}, we derive from \eqref{hat T1B T2B equation} that as $\B\nearrow\B^*$,
\begin{equation}\label{hat T1B T2B equation-1-1}
\left\{\begin{array}{lll}
\begin{split}
\widetilde{\mathcal{L}}_{1\B}\big(\hat T_{1\B},\hat T_{2\B}\big)
&=\eps_\B^8\rho_{1\B}\big[-V(x)\psi_0+3\psi_0^2w
                          +\Omega^2\,\big(x^\bot\cdot \nabla \phi_0\big)\big]
  +O(\eps_\B^{12}),\\[3mm]
\widetilde{\mathcal{L}}_{2\B}\big(\hat T_{2\B},\hat T_{1\B}\big)
&=\eps_\B^8\rho_{2\B}\big[-V(x)\psi_0+3\psi_0^2w
                          +\Omega^2\,\big(x^\bot\cdot \nabla \phi_0\big)\big]
  +O(\eps_\B^{12}).
\end{split}
\end{array}\right.
\end{equation}
Similar to \eqref{hat T1B T2B expan}, one can derive from \eqref{hat T1B T2B equation-1-1} that
\begin{equation*}
\hat T_{j\B}=\rho_{j\B}\eps_\B^8\psi_1+O(\eps_\B^{12})
\quad \hbox{in}\, \ \R^2 \ \ \hbox{as} \ \ \B\nearrow\B^*,\ \ j=1,2,
\end{equation*}
together with \eqref{hat TjB}, which yields that
\eqref{TjB expan} holds true.
Furthermore, the uniqueness of $\psi_{1}(x)$ follows from $\nabla \psi_{1}(0)=0$ and the property \eqref{ker line oper L}, see also \cite[Lemma 4.1]{Wei96}.
This completes the proof of Lemma \ref{lem 3}.
\qed

\subsection{Proof of Theorem \ref{thm-expan solution}}
This main purpose of this subsection is to complete the proof of Theorem \ref{thm-expan solution}. We first establish the following refined estimate of $\alpha_\B^2\mu_\B$ as $\B\nearrow\B^*$.

\begin{lem}\label{lem 4}
Under the assumptions of Theorem \ref{thm-expan solution}, the term $\mu_\B\alpha_\B^2$ in \eqref{eps2mu-1} satisfies
\begin{equation}\label{1 muB alphaB2 expan}
1+\mu_\B\alpha_\B^2
=\Big(\frac{3\langle\widetilde{\mathcal{L}}\psi_0,\psi_0\rangle}{2\lambda_0}
     +\frac{\lambda_0\big(4\gamma_1\gamma_2-1\big)}
           {\big(2\B^*-a_1-a_2\big)8\gamma_1^2\gamma_2^2}\Big)\alpha_\B^4
     +O(\alpha_\B^8)
\ \ \hbox{as}\ \ \B\nearrow\B^*,
\end{equation}
where $\alpha_\B=\Big[\frac{2\gam_1\gam_2}{\lam_0}(\B^*-\B)\Big]^\frac{1}{4}>0$ is given by \eqref{1.16def:beta.V.eps}, the operator $\widetilde{\mathcal{L}}$ is defined by \eqref{line oper L}, $\psi_{0}(x)\in C^2(\R^2)\cap L^\infty(\R^2)$ is as in \eqref{psi 0.equation}, $\lambda_0>0$ is defined by \eqref{1.14lam0}, and $0<\gamma_1,\gamma_2<1$ are as in \eqref{1.13gamma}.
\end{lem}
\noindent{\bf Proof.}
We first obtain from \eqref{v1B v2B-1} and Lemma \ref{lem 2} that
\begin{equation}\label{R1BR2B}
\inte \big(R_{1\B}^2+R_{2\B}^2\big)
=\inte \big(|v_{1\B}|^2+|v_{2\B}|^2-I_{1\B}^2-I_{2\B}^2\big)
=a^*-O(\eps_\B^{12})\ \ \hbox{as}\ \ \B\nearrow\B^*.
\end{equation}
On the other hand, it follows from \eqref{Q1BQ2B}, \eqref{v1B v2B} and Lemma \ref{lem 3} that
\begin{equation}\label{R1BR2B-1}
\begin{split}
\inte \big(R_{1\B}^2+R_{2\B}^2\big)
&=\inte \big(\rho_{1\B}^2+\rho_{2\B}^2\big)
  \Big[w^2+2\eps_\B^4w\psi_0+\eps_\B^8\big(\psi_0^2+2\psi_1w\big)\Big]dx+O(\eps_\B^{12})\\
&=a^*\big(\rho_{1\B}^2+\rho_{2\B}^2\big)
  +\inte \big(\rho_{1\B}^2+\rho_{2\B}^2\big)
  \Big[2\eps_\B^4w\psi_0+\eps_\B^8\big(\psi_0^2+2\psi_1w\big)\Big]\\
&\quad+O(\eps_\B^{12})\ \ \hbox{as} \ \ \B\nearrow\B^*,
\end{split}
\end{equation}
where $\psi_{0}(x)\in C^2(\R^2)\cap L^\infty(\R^2)$ and $\psi_{1}(x)\in C^2(\R^2)\cap L^\infty(\R^2)$ are as in \eqref{psi 0.equation} and \eqref{psi l.equation}, respectively.
We then get from \eqref{R1BR2B} and \eqref{R1BR2B-1} that
\begin{equation}\label{1-alpha12-alpha22}
\frac{(\rho_{1\B}^2+\rho_{2\B}^2)^{-1}-1}{\eps_\B^4}
=\frac{1}{a^*}\Big[\inte 2w\psi_0+\eps_\B^4\inte\big(\psi_0^2+2\psi_1w\big)\Big]
  +O(\eps_\B^8)
\ \ \hbox{as}\ \ \B\nearrow\B^*.
\end{equation}

We now simplify the coefficients of \eqref{1-alpha12-alpha22}.
We derive from \eqref{1.14lam0} and \eqref{psi 0.equation} that
\begin{equation}\label{1}
\begin{split}
\inte 2w\psi_0
&=-\inte \widetilde{\mathcal{L}}\big(w+x\cdot\nabla w\big)\psi_0
 =-\inte \big(w+x\cdot\nabla w\big)\widetilde{\mathcal{L}}\psi_0\\
&=\inte \big(w+x\cdot\nabla w\big)V(x)w
 =-\lambda_0,
\end{split}
\end{equation}
since $\widetilde{\mathcal{L}}\big(w+x\cdot\nabla w\big)=-2w$ and $x\cdot\nabla V(x)=2V(x)$. It follows from
\eqref{psi 0.equation}, \eqref{psi 0. deacy} and \eqref{psi l.equation} that
\begin{equation}\label{2}
\begin{split}
\inte\big(\psi_0^2+2\psi_1w\big)
&=\inte\big[\psi_0^2-\widetilde{\mathcal{L}}\big(w+x\cdot\nabla w\big)\psi_1\big]
 =\inte\big[\psi_0^2-\big(w+x\cdot\nabla w\big)\widetilde{\mathcal{L}}\psi_1\big]\\
&=\inte\Big[\psi_0^2-\big(w+x\cdot\nabla w\big)
       \Big(-V(x)\psi_0+3\psi_0^2w+\Om^2\big(x^\perp\cdot\nabla\phi_0\big)\Big)\Big]\\
&=\inte\Big[\psi_0^2-\big(w+x\cdot\nabla w\big)\Big(-V(x)\psi_0+3\psi_0^2w\Big)\Big]\\
&=\inte\Big[\psi_0^2+V(x)w\psi_0-3w^2\psi_0^2\\
&\qquad\quad-w\big(2V(x)\psi_0+x\cdot\nabla(V(x)\psi_0)\big)
            -\frac{3}{2}(x\cdot\nabla w^2)\psi_0^2\Big]\\
&=\inte\Big[\psi_0^2-3V(x)w\psi_0-3w^2\psi_0^2
  +(x\cdot\nabla \psi_0)\widetilde{\mathcal{L}}\psi_0
  -\frac{3}{2}(x\cdot\nabla w^2)\psi_0^2\Big]\\
&=\inte\Big[\psi_0^2-3V(x)w\psi_0-3w^2\psi_0^2-\psi_0^2\\
&\qquad\quad-\frac{3}{2}(x\cdot\nabla\psi_0^2)w^2
            -\frac{3}{2}(x\cdot\nabla w^2)\psi_0^2\Big]\\
&=-\inte3V(x)w\psi_0=3\langle\widetilde{\mathcal{L}}\psi_0,\psi_0\rangle.
\end{split}
\end{equation}
Using \eqref{1} and \eqref{2}, the estimate \eqref{1-alpha12-alpha22} is thus simplified into the following form
\begin{equation}\label{reduce.1-alpha12-alpha22}
\frac{(\rho_{1\B}^2+\rho_{2\B}^2)^{-1}-1}{\eps_\B^4}
=\frac{1}{a^*}\Big[-\lambda_0+3\eps_\B^4\langle\widetilde{\mathcal{L}}\psi_0,\psi_0\rangle\Big]
+O(\eps_\B^8) \ \ \hbox{as}\ \ \B\nearrow\B^*.
\end{equation}

On the other hand, the definition of $\rho_{j\B}$ in \eqref{rho jB define} yields that
\begin{equation}\label{cal 1}
(\rho_{1\B}^2+\rho_{2\B}^2)^{-1}-1
=\Big(\frac{a^*(\B-a_2)}{\B^2-a_1a_2}+\frac{a^*(\B-a_1)}{\B^2-a_1a_2}\Big)^{-1}-1
=\frac{1}{a^*}F(\B)-1,
\end{equation}
where $F(\B):=\ds\frac{\B^2-a_1a_2}{\B-a_1+\B-a_2}$.
Using Taylor's expansion, we deduce that
\begin{equation}\label{cal 2}
\begin{split}
F(\B)&=F(\B^*)+F'(\B^*)(\B-\B^*)+\frac{F''(\B^*)}{2}(\B-\B^*)^2+O(|\B-\B^*|^3)\\
     &=a^*+2\gamma_1\gamma_2(\B-\B^*)
       +\frac{1-4\gamma_1\gamma_2}{2\B^*-a_1-a_2}(\B-\B^*)^2+O(|\B-\B^*|^3) \ \ \hbox{as}\ \ \B\nearrow\B^*,\\
\end{split}
\end{equation}
where $\B^*:=a^*+\sqrt{(a^*-a_{1})(a^*-a_{2})}>0$ is as in \eqref{main 1.11def:beta*}.
It then follows from \eqref{cal 1} and \eqref{cal 2} that as $\B\nearrow\B^*$,
\begin{equation}\label{cal 3}
\frac{(\rho_{1\B}^2+\rho_{2\B}^2)^{-1}-1}{\eps_\B^4}
=-\frac{2\gamma_1\gamma_2}{a^*}\frac{\B^*-\B}{\eps_\B^4}
 +\frac{1-4\gamma_1\gamma_2}{a^*(2\B^*-a_1-a_2)}\frac{(\B^*-\B)^2}{\eps_\B^4}
 +\frac{O(|\B-\B^*|^3)}{\eps_\B^4}.
\end{equation}
We get from \eqref{reduce.1-alpha12-alpha22} and \eqref{cal 3} that
\begin{equation}\label{cal 4}
\begin{split}
\frac{\B^*-\B}{\eps_\B^4}
&=\frac{\lambda_0}{2\gamma_1\gamma_2}
-\frac{3\langle\widetilde{\mathcal{L}}\psi_0,\psi_0\rangle}{2\gamma_1\gamma_2}\eps_\B^4
+\frac{1-4\gamma_1\gamma_2}{(2\B^*-a_1-a_2)2\gamma_1\gamma_2}
\frac{(\B^*-\B)^2}{\eps_\B^4}\\
&\quad+\frac{O(|\B-\B^*|^3)}{\eps_\B^4}+O(\eps_\B^8)\\
&=\frac{\lambda_0}{2\gamma_1\gamma_2}\Big[
1-\frac{3\langle\widetilde{\mathcal{L}}\psi_0,\psi_0\rangle}{\lambda_0}\eps_\B^4
+\frac{\lambda_0\big(1-4\gamma_1\gamma_2\big)}{(2\B^*-a_1-a_2)4\gamma_1^2\gamma_2^2}
\eps_\B^4\Big]+O(\eps_\B^8) \ \ \hbox{as}\ \ \B\nearrow\B^*,
\end{split}
\end{equation}
where the relation \eqref{ep alpha B} is also used.
Using \eqref{ep alpha B} again, we obtain from \eqref{ep} and \eqref{cal 4} that
\begin{equation*}
\begin{split}
-\mu_\B\alpha_\B^2=\frac{\alpha_\B^2}{\eps_\B^2}
&=\Big(\frac{2\gamma_1\gamma_2}{\lambda_0}\Big)^{\frac{1}{2}}
  \Big(\frac{\B^*-\B}{\eps_\B^4}\Big)^{\frac{1}{2}}\\
&=1-\Big(\frac{3\langle\widetilde{\mathcal{L}}\psi_0,\psi_0\rangle}{2\lambda_0}
        +\frac{\lambda_0\big(4\gamma_1\gamma_2-1\big)}
              {(2\B^*-a_1-a_2)8\gamma_1^2\gamma_2^2}\Big)\eps_\B^4
   +O(\eps_\B^8)\\
&=1-\Big(\frac{3\langle\widetilde{\mathcal{L}}\psi_0,\psi_0\rangle}{2\lambda_0}
         +\frac{\lambda_0\big(4\gamma_1\gamma_2-1\big)}
               {(2\B^*-a_1-a_2)8\gamma_1^2\gamma_2^2}\Big)\alpha_\B^4
   +O(\alpha_\B^8) \ \ \hbox{as}\ \ \B\nearrow\B^*,
\end{split}
\end{equation*}
which implies that \eqref{1 muB alphaB2 expan} holds true, and Lemma \ref{lem 4} is thus proved.
\qed

We are now ready to complete the proof of Theorem \ref{thm-expan solution}.
\vskip 0.05truein

\noindent {\bf Proof of Theorem \ref{thm-expan solution}.}
We obtain from Lemma \ref{lem 4} that \eqref{alphaB2 muB expan} holds true, and hence the rest is to prove \eqref{solution u1B u2B expan}. It follows from \eqref{Q1BQ2B}, \eqref{v1B v2B}, \eqref{IjB expan-1} and Lemma \ref{lem 3} that for $j=1,2$,
\begin{equation}\label{ep v1B v2B expan}
v_{j\B}(x):=\sqrt{a^*}\eps_\B u_{j\B}(\eps_\B x)e^{i\widetilde{\theta}_{j\B}}
=\rho_{j\B}\Big\{w+\eps_\B^4\psi_0+O(\eps_\B^8)
+i\Big[\Omega\eps_\B^6\phi_0
+O(\eps_\B^{10})\Big]\Big\}
\end{equation}
in $L^\infty(\R^2, \mathbb{C})$ as $\B\nearrow\B^*$, where $\eps_\B:=\sqrt{\frac{1}{-\mu_\B}}>0$ is given by \eqref{ep}, $\widetilde{\theta}_{j\B}\in [0,2\pi)$ is chosen such that \eqref{wIjB} holds, and $\rho_{j\B}>0$ is defined by \eqref{rho jB define}. Here $\psi_0$ and $\phi_0$ are uniquely given by \eqref{psi 0.equation.thm} and \eqref{phi 0.equation.thm}, respectively.

Note from \eqref{alphaB2 muB expan} and \eqref{ep} that
\begin{equation}\label{frac alpha ep}
\frac{\alpha_\B}{\eps_\B}
=\sqrt{-\alpha_\B^2\mu_\B}
=1-\frac{1}{2}C(\lambda_0,a_1,a_2,\B^*)\alpha_\B^4+O(\alpha_\B^8)
\ \ \hbox{as}\ \ \B\nearrow\B^*,
\end{equation}
which further implies that
\begin{equation}\label{alpha ep 5}
|\eps_\B-\alpha_\B|\leq C\alpha_\B^5
\ \ \hbox{as}\ \ \B\nearrow\B^*.
\end{equation}
We then derive from \eqref{ep v1B v2B expan}--\eqref{alpha ep 5} that for $j=1,2$, \begin{equation}\label{alpha v1B v2B expan}
\begin{split}
&\quad\sqrt{a^*}\alpha_\B u_{j\B}(\alpha_\B x)e^{i\widetilde{\theta}_{j\B}}\\
&=\frac{\alpha_\B}{\eps_\B}v_{j\B}\Big(\frac{\alpha_\B}{\eps_\B}x\Big)\\
&=\Big[1-\frac{1}{2}C(\lambda_0,a_1,a_2,\B^*)\alpha_\B^4+O(\alpha_\B^8)\Big]\\
&\quad\cdot\rho_{j\B}\Big\{w\Big(\frac{\alpha_\B}{\eps_\B}x\Big)
+\eps_\B^4\psi_0\Big(\frac{\alpha_\B}{\eps_\B}x\Big)+O(\eps_\B^8)
+i\Big[\Omega\eps_\B^6\phi_0\Big(\frac{\alpha_\B}{\eps_\B}x\Big)+O(\eps_\B^{10})\Big]\Big\}\\
&=\Big[1-\frac{1}{2}C(\lambda_0,a_1,a_2,\B^*)\alpha_\B^4+O(\alpha_\B^8)\Big]\\
&\quad\cdot\rho_{j\B}\Big\{
       w-\frac{1}{2}C(\lambda_0,a_1,a_2,\B^*)\alpha_\B^4(x\cdot \nabla w)
        +\eps_\B^4\Big[\psi_0
        -\frac{1}{2}C(\lambda_0,a_1,a_2,\B^*)\alpha_\B^4(x\cdot \nabla \psi_0)\Big]\\
&\qquad\quad+O(\alpha_\B^8)
+i\Big[\Omega\eps_\B^6\Big(
   \phi_0-\frac{1}{2}C(\lambda_0,a_1,a_2,\B^*)\alpha_\B^4(x\cdot \nabla \phi_0)
   \Big)
+O(\alpha_\B^{10})\Big]\Big\}\\
&=\rho_{j\B}w+\rho_{j\B}\alpha_\B^4\Big[\psi_0
         -\frac{1}{2}C(\lambda_0,a_1,a_2,\B^*)\big(w+x\cdot\nabla w\big)\Big]
+O(\alpha_\B^{8})\\[3mm]
&\quad+i\Big[\rho_{j\B}\Omega\,\alpha_\B^6\phi_0+O(\alpha_\B^{10})\Big]
\ \ \hbox{as} \ \ \B\nearrow\B^*.
\end{split}
\end{equation}
Define
\begin{equation}\label{v1B v2B thm4.1}
w_{j\B}(x):=\sqrt{a^*}\alpha_\B u_{j\B}(\alpha_\B x)e^{i\theta_{j\B}},\ \ j=1,2,
\end{equation}
where $\theta_{j\B}\in [0,2\pi)$ is chosen such that \eqref{Sec 3 tilde theta-2-2} holds true.
It then follows from \eqref{Sec 3 tilde theta-2-2}, \eqref{v1B v2B-1} and \eqref{wIjB} that
\begin{equation}\label{tilde theta-2}
\inte wIm\big(w_{j\B}(x)-v_{j\B}(x)\big)
=0,\ \ j=1,2.
\end{equation}
On the other hand, we deduce from \eqref{ep v1B v2B expan} and
\eqref{alpha ep 5}--\eqref{v1B v2B thm4.1} that for $j=1,2$,
\begin{equation}\label{tilde theta-3}
\begin{split}
&\quad Im\big(w_{j\B}(x)-v_{j\B}(x)\big)\\
&=Im\Big\{\sqrt{a^*}\alpha_\B u_{j\B}(\alpha_\B x)e^{i\theta_{j\B}}
-\sqrt{a^*}\eps_\B u_{j\B}(\eps_\B x)e^{i\widetilde{\theta}_{j\B}}\Big\}\\
&=Im\Big\{\sqrt{a^*}\alpha_\B u_{j\B}(\alpha_\B x)e^{i\widetilde{\theta}_{j\B}}
e^{i(\theta_{j\B}-\widetilde{\theta}_{j\B})}
-\sqrt{a^*}\eps_\B u_{j\B}(\eps_\B x)e^{i\widetilde{\theta}_{j\B}}\Big\}\\
&=Im\Big\{\Big(\rho_{j\B}w+\rho_{j\B}\alpha_\B^4\Big[\psi_0
             -\frac{1}{2}C(\lambda_0,a_1,a_2,\B^*)\big(w+x\cdot\nabla w\big)\Big]
          +O(\alpha_\B^{8})\\
&\qquad\quad+i\Big[\rho_{j\B}\Omega\,\alpha_\B^6\phi_0+O(\alpha_\B^{10})\Big]\Big)
e^{i(\theta_{j\B}-\widetilde{\theta}_{j\B})}\\
&\qquad\quad-\rho_{j\B}\Big(w+\eps_\B^4\psi_0+O(\eps_\B^8)
+i\Big[\Omega\eps_\B^6\phi_0
+O(\eps_\B^{10})\Big]\Big)\Big\}\\
&=\big(\rho_{j\B}w+O(\alpha_\B^{4})\big)\sin(\theta_{j\B}-\widetilde{\theta}_{j\B})
    +\Big[\rho_{j\B}\Omega\,\alpha_\B^6\phi_0+O(\alpha_\B^{10})\Big]  \cos(\theta_{j\B}-\widetilde{\theta}_{j\B})\\
 &\quad   -\rho_{j\B}\Omega\,\alpha_\B^6\phi_0+O(\alpha_\B^{10})\ \ \hbox{as} \ \ \B\nearrow\B^*.
\end{split}
\end{equation}
Substituting \eqref{tilde theta-3} into \eqref{tilde theta-2} then yields that
\begin{equation}\label{tilde theta-4}
\big|\theta_{j\B}-\widetilde{\theta}_{j\B}\big|=O(\alpha_\B^{10})=o(\alpha_\B^{8})
\ \ \hbox{as} \ \ \B\nearrow\B^*,
\ \ j=1,2,
\end{equation}
where we have used the fact that $\inte \phi_0 w dx=0$ in view of \eqref{phi 0.equation.thm}.

Applying \eqref{alpha v1B v2B expan} and \eqref{tilde theta-4}, we conclude from \eqref{v1B v2B thm4.1} that \eqref{solution u1B u2B expan} holds true, and the proof of Theorem \ref{thm-expan solution} is therefore complete.
\qed

\section{Uniqueness of Solutions as $\B\nearrow\B^*$}
This section is to establish the following local uniqueness of solutions $(u_{1\B},u_{2\B})\in\m$ for the elliptic system \eqref{1.1equation system-1} satisfying \eqref{eps2mu-1} and \eqref{1.15lim:beta.V.u.exp-1} as $\B\nearrow\B^*$.

\begin{thm}\label{thm-equation.uniqueness}
Suppose $V(x)$ satisfies \eqref{V(x)} for some $\Lambda\ge1$, and assume $0<\Om<\Om^*:=2$ and $0<a_1,a_2<a^*$ are fixed. Then up to a constant phase, there exists a unique solution of the elliptic system \eqref{1.1equation system-1} satisfying \eqref{eps2mu-1} and \eqref{1.15lim:beta.V.u.exp-1} when $\B^*-\B>0$ is small enough.
\end{thm}

We remark that the local uniqueness, up to a constant phase, of Theorem \ref{thm-equation.uniqueness} holds in the following sense:
if $(u_{1,1\B},u_{1,2\B})\in\m$ and $(u_{2,1\B},u_{2,2\B})\in\m$ are two solutions of the system \eqref{1.1equation system-1} satisfying \eqref{eps2mu-1} and \eqref{1.15lim:beta.V.u.exp-1}, then
$(u_{1,1\B},u_{1,2\B})\equiv(u_{2,1\B}e^{i\phi_{1\B}},u_{2,2\B}e^{i\phi_{2\B}})$ for some constant phase $(\phi_{1\B},\phi_{2\B})\in[0,2\pi)\times[0,2\pi)$ when $\B^*-\B>0$ is small enough.

In order to prove Theorem \ref{thm-equation.uniqueness}, on the contrary, suppose that there exist two different solutions $(u_{1,1\B},u_{1,2\B})\in\m$ and $(u_{2,1\B},u_{2,2\B})\in\m$ of  the elliptic system \eqref{1.1equation system-1} satisfying \eqref{eps2mu-1} and \eqref{1.15lim:beta.V.u.exp-1}. Without loss of generality, we may assume
 \begin{equation}\label{5:17}
   u_{1,1\B}\not\equiv u_{2,1\B}e^{i\psi_{1\B}}\ \ \hbox{in $\R^2$ for any constant phase}\ \ \psi_{1\B}\in[0,2\pi).
 \end{equation}
We define for $j=1, 2$,
\begin{equation}\label{4u11.bar.ujBvjB.define}
\left\{\begin{array}{lll}
     \hat u_{j,1\B}(x):=\sqrt{a^*}\alpha_\B u_{j,1\B}(\alpha_\B x)
                        e^{i\theta_{j,1\B}}
=R_{j,1\B}(x)+iI_{j,1\B}(x),\\[3mm]
     \hat u_{j,2\B}(x):=\sqrt{a^*}\alpha_\B u_{j,2\B}(\alpha_\B x)
                        e^{i\theta_{j,2\B}}
=R_{j,2\B}(x)+iI_{j,2\B}(x),
   \end{array}\right.
   \end{equation}
where $\alpha_\B:=\Big[\frac{2\gamma_1\gamma_2}{\lambda_0}(\B^*-\B)\Big]^{\frac{1}{4}}
>0$ is as in \eqref{1.16def:beta.V.eps},
 $\big(R_{j,1\B},R_{j,2\B}\big)$ and
$\big(I_{j,1\B},I_{j,2\B}\big)$ denote the real and imaginary parts of
$\big(\hat u_{j,1\B},\hat u_{j,2\B}\big)$, respectively, and the constant phase $(\theta_{j,1\B},\theta_{j,2\B})\in [0,2\pi)\times[0,2\pi)$ is chosen properly such that
\begin{equation}\label{4u4}
\inte w(x)I_{j,1\B}(x)dx=\inte w(x)I_{j,2\B}(x)dx=0,\  \ j=1,2.
\end{equation}
We also define the following operator $\mathcal{L}_\B$:
\begin{equation}\label{LB}
\mathcal{L}_\B:=-\Delta+i\,\alpha^2_\B\Om\,(x^\perp\cdot\nabla)+\alpha_\B^4V(x)
\ \ \hbox{in} \ \ \R^2.
\end{equation}
Note from \eqref{1.1equation system-1}--\eqref{1.15lim:beta.V.u.exp-1} and \eqref{4u11.bar.ujBvjB.define}
that for $j=1,2$, $\big(\hat u_{j,1\B},\hat u_{j,2\B}\big)$ is a solution of the following system
\begin{equation}\label{4u14.bar.ujBvjB.equation}
\left\{\begin{array}{lll}
  \ds\quad \mathcal{L}_\B\hat u_{j,1\B}(x)
=\alpha^2_\B\mu_{j\B}\hat u_{j,1\B}(x)
 +\frac{a_1}{a^*}|\hat u_{j,1\B}|^2\hat u_{j,1\B}(x)
 +\frac{\B}{a^*}|\hat u_{j,2\B}|^2\hat u_{j,1\B}(x)
\quad \hbox{in}\  \ \R^2,\\[5mm]
  \ds\quad \mathcal{L}_\B\hat u_{j,2\B}(x)
=\alpha^2_\B\mu_{j\B}\hat u_{j,2\B}(x)
 +\frac{a_2}{a^*}|\hat u_{j,2\B}|^2\hat u_{j,2\B}(x)
 +\frac{\B}{a^*}|\hat u_{j,1\B}|^2\hat u_{j,2\B}(x)
\quad\hbox{in}\  \ \R^2,
\end{array}\right.
\end{equation}
satisfying
\begin{equation}\label{Sec 5 eps2mu-1}
\alpha_\B^2\mu_{j\B}\to-1\  \ \text{as}\  \ \B\nearrow\B^*,
\end{equation}
and
\begin{equation}\label{4u13.bar.ujBvjB.lim}
\big(\hat u_{j,1\B}(x),\hat u_{j,2\B}(x)\big)
\to\big(\sqrt{\gam_1}w(x),\sqrt{\gam_2}w(x)\big)
\ \ \hbox{strongly in} \ \ H^1(\R^2,\C)\cap L^\infty(\R^2,\C)
\end{equation}
as $\B\nearrow\B^*$, where $0<\gam_1,\gam_2<1$ are as in \eqref{1.13gamma}.

Similar to \eqref{v1Bv2B.decay} and \eqref{I1B I2B tidu decay},
one can deduce from \eqref{4u4} and \eqref{4u14.bar.ujBvjB.equation}--\eqref{4u13.bar.ujBvjB.lim} that there exists a constant $C>0$, independent of $0<\B<\B^*$, such that for $j,l=1,2$,
\begin{equation}\label{4u6.baruBvB.decay}
      |\hat u_{j,l\B}(x)|\leq Ce^{-\frac{2}{3}|x|} \quad \text{and} \quad
      |\nabla\hat u_{j,l\B}(x)|\leq Ce^{-\frac{1}{2}|x|}
      \  \ \text{uniformly in} \  \  \R^2\  \ \text{as}\  \ \B\nearrow\B^*,
     \end{equation}
     and
\begin{equation}\label{4u8.imaginary.decay}
      |I_{j,l\B}(x)|\leq C_{jl}(\alpha_\B)e^{-\frac{1}{4}|x|} \quad \text{and} \quad
      |\nabla I_{j,l\B}(x)|\leq C_{jl}(\alpha_\B)e^{-\frac{1}{8}|x|}
     \end{equation}
uniformly in $\R^2$ as $\B\nearrow\B^*$, where the constant $C_{jl}(\alpha_\B)$ satisfies $C_{jl}(\alpha_\B)=o(\alpha^2_\B)$ as $\B\nearrow\B^*$.
Inspired by \cite [Lemma 4.2] {GLWZ1}, we shall prove in Appendix A.2 that there exist constants $C_1>0$ and $C_2>0$, independent of $0<\B<\B^*$, such that
\begin{equation}\label{4u19.bar.ujBvjB.Linfty}
     C_1\|\hat u_{2,2\B}-\hat u_{1,2\B}\|_{L^\infty(\R^2)}
   \leq \|\hat u_{2,1\B}-\hat u_{1,1\B}\|_{L^\infty(\R^2)}
\leq C_2\|\hat u_{2,2\B}-\hat u_{1,2\B}\|_{L^\infty(\R^2)}\  \ \text{as}\  \ \B\nearrow\B^*.
     \end{equation}

Following the assumption \eqref{5:17}, one can deduce from (5.2) and (5.10) that \begin{equation}\label{5.18}
 \hat{u}_{1,1\B}\not\equiv \hat{u}_{2,1\B}\ \ \hbox{and}\ \ \hat{u}_{1,2\B}\not\equiv \hat{u}_{2,2\B}\ \ \hbox{in}\ \ \R^2.
\end{equation}
We next use \eqref{5.18} to derive a contradiction, which then implies that the assumption (5.1) is not true, and Theorem \ref{thm-equation.uniqueness} is thus proved.

\subsection{Analysis of a linearized problem}
In this subsection, we focus on the analysis of the linearized problem \eqref{4u29.xi1B.xi2B.equation} defined below.
In view of \eqref{5.18}, we
define
\begin{equation}\label{4u28.xi1B.xi2B.define}
 \left\{\begin{array}{lll}
  \xi_{1\B}(x):=\ds\frac{\hat u_{2,1\B}(x)-\hat u_{1,1\B}(x)}
  {\|\hat u_{2,1\B}-\hat u_{1,1\B}\|^{\frac{1}{2}}_{L^\infty(\R^2)}
   \|\hat u_{2,2\B}-\hat u_{1,2\B}\|^{\frac{1}{2}}_{L^\infty(\R^2)}}
  =R_{\xi_{1\B}}(x)+iI_{\xi_{1\B}}(x),\\[6mm]
 \xi_{2\B}(x):=\ds\frac{\hat u_{2,2\B}(x)-\hat u_{1,2\B}(x)}
  {\|\hat u_{2,1\B}-\hat u_{1,1\B}\|^{\frac{1}{2}}_{L^\infty(\R^2)}
   \|\hat u_{2,2\B}-\hat u_{1,2\B}\|^{\frac{1}{2}}_{L^\infty(\R^2)}}
  =R_{\xi_{2\B}}(x)+iI_{\xi_{2\B}}(x),
   \end{array}\right.
\end{equation}
where $\big(R_{\xi_{1\B}},R_{\xi_{2\B}}\big)$ and $\big(I_{\xi_{1\B}},I_{\xi_{2\B}}\big)$ denote the real and imaginary parts of $\big(\xi_{1\B},\xi_{2\B}\big)$, respectively.
We then derive from \eqref{4u14.bar.ujBvjB.equation} that $\big(\xi_{1\B},\xi_{2\B}\big)$ satisfies
\begin{equation}\label{4u29.xi1B.xi2B.equation}
 \left\{\begin{array}{lll}
\mathcal{L}_\B\xi_{1\B}
=\alpha^2_\B\mu_{2\B}\xi_{1\B}
+\ds\frac{\alpha^2_\B(\mu_{2\B}-\mu_{1\B})}
 {\|\hat u_{2,1\B}-\hat u_{1,1\B}\|^{\frac{1}{2}}_{L^\infty(\R^2)}
  \|\hat u_{2,2\B}-\hat u_{1,2\B}\|^{\frac{1}{2}}_{L^\infty(\R^2)}}
  \hat u_{1,1\B}\\[5mm]
\qquad\qquad+\ds\frac{a_1}{a^*}|\hat u_{2,1\B}|^2\xi_{1\B}
            +\frac{a_1}{a^*}\Big[R_{\xi_{1\B}}\big(R_{2,1\B}+R_{1,1\B}\big)
              +I_{\xi_{1\B}}\big(I_{2,1\B}+I_{1,1\B}\big)\Big]\hat u_{1,1\B}\\[3mm]
\qquad\qquad+\ds\frac{\B}{a^*}|\hat u_{1,2\B}|^2\xi_{1\B}
            +\frac{\B}{a^*}\Big[R_{\xi_{2\B}}\big(R_{2,2\B}+R_{1,2\B}\big)
            +I_{\xi_{2\B}}\big(I_{2,2\B}+I_{1,2\B}\big)\Big]\hat u_{2,1\B}
\ \ \hbox{in} \ \ \R^2,\\[5mm]
\mathcal{L}_\B\xi_{2\B}
=\alpha^2_\B\mu_{2\B}\xi_{2\B}
+\ds\frac{\alpha^2_\B(\mu_{2\B}-\mu_{1\B})}
 {\|\hat u_{2,1\B}-\hat u_{1,1\B}\|^{\frac{1}{2}}_{L^\infty(\R^2)}
  \|\hat u_{2,2\B}-\hat u_{1,2\B}\|^{\frac{1}{2}}_{L^\infty(\R^2)}}
  \hat u_{1,2\B}\\[5mm]
\qquad\qquad+\ds\frac{a_2}{a^*}|\hat u_{2,2\B}|^2\xi_{2\B}
            +\frac{a_2}{a^*}\Big[R_{\xi_{2\B}}\big(R_{2,2\B}+R_{1,2\B}\big)
             +I_{\xi_{2\B}}\big(I_{2,2\B}+I_{1,2\B}\big)\Big]\hat u_{1,2\B}\\[3mm]
\qquad\qquad+\ds\frac{\B}{a^*}|\hat u_{1,1\B}|^2\xi_{2\B}
            +\frac{\B}{a^*}\Big[R_{\xi_{1\B}}\big(R_{2,1\B}+R_{1,1\B}\big)
            +I_{\xi_{1\B}}\big(I_{2,1\B}+I_{1,1\B}\big)\Big]\hat u_{2,2\B}
\ \ \hbox{in} \ \ \R^2,
\end{array}\right.
\end{equation}
where the operator $\mathcal{L}_\B$ is given by \eqref{LB}.

One can obtain from \eqref{4u19.bar.ujBvjB.Linfty} that there exists a constant $C>0$, independent of $\B$, such that
\begin{equation}\label{4u31-1}
0\leq|\xi_{1\B}(x)|,\ |\xi_{2\B}(x)|\leq C<\infty\quad \text{and}
\quad |\xi_{1\B}(x)\xi_{2\B}(x)|\leq1\  \ \text{in}\  \ \R^2.
\end{equation}
By \eqref{Sec 5 eps2mu-1}, \eqref{4u6.baruBvB.decay} and \eqref{4u31-1}, the same argument of proving \eqref{4u25.eta.decay} in the Appendix yields that there exists a constant $C>0$, independent of $0<\B<\B^*$, such that for $j=1,2$,
\begin{equation}\label{4u31.xijB.estimates}
|\xi_{j\B}(x)|\leq Ce^{-\frac{2}{3}|x|} \quad \text{and} \quad
|\nabla \xi_{j\B}(x)|\leq Ce^{-\frac{1}{2}|x|}
\ \ \text{uniformly in} \  \  \R^2\  \ \text{as}\  \ \B\nearrow\B^*.
\end{equation}
Applying \eqref{4u31.xijB.estimates} of $\big(\xi_{1\B},\xi_{2\B}\big)$, we first study the following $L^\infty-$uniform convergence of $\big(\xi_{1\B},\xi_{2\B}\big)$ as $\B\nearrow\B^*$.

\begin{lem}\label{lem xijB.limit}
Assume $\big(\xi_{1\B},\xi_{2\B}\big)$ is defined by \eqref{4u28.xi1B.xi2B.define}. Then there exist a subsequence, still denoted by $\big\{\big(\xi_{1\B},\xi_{2\B}\big)\big\}$, of $\big\{\big(\xi_{1\B},\xi_{2\B}\big)\big\}$ and some constants $b_0$, $b_1$ and $b_2$ such that  for $j=1,2$,
\begin{equation}\label{4u32.xijB.limi}
\xi_{j\B}(x)\to
\sqrt{\gamma_j}
\Big(b_0(w+x\cdot\nabla w)+\sum^2_{l=1}b_l\frac{\partial w}{\partial x_l}\Big)
\ \ \hbox{uniformly in}\ \ \R^2\ \ \hbox{as}\ \ \B\nearrow\B^*,
\end{equation}
where $0<\gam_j<1$ is as in \eqref{1.13gamma}.
\end{lem}

{\noindent \bf Proof.}
We first claim that up to a subsequence if necessary, there exist constants $b_0$, $b_1$ and $b_2$ such that for $j=1,2$,
\begin{equation}\label{4u34}
\xi_{j\B}(x)\to
\sqrt{\gamma_j}
\Big(b_0(w+x\cdot\nabla w)+\sum^2_{l=1}b_l\frac{\partial w}{\partial x_l}\Big)
\ \ \hbox{uniformly in}\ \ C^1_{loc}(\R^2,\C)\ \ \hbox{as}\ \ \B\nearrow\B^*,
\end{equation}
where $0<\gam_j<1$ is as in \eqref{1.13gamma}.

In fact, using \eqref{Sec 5 eps2mu-1}, \eqref{4u6.baruBvB.decay}, \eqref{4u29.xi1B.xi2B.equation}, \eqref{4u31.xijB.estimates} and the standard elliptic regularity, the similar argument of proving \eqref{4u23-8} yields that up to a subsequence if necessary,
\begin{equation}\label{4u36}
   \xi_{j\B}=R_{\xi_{j\B}}+iI_{\xi_{j\B}}
\to\xi_{j0}=R_{\xi_{j0}}+iI_{\xi_{j0}}
\ \ \text{uniformly in}\  \ C^1_{loc}(\R^2,\C)\  \ \text{as}\  \ \B\nearrow\B^*,
\ \ j=1,2.
\end{equation}
Similar to \eqref{4u23-4}, we deduce from
\eqref{Sec 5 eps2mu-1}--\eqref{4u6.baruBvB.decay}, \eqref{4u31.xijB.estimates} and \eqref{4u36} that
\begin{equation}\label{4up51-1}
\begin{split}
&\quad\lim_{\B\nearrow\B^*}\frac{\alpha^2_\B(\mu_{2\B}-\mu_{1\B})}
 {\|\hat u_{2,1\B}-\hat u_{1,1\B}\|^{\frac{1}{2}}_{L^\infty(\R^2)}
  \|\hat u_{2,2\B}-\hat u_{1,2\B}\|^{\frac{1}{2}}_{L^\infty(\R^2)}}\\
&=-\frac{2a_1\gamma_1^{\frac{3}{2}}}{(a^*)^2}\inte w^3R_{\xi_{10}}dx
  -\frac{2a_2\gamma_2^{\frac{3}{2}}}{(a^*)^2}\inte w^3R_{\xi_{20}}dx\\
&\quad-2\B^*\frac{\gamma_2\sqrt{\gamma_1}}{(a^*)^2}\inte w^3R_{\xi_{10}}dx
      -2\B^*\frac{\gamma_1\sqrt{\gamma_2}}{(a^*)^2}\inte w^3R_{\xi_{20}}dx\\
&=-\Big(2\frac{\sqrt{\gamma_1}}{a^*}\inte w^3R_{\xi_{10}}dx
  +2\frac{\sqrt{\gamma_2}}{a^*}\inte w^3R_{\xi_{20}}dx\Big),
\end{split}
\end{equation}
where $0<\gam_1,\gam_2<1$ are as in \eqref{1.13gamma}, and we have used the facts that $a_1\gamma_1+\B^*\gamma_2=a^*$ and $a_2\gamma_2+\B^*\gamma_1=a^*$.
By \eqref{Sec 5 eps2mu-1}--\eqref{4u13.bar.ujBvjB.lim},
\eqref{4u31.xijB.estimates}, \eqref{4u36} and \eqref{4up51-1}, we derive from \eqref{4u29.xi1B.xi2B.equation} that  $(\xi_{10},\xi_{20})$ solves the following system
\begin{equation*}
 \left\{\begin{array}{lll}
\quad-\ds\Delta\xi_{10}+\xi_{10}
     -\frac{2a_1\gamma_1}{a^*}w^2R_{\xi_{10}}-\frac{a_1\gamma_1}{a^*}w^2\xi_{10}
     -\frac{\B^*\gamma_2}{a^*}w^2\xi_{10}
     -\frac{2\B^*\sqrt{\gamma_2\gamma_1}}{a^*}w^2R_{\xi_{20}}\\[3mm]
 =\ds-\sqrt{\gamma_1}w\Big(\frac{2\sqrt{\gamma_1}}{a^*}\inte w^3R_{\xi_{10}}dx
                          +\frac{2\sqrt{\gamma_2}}{a^*}\inte w^3R_{\xi_{20}}dx\Big)
\quad \hbox{in}\  \ \R^2,\\[5mm]
\quad-\ds\Delta\xi_{20}+\xi_{20}
     -\frac{2a_2\gamma_2}{a^*}w^2R_{\xi_{20}}-\frac{a_2\gamma_2}{a^*}w^2\xi_{20}
     -\frac{\B^*\gamma_1}{a^*}w^2\xi_{20}
     -\frac{2\B^*\sqrt{\gamma_2\gamma_1}}{a^*}w^2R_{\xi_{10}}\\[3mm]
 =\ds-\sqrt{\gamma_2}w\Big(\frac{2\sqrt{\gamma_1}}{a^*}\inte w^3R_{\xi_{10}}dx
                          +\frac{2\sqrt{\gamma_2}}{a^*}\inte w^3R_{\xi_{20}}dx\Big)
\quad \hbox{in}\  \ \R^2.
\end{array}\right.
\end{equation*}
This further implies that $(R_{\xi_{10}},R_{\xi_{20}})$ and
$(I_{\xi_{10}},I_{\xi_{20}})$ satisfy
\begin{equation}\label{4u37}
 \left\{\begin{array}{lll}
\quad-\ds\Delta R_{\xi_{10}}+R_{\xi_{10}}
     -\frac{3a_1\gamma_1}{a^*}w^2R_{\xi_{10}}
     -\frac{\B^*\gamma_2}{a^*}w^2R_{\xi_{10}}
     -\frac{2\B^*\sqrt{\gamma_2\gamma_1}}{a^*}w^2R_{\xi_{20}}\\[3mm]
    =-\ds\sqrt{\gamma_1}w\Big(\frac{2\sqrt{\gamma_1}}{a^*}\inte w^3R_{\xi_{10}}dx
     +\frac{2\sqrt{\gamma_2}}{a^*}\inte w^3R_{\xi_{20}}dx\Big)
\quad \hbox{in}\  \ \R^2,\\[5mm]
\quad-\ds\Delta R_{\xi_{20}}+R_{\xi_{20}}
     -\frac{3a_2\gamma_2}{a^*}w^2R_{\xi_{20}}
     -\frac{\B^*\gamma_1}{a^*}w^2R_{\xi_{20}}
     -\frac{2\B^*\sqrt{\gamma_2\gamma_1}}{a^*}w^2R_{\xi_{10}}\\[3mm]
    =-\ds\sqrt{\gamma_2}w\Big(\frac{2\sqrt{\gamma_1}}{a^*}\inte w^3R_{\xi_{10}}dx
     +\frac{2\sqrt{\gamma_2}}{a^*}\inte w^3R_{\xi_{20}}dx\Big)
\quad \hbox{in}\  \ \R^2,\\[5mm]
\quad-\Delta I_{\xi_{10}}+I_{\xi_{10}}
-w^2I_{\xi_{10}}=0
\quad \hbox{in}\  \ \R^2,\\[5mm]
\quad-\Delta I_{\xi_{20}}+I_{\xi_{20}}-w^2I_{\xi_{20}}=0
\quad \hbox{in}\  \ \R^2.
\end{array}\right.
\end{equation}

Moreover, it follows from \cite [Lemma 2.2 and Theorem 3.1] {DW} that the solution of the following linearized system
\begin{equation}\label{4u39}
 \left\{\begin{array}{lll}
\mathcal{L}_1(\phi_1,\phi_2)
:=\!-\ds\Delta \phi_1+\phi_1-\frac{3a_1\gamma_1}{a^*}w^2\phi_1
    -\frac{\B^*\gamma_2}{a^*}w^2\phi_1
    -\frac{2\B^*\sqrt{\gamma_2\gamma_1}}{a^*}w^2\phi_2\!=\!0
\  \ \hbox{in}\  \ \R^2,\\[5mm]
\mathcal{L}_2(\phi_2,\phi_1)
:=\!-\ds\Delta \phi_2+\phi_2-\frac{3a_2\gamma_2}{a^*}w^2\phi_2
    -\frac{\B^*\gamma_1}{a^*}w^2\phi_2
    -\frac{2\B^*\sqrt{\gamma_2\gamma_1}}{a^*}w^2\phi_1\!=\!0
\  \ \hbox{in}\  \ \R^2,
\end{array}\right.
\end{equation}
is given by
\begin{equation}\label{4u40}
\begin{pmatrix}
  \phi_1\\[5mm]
  \phi_2
\end{pmatrix}
=\sum^2_{l=1}b_l
\begin{pmatrix}
\ds\sqrt{\gamma_1}\,\frac{\partial w}{\partial x_l}\\[5mm]
\ds\sqrt{\gamma_2}\,\frac{\partial w}{\partial x_l}
\end{pmatrix}
\end{equation}
for some constants $b_l\in\R$. Furthermore, one can check that
\begin{equation}\label{4u41}
 \left\{\begin{array}{lll}
\mathcal{L}_1
\big(\sqrt{\gamma_1}(w+x\cdot\nabla w),\sqrt{\gamma_2}(w+x\cdot\nabla w)\big)
=-2\sqrt{\gamma_1}w,\\[3mm]
\mathcal{L}_2
\big(\sqrt{\gamma_2}(w+x\cdot\nabla w),\sqrt{\gamma_1}(w+x\cdot\nabla w)\big)
=-2\sqrt{\gamma_2}w.
\end{array}\right.
\end{equation}
Note from \eqref{4u4} and \eqref{4u28.xi1B.xi2B.define} that $\int_{\R^2}w(x)I_{\xi_{j\B}}(x)dx=0$ for $j=1,2$, together with \eqref{4u31-1} and \eqref{4u36}, which further implies that
     \begin{equation}\label{4u43}
     \int_{\R^2}w(x)I_{\xi_{j0}}(x)dx=0,\ \ j=1,2.
     \end{equation}
Applying \eqref{G5}, we then conclude from \eqref{4u37}--\eqref{4u43} that there exist some constants $b_0$, $b_1$ and $b_2$ such that
\begin{equation*}
R_{\xi_{j0}}
=\sqrt{\gamma_j}\Big(b_0(w+x\cdot\nabla w)
+\sum^2_{l=1}b_l
\ds\,\frac{\partial w}{\partial x_l}\Big),
\quad\text{and}\quad
I_{\xi_{j0}}
\equiv0
\quad\text{in}\  \ \R^2,\ \ j=1,2,
\end{equation*}
and the claim \eqref{4u34} is therefore proved.

On the other hand, we deduce from  \eqref{1.10decay:w} and \eqref{4u31.xijB.estimates} that for any fixed sufficiently large $R>0$,
\begin{equation}\label{4u44}
\Big|\xi_{j\B}(x)
-\sqrt{\gamma_j}\Big(b_0(w+x\cdot\nabla w)
                     +\sum^2_{l=1}b_l\frac{\partial w}{\partial x_l}\Big)
\Big|\leq Ce^{-\frac{1}{12}R}e^{-\frac{1}{2}|x|}\ \ \hbox{in}\ \ \R^2/B_R(0)
\ \ \hbox{as}\ \ \B\nearrow\B^*,
\end{equation}
where $j=1,2$. Since $R>0$ is arbitrary,  we conclude from  \eqref{4u34} and \eqref{4u44}
that \eqref{4u32.xijB.limi} holds true, and Lemma \ref{lem xijB.limit} is thus proved.\qed

Applying the refined expansions of Theorem \ref{thm-expan solution} in Section 4, one is able to establish the following lemma.

\begin{lem}\label{lem 5}
Suppose $(u_{j,1\B},u_{j,2\B})\in\m$ is a normalized concentration solution of \eqref{1.1equation system-1} satisfying \eqref{eps2mu-1} and \eqref{1.15lim:beta.V.u.exp-1}, and define $\big(\hat u_{j,1\B},\hat u_{j,2\B}\big)$ by \eqref{4u11.bar.ujBvjB.define}, where $j=1,2$. Then the parameter $\mu_{j\B}$ of \eqref{4u14.bar.ujBvjB.equation} satisfies
\begin{equation}\label{alpha_B2 mu_jB expan}
\begin{split}
&\quad\frac{\alpha^2_\B(\mu_{2\B}-\mu_{1\B})}
{\|\hat u_{2,1\B}-\hat u_{1,1\B}\|^{\frac{1}{2}}_{L^\infty(\R^2)}
  \|\hat u_{2,2\B}-\hat u_{1,2\B}\|^{\frac{1}{2}}_{L^\infty(\R^2)}}\\
&=-\frac{a_1}{2(a^*)^2}\inte\big(R_{2,1\B}^2+R_{1,1\B}^2\big)
                            \big(R_{2,1\B}+R_{1,1\B}\big)R_{\xi_{1\B}}\\
&\quad-\frac{a_2}{2(a^*)^2}\inte\big(R_{2,2\B}^2+R_{1,2\B}^2\big)
                                \big(R_{2,2\B}+R_{1,2\B}\big)R_{\xi_{2\B}}\\
&\quad-\frac{\B}{(a^*)^2}\inte
\Big[\big(R_{2,1\B}+R_{1,1\B}\big)R_{2,2\B}^2R_{\xi_{1\B}}
    +\big(R_{2,2\B}+R_{1,2\B}\big)R_{1,1\B}^2R_{\xi_{2\B}}\Big]\\
&\quad+o(\alpha_\B^4)\ \ \hbox{as}\ \ \B\nearrow\B^*.
\end{split}
\end{equation}
\end{lem}
Since the proof of Lemma \ref{lem 5} is involved with complicated calculations, for simplicity we leave it to Appendix A.2.

\subsection{Proof of Theorem \ref{thm-equation.uniqueness}}
Following the estimates of the previous subsection, in this subsection we shall complete the proof of Theorem \ref{thm-equation.uniqueness}.

\vskip 0.05truein

\noindent {\bf Proof of Theorem \ref{thm-equation.uniqueness}.}
We shall carry out the rest proof of Theorem \ref{thm-equation.uniqueness} by the following two steps:
\vskip0.2cm

\noindent $\mathbf{Step\  \ 1}$. We claim that the constants $b_1$ and $b_2$ in \eqref{4u32.xijB.limi} satisfy
\begin{equation}\label{b0 b1 b2 0}
b_1=b_2=0.
\end{equation}

To prove the claim \eqref{b0 b1 b2 0}, we first rewrite \eqref{4u14.bar.ujBvjB.equation} as
\begin{equation*}
\left\{\begin{array}{lll}
  \quad -\ds\Big(\nabla-i\frac{\Om }{2}\alpha_\B^2x^\perp\Big)^2 \hat u_{j,1\B}(x)
        +\alpha_\B^4V_\Om(x)\hat u_{j,1\B}(x)\\[3mm]
=\ds\alpha^2_\B\mu_{j\B}\hat u_{j,1\B}(x)
 +\frac{a_1}{a^*}|\hat u_{j,1\B}|^2\hat u_{j,1\B}(x)
 +\frac{\B}{a^*}|\hat u_{j,2\B}|^2\hat u_{j,1\B}(x)
\quad \hbox{in}\  \ \R^2,\\[3mm]
  \quad -\ds\Big(\nabla-i\frac{\Om}{2}\alpha_\B^2 x^\perp\Big)^2 \hat u_{j,2\B}(x)
        +\alpha_\B^4V_\Om( x)\hat u_{j,2\B}(x)\\[3mm]
=\ds\alpha^2_\B\mu_{j\B}\hat u_{j,2\B}(x)
 +\frac{a_2}{a^*}|\hat u_{j,2\B}|^2\hat u_{j,2\B}(x)
 +\frac{\B}{a^*}|\hat u_{j,1\B}|^2\hat u_{j,2\B}(x)
\quad\hbox{in}\  \ \R^2,
\end{array}\right.
\end{equation*}
where $j=1,2$.
Similar to the argument of Lemma \ref{lem.A.1}, we derive from \eqref{Sec 5 eps2mu-1} and \eqref{4u6.baruBvB.decay} that for $j,k=1,2$,
\begin{equation}\label{pohozaev uj1B plus uj2B}
\inte\frac{\partial V_\Om( x)}{\partial x_k}
\big(|\hat u_{j,1\B}|^2+|\hat u_{j,2\B}|^2\big)=0\ \ \hbox{as}\ \ \B\nearrow\B^*.
\end{equation}
Since $V_{\Om}(x)$ is even in $x$, by \eqref{4u11.bar.ujBvjB.define}, \eqref{4u13.bar.ujBvjB.lim}, \eqref{4u8.imaginary.decay}, \eqref{4u31.xijB.estimates} and Lemma \ref{lem xijB.limit}, we conclude from
\eqref{pohozaev uj1B plus uj2B} that
as $\B\nearrow\B^*$,
\begin{equation*}
\begin{split}
0&= \inte \frac{\partial V_\Om( x)}{\partial x_k}
\Big[\big(R_{2,1\B}+R_{1,1\B}\big)R_{\xi_{1\B}}
    +\big(R_{2,2\B}+R_{1,2\B}\big)R_{\xi_{2\B}}
\Big]\\
&\quad+\inte \frac{\partial V_\Om( x)}{\partial x_k}
\Big[\big(I_{2,1\B}+I_{1,1\B}\big)I_{\xi_{1\B}}
    +\big(I_{2,2\B}+I_{1,2\B}\big)I_{\xi_{2\B}}
\Big]\\
&=2\inte\frac{\partial V_{\Om}(x)}{\partial x_k}
w\Big[b_0\big(w+x\cdot\nabla w\big)
      +\sum^2_{l=1}b_l\frac{\partial w}{\partial x_l}\Big]+o(1)\\
&= \sum^2_{l=1}b_l\inte\frac{\partial V_{\Om}(x)}{\partial x_k}
\frac{\partial w^2}{\partial x_l}+o(1),
      \  \ k=1,2,
\end{split}
\end{equation*}
which yields that the claim \eqref{b0 b1 b2 0} holds true in view of the non-degeneracy in \eqref{1.14nondege}.

\vskip0.05cm
\noindent $\mathbf{Step\  \ 2}$. The constant $b_0$ in \eqref{4u32.xijB.limi} satisfies
\begin{equation}\label{4up52-2}
 b_0=0.
\end{equation}

Multiplying those two equations of \eqref{4u14.bar.ujBvjB.equation} by
$\big(x\cdot\nabla \overline{\hat u_{j,1\B}}\big)$ and $\big(x\cdot\nabla \overline{\hat u_{j,2\B}}\big)$, respectively, integrating over $\R^2$ and taking their real parts, we have for $j=1,2$,
\begin{equation}\label{4up61}
\left\{\begin{array}{lll}
\quad Re\Big(\ds\inte\Big[-\Delta \hat u_{j,1\B}
               +i\,\alpha^2_\B\,\Om\big(x^\perp\cdot\nabla \hat u_{j,1\B}\big)\Big]
\big(x\cdot\nabla \overline{\hat u_{j,1\B}}\big)\Big)\\[3mm]
=Re\Big(\ds\inte\Big[\Big(\alpha^2_\B\mu_{j\B}-\alpha_\B^4V(x)\Big)\hat u_{j,1\B}
         +\Big(\frac{a_1}{a^*}|\hat u_{j,1\B}|^2+\frac{\B}{a^*}|\hat u_{j,2\B}|^2\Big)
        \hat u_{j,1\B}\Big]\big(x\cdot\nabla \overline{\hat u_{j,1\B}}\big)\Big)\\[5mm]
\quad Re\Big(\ds\inte\Big[-\Delta \hat u_{j,2\B}
               +i\,\alpha^2_\B\,\Om\big(x^\perp\cdot\nabla \hat u_{j,2\B}\big)\Big]
\big(x\cdot\nabla \overline{\hat u_{j,2\B}}\big)\Big)\\[3mm]
=Re\Big(\ds\inte\Big[\Big(\alpha^2_\B\mu_{j\B}-\alpha_\B^4V(x)\Big)\hat u_{j,2\B}
         +\Big(\frac{a_2}{a^*}|\hat u_{j,2\B}|^2+\frac{\B}{a^*}|\hat u_{j,1\B}|^2\Big)
        \hat u_{j,2\B}\Big]\big(x\cdot\nabla \overline{\hat u_{j,2\B}}\big)\Big).
\end{array}\right.
\end{equation}
Using integration by parts, we obtain from \eqref{4u6.baruBvB.decay} and \eqref{4up61} that for $j=1,2$,
\begin{align*}
A_{j\B}:&=-Re\Big(
\inte\Big[\Delta \hat u_{j,1\B}\big(x\cdot\nabla \overline{\hat u_{j,1\B}}\big)
         +\Delta \hat u_{j,2\B}\big(x\cdot\nabla \overline{\hat u_{j,2\B}}\big)\Big]\Big)\\
        &=-Re\lim\limits_{R\to\infty}\int_{\partial B_R(0)}
\Big[\frac{\partial \hat u_{j,1\B}}{\partial \nu}
\big(x\cdot\nabla \overline{\hat u_{j,1\B}}\big)
    -\frac{1}{2}(x\cdot \nu)|\nabla \hat u_{j,1\B}|^2\Big]dS\\
        &\quad-Re\lim\limits_{R\to\infty}\int_{\partial B_R(0)}
\Big[\frac{\partial \hat u_{j,2\B}}{\partial \nu}
\big(x\cdot\nabla \overline{\hat u_{j,2\B}}\big)
    -\frac{1}{2}(x\cdot \nu)|\nabla \hat u_{j,2\B}|^2\Big]dS
        =0,
\end{align*}
where $\nu=(\nu_1,\nu_2)$ denotes the outward unit normal of $\partial B_R(0)$, and
\begin{align*}
B_{j\B}:&=Re
\Big(\inte\Big\{\Big[\alpha^2_\B\mu_{j\B}-\alpha_\B^4V(x)\Big]\cdot
           \Big[\hat u_{j,1\B}\big(x\cdot\nabla \overline{\hat u_{j,1\B}}\big)
        +\hat u_{j,2\B}\big(x\cdot\nabla \overline{\hat u_{j,2\B}}\big)\Big]\Big\}\Big)\\
&=Re\Big(
\inte\Big\{\Big[\alpha^2_\B\mu_{j\B}-\alpha_\B^4V(x)\Big]\cdot
           \Big[\frac{1}{2}\big(x\cdot\nabla|\hat u_{j,1\B}|^2\big)
        +\frac{1}{2}\big(x\cdot\nabla |\hat u_{j,2\B}|^2\big)\Big]\Big\}\Big)\\
&=Re\Big(
\inte\Big\{\Big[-\alpha^2_\B\mu_{j\B}+\frac{1}{2}\alpha_\B^4\big(2V(x)+x\cdot\nabla V(x)\big)\Big]\cdot
           \big(|\hat u_{j,1\B}|^2
        +|\hat u_{j,2\B}|^2\big)\Big\}\Big)\\
   &=-\inte\Big[\alpha^2_\B\mu_{j\B}-2\alpha^4_\B V(x)\Big]
           \big(|\hat u_{j,1\B}|^2+|\hat u_{j,2\B}|^2\big),
\end{align*}
due to the fact that $x\cdot\nabla V(x)=2V(x)$. We also have
\begin{align*}
C_{j\B}:&=Re\Big(\inte\Big[\Big(\frac{a_1}{a^*}|\hat u_{j,1\B}|^2
+\frac{\B}{a^*}|\hat u_{j,2\B}|^2\Big)
        \hat u_{j,1\B}\big(x\cdot\nabla \overline{\hat u_{j,1\B}}\big)\\
        &\qquad\qquad+\Big(\frac{a_2}{a^*}|\hat u_{j,2\B}|^2+\frac{\B}{a^*}|\hat u_{j,1\B}|^2\Big)
        \hat u_{j,2\B}\big(x\cdot\nabla \overline{\hat u_{j,2\B}}\big)\Big]\Big)\\
        &=\frac{a_1}{4a^*}\inte\big(x\cdot\nabla |\hat u_{j,1\B}|^4\big)
         +\frac{a_2}{4a^*}\inte\big(x\cdot\nabla |\hat u_{j,2\B}|^4\big)\\
&\quad+\frac{\B}{2a^*}\inte\Big[|\hat u_{j,2\B}|^2\big(x\cdot\nabla |\hat u_{j,1\B}|^2\big)
      +|\hat u_{j,1\B}|^2\big(x\cdot\nabla |\hat u_{j,2\B}|^2\big)\Big]\\
        &=-\frac{a_1}{2a^*}\inte |\hat u_{j,1\B}|^4
         -\frac{a_2}{2a^*}\inte  |\hat u_{j,2\B}|^4
         -\frac{\B}{a^*}\inte |\hat u_{j,1\B}|^2|\hat u_{j,2\B}|^2,
\end{align*}
and
\begin{equation*}
\begin{split}
D_{j\B}:=&Re\Big(\inte \Big[i\,\alpha^2_\B\Om\big(x^\perp\cdot\nabla \hat u_{j,1\B}\big)
                        \big(x\cdot\nabla \overline{\hat u_{j,1\B}}\big)
+i\,\alpha^2_\B\Om\big(x^\perp\cdot\nabla \hat u_{j,2\B}\big)
                        \big(x\cdot\nabla \overline{\hat u_{j,2\B}}\big)
\Big]\Big)\\
=&\alpha^2_\B\Om\inte\Big[\big(x^\perp\cdot\nabla R_{j,1\B}\big)
                        \big(x\cdot\nabla I_{j,1\B}\big)
                        -\big(x^\perp\cdot\nabla I_{j,1\B}\big)
                        \big(x\cdot\nabla R_{j,1\B}\big)\Big]\\
&+\alpha^2_\B\Om\inte\Big[\big(x^\perp\cdot\nabla R_{j,2\B}\big)
                        \big(x\cdot\nabla I_{j,2\B}\big)
                        -\big(x^\perp\cdot\nabla I_{j,2\B}\big)
                       \big (x\cdot\nabla R_{j,2\B}\big)\Big].
\end{split}
\end{equation*}
Therefore, we derive from above that
\begin{equation}\label{4up62}
\begin{split}
&\quad\frac{D_{2\B}-D_{1\B}}
{\|\hat u_{2,1\B}-\hat u_{1,1\B}\|^{\frac{1}{2}}_{L^\infty(\R^2)}
  \|\hat u_{2,2\B}-\hat u_{1,2\B}\|^{\frac{1}{2}}_{L^\infty(\R^2)}}\\
  &=\frac{(B_{2\B}-B_{1\B})+(C_{2\B}-C_{1\B})}
{\|\hat u_{2,1\B}-\hat u_{1,1\B}\|^{\frac{1}{2}}_{L^\infty(\R^2)}
  \|\hat u_{2,2\B}-\hat u_{1,2\B}\|^{\frac{1}{2}}_{L^\infty(\R^2)}}.
\end{split}
\end{equation}
We next estimate the terms containing $D_{j\B}$, $B_{j\B}$ and $C_{j\B}$ of \eqref{4up62} as follows.

Following Theorem \ref{thm-expan solution}, we obtain from \eqref{4u11.bar.ujBvjB.define} that for $j,l=1,2$,
\begin{equation}\label{expan Rj1B Rj2B}
R_{j,l\B}=\rho_{l\B}w+O(\alpha_\B^{4})
\ \ \hbox{in} \ \ \R^2 \ \ \hbox{as} \ \ \B\nearrow\B^*,
   \end{equation}
and
\begin{equation}\label{expan Ij1B Ij2B}
I_{j,l\B}=O(\alpha_\B^{6})
\ \ \hbox{in} \ \ \R^2
\ \ \hbox{as} \ \ \B\nearrow\B^*,
   \end{equation}
where $\rho_{l\B}>0$ is as in \eqref{rho jB define}.

As for the term containing $D_{j\B}$, applying \eqref{4u8.imaginary.decay} and \eqref{4u31.xijB.estimates}, we derive from \eqref{expan Rj1B Rj2B} that
\begin{equation}\label{4up63}
\begin{split}
&\quad\frac{D_{2\B}-D_{1\B}}
{\|\hat u_{2,1\B}-\hat u_{1,1\B}\|^{\frac{1}{2}}_{L^\infty(\R^2)}
  \|\hat u_{2,2\B}-\hat u_{1,2\B}\|^{\frac{1}{2}}_{L^\infty(\R^2)}}\\
&=\alpha^2_\B\Om\inte\Big[\big(x^\perp\cdot\nabla R_{\xi_{1\B}}\big)
                        \big(x\cdot\nabla I_{2,1\B}\big)
                        +\big(x^\perp\cdot\nabla R_{1,1\B}\big)
                        \big(x\cdot\nabla I_{\xi_{1\B}}\big)\Big]\\
&\quad+\alpha^2_\B\Om\inte\Big[(x^\perp\cdot\nabla R_{\xi_{2\B}}\big)
                        \big(x\cdot\nabla I_{2,2\B}\big)
                        +\big(x^\perp\cdot\nabla R_{1,2\B}\big)
                        \big(x\cdot\nabla I_{\xi_{2\B}}\big)\Big]\\
&\quad-\alpha^2_\B\Om\inte\Big[\big(x^\perp\cdot\nabla I_{\xi_{1\B}}\big)
                        \big(x\cdot\nabla R_{2,1\B}\big)
                        +\big(x^\perp\cdot\nabla I_{1,1\B}\big)
                        \big(x\cdot\nabla R_{\xi_{1\B}}\big)\Big]\\
&\quad-\alpha^2_\B\Om\inte\Big[\big(x^\perp\cdot\nabla I_{\xi_{2\B}}\big)
                        \big(x\cdot\nabla R_{2,2\B}\big)
                        +\big(x^\perp\cdot\nabla I_{1,2\B}\big)
                        \big(x\cdot\nabla R_{\xi_{2\B}}\big)\Big]\\
&=\alpha^2_\B\Om\inte \Big(x^\perp\cdot\nabla\big(\rho_{1\B}w+O(\alpha_\B^{4})\big)\Big)
    \big(x\cdot\nabla I_{\xi_{1\B}}\big)
    +\alpha^2_\B\Om\inte \Big(x^\perp\cdot\nabla\big(\rho_{2\B}w+O(\alpha_\B^{4})\big)\Big)
    \big(x\cdot\nabla I_{\xi_{2\B}}\big)\\
&\quad-\alpha^2_\B\Om\inte(x^\perp\cdot\nabla I_{\xi_{1\B}})
           \Big(x\cdot\nabla\big(\rho_{1\B}w+O(\alpha_\B^{4})\big)\Big)
           -\alpha^2_\B\Om\inte(x^\perp\cdot\nabla I_{\xi_{2\B}})
           \Big(x\cdot\nabla\big(\rho_{2\B}w+O(\alpha_\B^{4})\big)\Big)\\
&\quad+o(\alpha^4_\B)\\
&=\alpha^2_\B\Om\rho_{1\B}
\inte\Big(x^\perp\cdot\nabla\big(x\cdot\nabla w\big)\Big)I_{\xi_{1\B}}
+\alpha^2_\B\Om\rho_{2\B}
\inte\Big(x^\perp\cdot\nabla\big(x\cdot\nabla w\big)\Big)I_{\xi_{2\B}}
                      +o(\alpha^4_\B)\\
&=o(\alpha^4_\B)\  \ \text{as}\  \  \B\nearrow\B^*,
\end{split}
\end{equation}
where $\rho_{1\B}>0$ and $\rho_{2\B}>0$ are defined by \eqref{rho jB define}.
As for the term containing $B_{j\B}$, we deduce from \eqref{4u8.imaginary.decay}  and \eqref{4u31-1} that
\begin{equation}\label{4up64}
  \begin{split}
&\quad\frac{B_{2\B}-B_{1\B}}
 {\|\hat u_{2,1\B}-\hat u_{1,1\B}\|^{\frac{1}{2}}_{L^\infty(\R^2)}
  \|\hat u_{2,2\B}-\hat u_{1,2\B}\|^{\frac{1}{2}}_{L^\infty(\R^2)}}\\
&=2\alpha^{4}_\B \inte V(x)\cdot\Big[\big(R_{2,1\B}+R_{1,1\B}\big)R_{\xi_{1\B}}
                                    +\big(R_{2,2\B}+R_{1,2\B}\big)R_{\xi_{2\B}}\\
&\qquad\qquad+\big(I_{2,1\B}+I_{1,1\B}\big)I_{\xi_{1\B}}
             +\big(I_{2,2\B}+I_{1,2\B}\big)I_{\xi_{2\B}}\Big]\\
&\quad-\frac{\alpha^2_\B(\mu_{2\B}-\mu_{1\B})a^*}
            {\|\hat u_{2,1\B}-\hat u_{1,1\B}\|^{\frac{1}{2}}_{L^\infty(\R^2)}
             \|\hat u_{2,2\B}-\hat u_{1,2\B}\|^{\frac{1}{2}}_{L^\infty(\R^2)}}\\
&=2\alpha^{4}_\B \inte V(x)\cdot\Big[\big(R_{2,1\B}+R_{1,1\B}\big)R_{\xi_{1\B}}
                                    +\big(R_{2,2\B}+R_{1,2\B}\big)R_{\xi_{2\B}}\Big]\\
&\quad-\frac{\alpha^2_\B(\mu_{2\B}-\mu_{1\B})a^*}
            {\|\hat u_{2,1\B}-\hat u_{1,1\B}\|^{\frac{1}{2}}_{L^\infty(\R^2)}
             \|\hat u_{2,2\B}-\hat u_{1,2\B}\|^{\frac{1}{2}}_{L^\infty(\R^2)}}
  +o(\alpha^{4}_\B)\  \ \text{as}\  \ \B\nearrow\B^*.
 \end{split}
\end{equation}
Using Lemma \ref{lem 5} and \eqref{expan Ij1B Ij2B}, we obtain from \eqref{4u6.baruBvB.decay} and \eqref{4u31-1} that
\begin{equation}\label{4up65}
  \begin{split}
&\quad\frac{C_{2\B}-C_{1\B}}
{\|\hat u_{2,1\B}-\hat u_{1,1\B}\|^{\frac{1}{2}}_{L^\infty(\R^2)}
  \|\hat u_{2,2\B}-\hat u_{1,2\B}\|^{\frac{1}{2}}_{L^\infty(\R^2)}}\\
&=-\frac{a_1}{2a^*}\inte\big(|\hat u_{2,1\B}|^2+|\hat u_{1,1\B}|^2\big)
          \Big[\big(R_{2,1\B}+R_{1,1\B}\big)R_{\xi_{1\B}}
                +\big(I_{2,1\B}+I_{1,1\B}\big)I_{\xi_{1\B}}\Big]\\
&\quad-\frac{a_2}{2a^*}\inte\big(|\hat u_{2,2\B}|^2+|\hat u_{1,2\B}|^2\big)
          \Big[\big(R_{2,2\B}+R_{1,2\B}\big)R_{\xi_{2\B}}
                 +\big(I_{2,2\B}+I_{1,2\B}\big)I_{\xi_{2\B}}\Big]\\
&\quad-\frac{\B}{a^*}\inte|\hat u_{2,2\B}|^2
           \Big[\big(R_{2,1\B}+R_{1,1\B}\big)R_{\xi_{1\B}}
           +\big(I_{2,1\B}+I_{1,1\B}\big)I_{\xi_{1\B}}\Big]\\
&\quad-\frac{\B}{a^*}\inte|\hat u_{1,1\B}|^2
           \Big[\big(R_{2,2\B}+R_{1,2\B}\big)R_{\xi_{2\B}}
            +\big(I_{2,2\B}+I_{1,2\B}\big)I_{\xi_{2\B}}\Big]\\
&=-\frac{a_1}{2a^*}\inte\big(R_{2,1\B}^2+R_{1,1\B}^2\big)
            \big(R_{2,1\B}+R_{1,1\B}\big)R_{\xi_{1\B}}\\
&\quad-\frac{a_2}{2a^*}\inte\big(R_{2,2\B}^2+R_{1,2\B}^2\big)
            \big(R_{2,2\B}+R_{1,2\B}\big)R_{\xi_{2\B}}\\
&\quad-\frac{\B}{a^*}\inte\big[(R_{2,1\B}+R_{1,1\B})R_{2,2\B}^2R_{\xi_{1\B}}
                +(R_{2,2\B}+R_{1,2\B})R_{1,1\B}^2R_{\xi_{2\B}}\big]+o(\alpha^4_\B)\\
&=\frac{\alpha^2_\B(\mu_{2\B}-\mu_{1\B})a^*}
{\|\hat u_{2,1\B}-\hat u_{1,1\B}\|^{\frac{1}{2}}_{L^\infty(\R^2)}
  \|\hat u_{2,2\B}-\hat u_{1,2\B}\|^{\frac{1}{2}}_{L^\infty(\R^2)}}+o(\alpha^4_\B)\  \ \text{as}\  \  \B\nearrow\B^*.
\end{split}
\end{equation}
By \eqref{4u13.bar.ujBvjB.lim}, \eqref{4u31.xijB.estimates} and Lemma \ref{lem xijB.limit}, we conclude from \eqref{4up62} and \eqref{4up63}--\eqref{4up65} that
\begin{equation}\label{4up66}
\begin{split}
o(\alpha^4_\B)&=
\frac{(B_{2\B}-B_{1\B})+(C_{2\B}-C_{1\B})}
     {\|\hat u_{2,1\B}-\hat u_{1,1\B}\|^{\frac{1}{2}}_{L^\infty(\R^2)}
      \|\hat u_{2,2\B}-\hat u_{1,2\B}\|^{\frac{1}{2}}_{L^\infty(\R^2)}}\\
  &=2\alpha^{4}_\B\inte V(x)\cdot
  \Big[\big(R_{2,1\B}+R_{1,1\B}\big)R_{\xi_{1\B}}
      +\big(R_{2,2\B}+R_{1,2\B}\big)R_{\xi_{2\B}}\Big]+o(\alpha^4_\B)
\\
 &=4\alpha^{4}_\B\sqrt{\gamma_1}\inte V(x)w R_{\xi_{10}}
  +4\alpha^{4}_\B\sqrt{\gamma_2}\inte V(x)w R_{\xi_{20}}+o(\alpha^4_\B)\\
 &=4\alpha^{4}_\B \Big[b_0\inte V(x)w\big(w+x\cdot\nabla w\big)
                       +\sum^2_{l=1}b_l\inte\frac{1}{2}V(x)
                       \frac{\partial w^2}{\partial x_l}\Big]+o(\alpha^4_\B)
 \  \ \text{as}\  \  \B\nearrow\B^*.
\end{split}
\end{equation}
Applying Step 1, we derive from \eqref{4up66} that
    \begin{align*}
    0
    &=b_0\inte V(x)w\big(w+x\cdot\nabla w\big)
      +\sum^2_{l=1}b_l\inte\frac{1}{2}V(x)\frac{\partial w^2}{\partial x_l}\\
    &=b_0\inte V(x)w^2-\frac{b_0}{2}\inte\Big(x\cdot\nabla V(x)w^2+2V(x)w^2\Big)\\
    &=-\frac{b_0}{2}\inte\Big(x\cdot\nabla V(x)w^2\Big)
    =-b_0\inte V(x)w^2,
    \end{align*}
which further implies that $b_0=0$, 
and \eqref{4up52-2} is therefore proved.

We are now ready to finish the proof of Theorem \ref{thm-equation.uniqueness}.
Let $(x_\B,y_\B)$ satisfy $|\xi_{1\B}(x_\B)\xi_{2\B}(y_\B)|=\|\xi_{1\B}\xi_{2\B}\|_{L^\infty(\R^2)}=1$. By the exponential decay of $\xi_{1\B}$ and $\xi_{2\B}$ in \eqref{4u31.xijB.estimates}, we obtain that $|x_\B|\leq C$ and $|y_\B|\leq C$ uniformly in $\B$. We thus conclude that $\xi_{j\B}\to\xi_{j0}=R_{\xi_{j0}}+iI_{\xi_{j0}}\not\equiv0$ uniformly in $C^1(\R^2)$ as $\B\nearrow\B^*$, where $j=1,2$. However, we conclude from Step 2 that $\xi_{j0}\equiv0$, a contradiction, and the assumption \eqref{5:17} is false. The proof of Theorem \ref{thm-equation.uniqueness} is therefore complete.
\qed

\section{Nonexistence of Vortices}

In this section, we shall complete the proof of Theorem \ref{thm-nonvorex} on the non-existence of vortices for $e(\Om,\f)$. We first prove Theorem \ref{thm-nonvorex} (1) by the uniqueness of Theorem \ref{thm-equation.uniqueness}.
\vskip 0.05truein

\noindent {\bf Proof of Theorem \ref{thm-nonvorex} (1).}  In this case, $V(x)$ satisfies \eqref{V(x)} for $\Lambda=1$, i.e., $V(x)=|x|^2$. Let $(u_{1\B},u_{2\B})$ be a complex-valued minimizer of $e(\Om,\f)$, where $0<\Om<\Om^*:=2$ and $0<a_1,a_2<a^*$ are fixed. By Theorems \ref{thm-1.3} and \ref{thm-equation.uniqueness}, we obtain from \eqref{1.1-1equation system} that when $\B^*-\B>0$ is small enough, up to a constant phase, $(u_{1\B},u_{2\B})$ is the unique solution of the following system
\begin{equation}\label{1.1equation system-1-1}
\left\{\begin{array}{lll}
-\Delta u_{1\B}+|x|^2u_{1\B}+i\,\Om \big(x^{\perp}\cdot \nabla u_{1\B}\big)
=\mu_\B u_{1\B}+a_1|u_{1\B}|^2u_{1\B}+\B |u_{2\B}|^2u_{1\B}
\quad \hbox{in}\  \ \R^2,\\[5mm]
-\Delta u_{2\B}+|x|^2u_{2\B}+i\,\Om \big(x^{\perp}\cdot \nabla u_{2\B}\big)
=\mu_\B u_{2\B}+a_2|u_{2\B}|^2u_{2\B}+\B |u_{1\B}|^2u_{2\B}
\quad \hbox{in}\  \ \R^2,
\end{array}\right.
\end{equation}
satisfying
\begin{equation}\label{eps2mu-1-1}
\alpha_\B^2\mu_\B\to-1\  \ \text{as}\  \ \B\nearrow\B^*,
\end{equation}
and for $j=1,2$,
\begin{equation}\label{1.15lim:beta.V.u.exp-1-1}
     w_{j\B}:=\sqrt{a^*}\alpha_\B u_{j\B}(\alpha_\B x)
     e^{i\theta_{j\B}}
\to\sqrt{\gam_j}w(x)
\ \ \text{strongly in}\  \ H^1(\R^2,\C)\cap L^\infty(\R^2,\C)\  \ \text{as}\  \ \B\nearrow\B^*,
   \end{equation}
where $\B^*>0$ is as in \eqref{main 1.11def:beta*}, $\alpha_\B>0$ is defined by \eqref{1.16def:beta.V.eps}, $0<\gam_j<1$ is as in \eqref{1.13gamma}, and $\theta_{j\B}\in [0,2\pi)$ is a suitable constant.

Let $(u_{10},u_{20})$ be a real-valued positive minimizer (i.e., $u_{10}>0$ and $u_{20}>0$) of the following variational problem
\begin{equation}\label{1.3miniproble-1}
e(0,\f):=\inf_{(u_1,u_2)\in\mathbb{M}}F_{0,\f}(u_1,u_2),
\  \ a_1>0,\  \ a_2>0,\  \ \B>0,
\end{equation}
where
\begin{equation}\label{1.4energyfunctional-1}
\begin{split}
 F_{0,\f}(u_1,u_2):=
 &\sum_{j=1}^2\inte\Big(|\nabla u_j|^2+|x|^2|u_j|^2-\frac{a_j}{2}|u_j|^4\Big)dx
           -\B\inte|u_1|^2|u_2|^2dx,
\end{split}
\end{equation}
and the space $\mathbb{M}$ is defined by
\begin{equation*}
\mathbb{M}:=\Big\{(u_1,u_2)\in\mathbb{H}\times\mathbb{H}: \ \ \inte (|u_{1}|^2+|u_{2}|^2)dx=1\Big\}.
\end{equation*}
Here $\mathbb{H}$ is defined as
\begin{equation*}
\mathbb{H}:=\Big\{u\in  H^1(\R^2,\R):\ \inte |x|^2|u(x)|^2dx<\infty \Big\}.
\end{equation*}
Since $(u_{10},u_{20})$ is a minimizer of $e(0,\f)$, it satisfies the following Euler-Lagrange system
\begin{equation}\label{1.1equation system-1-2}
\left\{\begin{array}{lll}
-\Delta u_{10}+|x|^2u_{10}=\lambda_\B u_{10}+a_1u_{10}^3+\B u_{20}^2u_{10}\  \ \hbox{in}\  \ \R^2,\\[3mm]
-\Delta u_{20}+|x|^2u_{20}=\lambda_\B u_{20}+a_2u_{20}^3+\B u_{10}^2u_{20}\  \ \hbox{in}\  \ \R^2,
\end{array}\right.
\end{equation}
where $\lambda_\B\in\R$ is a Lagrange multiplier and satisfies
\begin{equation*}
\lambda_\B=e(0,\f)-\inte\Big(\frac{a_1}{2}|u_{10}|^4+\frac{a_2}{2}|u_{20}|^4
                  +\B|u_{10}|^2|u_{20}|^2\Big)dx.
\end{equation*}
In view of \cite [Theorems 1.4 and 1.5] {GLWZ1}, we obtain that $(u_{10},u_{20})$ is a unique positive minimizer of $e(0,\f)$ as $\B\nearrow\B^*$, and
\begin{equation}\label{1.5Lmu-1-1}
\alpha_{\B}^2\lambda_\B\to -1\  \ \text{as}\  \ \B\nearrow \B^*,
\end{equation}
and
\begin{equation}\label{1.15lim:beta.V.u.exp-1-2}
\lim\limits_{\B\nearrow\B^*}\sqrt{a^*}\alpha_\B u_{j0}(\alpha_\B x)
=\sqrt{\gam_j}w(x)
\ \ \text{strongly in}\  \ H^1(\R^2,\R)\cap L^\infty(\R^2,\R),\ \ j=1,2,
   \end{equation}
where $0<\gam_j<1$ is as in \eqref{1.13gamma}, and $\alpha_\B>0$ is defined by \eqref{1.16def:beta.V.eps}.
By the symmetric-decreasing rearrangement, we derive from \eqref{1.3miniproble-1} and \eqref{1.4energyfunctional-1} that $(u_{10},u_{20})$ is radially symmetric. Since $(u_{10},u_{20})$ is real-valued and radially symmetric, we get that $x^{\perp}\cdot\nabla u_{j0}=0$ for $j=1,2$. Therefore, we obtain from \eqref{1.1equation system-1-2}--\eqref{1.15lim:beta.V.u.exp-1-2} that $(u_{10},u_{20})$ is essentially a radially symmetric positive solution of \eqref{1.1equation system-1-1} satisfying \eqref{eps2mu-1-1} and \eqref{1.15lim:beta.V.u.exp-1-1}.

Following the uniqueness of Theorem \ref{thm-equation.uniqueness}, we conclude from above that up to a constant phase $(\theta_{1\B}, \theta_{2\B})\in [0,2\pi)\times [0,2\pi)$,
\begin{equation*}
u_{j\B}(x)e^{i\theta_{j\B}}=u_{j0}(x)>0\  \ \text{in}\  \ \R^2
\  \ \text{as}\  \ \B\nearrow \B^*,\  \ j=1,2,
\end{equation*}
which implies that up to a constant phase, $(u_{1\B},u_{2\B})$ is real-valued, unique and free of vortices as $\B\nearrow\B^*$. This therefore completes the proof of Theorem \ref{thm-nonvorex} (1).\qed

We are now ready to prove Theorem \ref{thm-nonvorex} (2) by the comparison principle.

\vskip 0.05truein

\noindent {\bf Proof of Theorem \ref{thm-nonvorex} (2).}
In this case, $V(x)$ satisfies \eqref{V(x)} for $\Lambda>1$.
Let $(u_{1\beta},u_{2\beta})$ be a complex-valued minimizer of $e(\Omega,a_{1},a_{2},\beta)$, where $0<\Om<\Om^*:=2$ and $0<a_1,a_2<a^*$ are fixed. Following Theorem \ref{thm-1.3}, we obtain from Theorem \ref{thm-expan solution} that for any fixed and sufficiently large $R>1$, there exists a constant $C_R:=C(R)>0$ such that for $j=1,2$,
\begin{equation}\label{5:5}
\big|w_{j\B}(x)\big|\ge \rho_{j\beta}w(x)-C_R\alpha_\B^4>0
\ \ \mbox{in}\ \ \big\{x\in\R^2:\, |x|\le R\big\}\ \ \hbox{as}\ \ \B\nearrow\B^*,
\end{equation}
where $w_{j\B}$ and $\rho_{j\beta}>0$ are as in \eqref{hatw.define} and \eqref{rho jB define}, respectively.

Define
\begin{equation*}
\tilde{w}_{j\B}=w_{j\B}(x)-\rho_{j\beta}w(x),\ \ j=1,2.
\end{equation*}
We claim that there exist constants $C_{1}>0$ and $C_{2}>0$, independent of $0<\B<\B^*$, such that
\begin{equation}\label{5:6}
|\tilde{w}_{j\B}|
\leq C_{1}\alpha_\B^{4}|x|^{\frac{5}{2}}e^{-\sqrt{1-C_{2}\alpha_\B^{4}}\,|x|}
    \ \ \hbox{uniformly in}\ \ \R^2 \ \ \mbox{as} \ \ \B\nearrow \B^*,\ \ j=1,2.
\end{equation}
To prove \eqref{5:6}, we note from \eqref{1.1-1equation system} and \eqref{hatw.define} that $(\tilde{w}_{1\B},\tilde{w}_{2\B})$ satisfies
\begin{equation}\label{5:7}
\left\{\begin{array}{lll}
    \big(-\Delta+\hat{V}_{1\B}\big)\tilde{w}_{1\B}
    +i\,\alpha_\B^2\,\Omega \big(x^{\bot}\cdot\nabla\tilde{w}_{1\B}\big)
    +\big(-\Delta+\hat{V}_{1\B}\big)\rho_{1\beta}w=0\ \ \hbox{in}\ \ \R^2,\\[5mm]
    \big(-\Delta+\hat{V}_{2\B}\big)\tilde{w}_{2\B}
    +i\,\alpha_\B^2\,\Omega \big(x^{\bot}\cdot\nabla\tilde{w}_{2\B}\big)
    +\big(-\Delta+\hat{V}_{2\B}\big)\rho_{2\beta}w=0\ \ \hbox{in}\ \ \R^2,
\end{array}\right.
\end{equation}
where
\begin{equation}\label{5:8}
\left\{\begin{array}{lll}
    \ds\hat{V}_{1\B}(x)
    =\alpha_\B^{4}V(x)-\alpha_\B^2\mu_\B-\frac{a_1}{a^*}|w_{1\B}|^2-\frac{\B}{a^*}|w_{2\B}|^2
    \ \ \hbox{in}\ \ \R^2,\\[5mm]
    \ds\hat{V}_{2\B}(x)
    =\alpha_\B^{4}V(x)-\alpha_\B^2\mu_\B-\frac{a_2}{a^*}|w_{2\B}|^2-\frac{\B}{a^*}|w_{1\B}|^2
    \ \ \hbox{in}\ \ \R^2.
\end{array}\right.
\end{equation}
One can derive from \eqref{5:7} that
\begin{equation}\label{5:9}
\left\{\begin{array}{lll}
    -\ds\frac{1}{2}\Delta|\tilde{w}_{1\B}|^{2}+\Big[\alpha_\B^{4}V(x)-\alpha_\B^2\mu_\B
      -\frac{a_1}{a^*}|w_{1\B}|^2-\frac{\B}{a^*}|w_{2\B}|^2\Big]|\tilde{w}_{1\B}|^{2}
    +|\nabla\tilde{w}_{1\B}|^{2}\\[5mm]
    -\alpha_\B^{2}\Omega\,x^{\bot}\cdot(i\tilde{w}_{1\B},\nabla\tilde{w}_{1\B})
    +\big(-\Delta+\hat{V}_{1\B}\big)(\rho_{1\beta}w,\tilde{w}_{1\B})=0\ \ \hbox{in}\ \ \R^2,\\[5mm]
    -\ds\frac{1}{2}\Delta|\tilde{w}_{2\B}|^{2}+\Big[\alpha_\B^{4}V(x)-\alpha_\B^2\mu_\B
      -\frac{a_2}{a^*}|w_{2\B}|^2-\frac{\B}{a^*}|w_{1\B}|^2\Big]|\tilde{w}_{2\B}|^{2}
    +|\nabla\tilde{w}_{2\B}|^{2}\\[5mm]
    -\alpha_\B^{2}\Omega\, x^{\bot}\cdot(i\tilde{w}_{2\B},\nabla\tilde{w}_{2\B})
    +\big(-\Delta+\hat{V}_{2\B}\big)(\rho_{2\beta}w,\tilde{w}_{2\B})=0\ \ \hbox{in}\ \ \R^2.
\end{array}\right.
\end{equation}
By the diamagnetic inequality \eqref{2.3ineq:diam}, we have
\begin{equation}\label{5:10}
    |\nabla\tilde{w}_{j\B}|^{2}+\frac{\alpha_\B^{4}\Omega^{2}}{4}|x|^2|\tilde{w}_{j\B}|^{2}
    -\alpha_\B^{2}\Omega\, x^{\bot}\cdot(i\tilde{w}_{j\B},\nabla\tilde{w}_{j\B})
    \geq \big|\nabla|\tilde{w}_{j\B}|\big|^{2}\ \ \hbox{in}\ \ \R^2,\ \ j=1,2.
\end{equation}
Since
\[
 \frac{1}{2}\Delta|\tilde{w}_{j\B}|^{2}
 =|\tilde{w}_{j\B}|\Delta |\tilde{w}_{j\B}|+\big|\nabla |\tilde{w}_{j\B}|\big|^2,\ \ j=1,2,
\]
we deduce from \eqref{5:9} and \eqref{5:10} that
\begin{equation}\label{5:11}
\left\{\begin{array}{lll}
    -\ds\Delta|\tilde{w}_{1\B}|-\alpha_\B^2\mu_\B|\tilde{w}_{1\B}|
    \leq\big|(-\Delta+\hat{V}_{1\B})\rho_{1\beta}w\big|
    +\Big(\frac{a_1}{a^*}|w_{1\B}|^2+\frac{\B}{a^*}|w_{2\B}|^2\Big)|\tilde{w}_{1\B}|
    \ \ \hbox{in}\ \ \R^2,\\[5mm]
    -\ds\Delta|\tilde{w}_{2\B}|-\alpha_\B^2\mu_\B|\tilde{w}_{2\B}|
    \leq\big|(-\Delta+\hat{V}_{2\B})\rho_{2\beta}w\big|
    +\Big(\frac{a_2}{a^*}|w_{2\B}|^2+\frac{\B}{a^*}|w_{1\B}|^2\Big)|\tilde{w}_{2\B}|
    \ \ \hbox{in}\ \ \R^2.
\end{array}\right.
\end{equation}
Similar to the argument of proving  \eqref{3.59}, one can deduce that for $j=1,2,$
\[|w_{j\B}|^{2}\le Ce^{-\frac{4}{3}|x|}\ \ \hbox{in}\ \ \R^2\ \ \hbox{as}\ \ \B\nearrow\B^*,\]
which implies that for $j,m=1,2$ and $j\neq m$,
\begin{equation}\label{5:12}
    \Big(\frac{a_j}{a^*}|w_{j\B}|^2+\frac{\B}{a^*}|w_{m\B}|^2\Big)|\tilde{w}_{j\B}|
    \leq C\alpha_\B^{4}e^{-\frac{4}{3}|x|}\ \ \hbox{in}\ \ \R^2
    \ \ \hbox{as}\ \ \B\nearrow\B^*,
\end{equation}
due to Theorem \ref{thm-expan solution}.

Note also from Theorem \ref{thm-expan solution} that
\begin{equation*}
  |1+\alpha_\B^2\mu_\B|\leq C\alpha_\B^4\ \ \hbox{and}\ \ w_{j\B}(x)=\rho_{j\B}w+O(\alpha_\B^{4})
  \ \ \hbox{as}\ \ \B\nearrow\B^*,\ \ j=1,2.
\end{equation*}
We then calculate from \eqref{1.10decay:w} and \eqref{5:8} that
\begin{equation}\label{5:13}
\left\{\begin{array}{lll}
\begin{split}
 \big|(-\Delta+\hat{V}_{1\B})\rho_{1\beta}w\big|
   &=\ds\Big|\Big(\alpha_\B^4 V(x)+w^2-1-\alpha_\B^2\mu_\B
      -\frac{a_1}{a^*}|w_{1\B}|^2-\frac{\B}{a^*}|w_{2\B}|^2\Big)\rho_{1\beta}w\Big|\\[5mm]
    &\leq C \alpha_\B^4|x|^{\frac{3}{2}}e^{-|x|}
    \ \ \hbox{in}\ \ \R^2/B_R(0)\ \ \hbox{as}\ \ \B\nearrow\B^*,\\[5mm]
   \big|(-\Delta+\hat{V}_{2\B})\rho_{2\beta}w\big|
   &=\ds\Big|\Big(\alpha_\B^4 V(x)+w^2-1-\alpha_\B^2\mu_\B
      -\frac{a_2}{a^*}|w_{2\B}|^2-\frac{\B}{a^*}|w_{1\B}|^2\Big)\rho_{2\beta}w\Big|\\[5mm]
    &\leq C \alpha_\B^4|x|^{\frac{3}{2}}e^{-|x|}
    \ \ \hbox{in}\ \ \R^2/B_R(0)\ \ \hbox{as}\ \ \B\nearrow\B^*,
\end{split}
\end{array}\right.
\end{equation}
where the sufficiently large constant $R>1$ is as in \eqref{5:5}.
We thus deduce from \eqref{5:11}--\eqref{5:13} that
\begin{equation}\label{5:14}
    -\Delta|\tilde{w}_{j\B}|-\alpha_\B^2\mu_\B|\tilde{w}_{j\B}|
    \leq C_0\alpha_\B^4|x|^{\frac{3}{2}}e^{-|x|}\ \, \ \hbox{in}\ \ \R^2/B_R(0)\ \ \hbox{as}\ \ \B\nearrow\B^*,\ \ j=1,2,
\end{equation}
where the constant $C_0>0$ is independent of $0<\B<\B^*$.
Since Theorem \ref{thm-expan solution} gives that $|\tilde{w}_{j\B}|=O(\alpha_\B^4)$ as $\B\nearrow\B^*,$ we have
\begin{equation}\label{5:15}
|\tilde{w}_{j\B}|
\le C\alpha_\B^4|x|^{\frac{5}{2}}e^{-\sqrt{1-|C^*|\alpha_\B^4}\,|x|}\ \ \hbox{at}\ \ |x|=R>1\ \ \hbox{as}\ \ \B\nearrow\B^*,
\end{equation}
where $R>1$ is as in \eqref{5:5}, and $C>0$ is large enough and independent of $0<\B<\B^*$.
By the comparison principle, we thus derive from \eqref{5:14} and \eqref{5:15} that the claim \eqref{5:6} holds true, in view of the fact that $|\tilde{w}_{j\B}|=O(\alpha_\B^4)$ as $\B\nearrow\B^*$.

Applying \eqref{1.10decay:w} and \eqref{5:6}, we now have as $\B\nearrow\B^*,$
\begin{equation}\label{5:16}
\left\{\begin{array}{lll}
\begin{split}
|w_{1\B}|
&\ge |\rho_{1\beta}w|-|w_{1\B}-\rho_{1\beta}w|\\[3mm]
&\ge C_{3}|x|^{-\frac{1}{2}}e^{-|x|}
     -C_{1}\alpha_\B^{4}|x|^{\frac{5}{2}}e^{-\sqrt{1-C_2\alpha_\B^{4}}\,|x|}\\[3mm]
&\ge|x|^{-\frac{1}{2}}e^{-|x|}\big(C_3-C_1\alpha_\B^{4}|x|^{3}
e^{C\alpha_\B^{4}\,|x|}\big)>0,\ \ \mbox{if}\ \ R\le|x|\le \Big(\frac{C_3}{2 C_1}\Big)^\frac{1}{3}\alpha_\B^{-\frac{4}{3}},\\[3mm]
|w_{2\B}|
&\ge |\rho_{2\beta}w|-|w_{2\B}-\rho_{2\beta}w|\\[3mm]
&\ge C_{3}|x|^{-\frac{1}{2}}e^{-|x|}
     -C_{1}\alpha_\B^{4}|x|^{\frac{5}{2}}e^{-\sqrt{1-C_2\alpha_\B^{4}}\,|x|}\\[3mm]
&\ge|x|^{-\frac{1}{2}}e^{-|x|}\big(C_3-C_1\alpha_\B^{4}|x|^{3}
e^{C\alpha_\B^{4}\,|x|}\big)>0,\ \ \mbox{if}\ \ R\le|x|\le \Big(\frac{C_3}{2 C_1}\Big)^\frac{1}{3}\alpha_\B^{-\frac{4}{3}}.
\end{split}
\end{array}\right.
\end{equation}
We thus conclude from \eqref{5:5} and \eqref{5:16} that as $\B\nearrow\B^*,$
\begin{equation*}
\left\{\begin{array}{lll}
\begin{split}
|w_{1\B}(x)|
&=\big|\sqrt{a^*}\alpha_\B u_{1\B}(\alpha_\B x)e^{i\theta_{1\B}}\big|>0\\[3mm]
|w_{2\B}(x)|
&=\big|\sqrt{a^*}\alpha_\B u_{2\B}(\alpha_\B x)e^{i\theta_{2\B}}\big|>0
\end{split}
\end{array}\right.
\ \ \mbox{in}
\ \ \Big\{x\in\R^2:\, |x|\le \Big(\frac{C_3}{2 C_1}\Big)^\frac{1}{3}\alpha_\B^{-\frac{4}{3}}\Big\}.
\end{equation*}
Therefore, there exists a small constant  $C_*>0$, independent of $0<\B<\B^*$, such that
\begin{equation*}
    |u_{1\B}(y)|,\ \ |u_{2\B}(y)|>0,
    \ \ \hbox{if}\ \  |y|\leq C_{4}\alpha_{\B}^{-\frac{1}{3}}:=\frac{C_*}{(\B^*-\B)^{\frac{1}{12}}}\ \ \hbox{as}\ \ \B\nearrow\B^*,
\end{equation*}
i.e., $(u_{1\B},u_{2\B})$ does not admit any vortex in the region
$R(\B):=\big\{x\in\R^2:\, |x|\le C_*(\B^*-\B)^{-\frac{1}{12}} \big\}$ as $\B\nearrow\B^*$. This completes the proof of Theorem \ref{thm-nonvorex} (2), and we are done.
\qed

\vskip 0.17truein
\noindent {\bf Acknowledgements:} The authors are very grateful to Professor Amandine Aftalion for her fruitful discussions on the present paper.

\appendix
\section{Appendix}
\renewcommand{\theequation}{A.\arabic{equation}}
\setcounter{equation}{0}

\subsection{An integral identity}
In this appendix, we shall establish Lemma \ref{lem.A.1} on an integral identity, which is often used in Sections 3 and 4.
\begin{lem}\label{lem.A.1}
Suppose $V(x)$ satisfies \eqref{V(x)} for some $\Lambda\ge1$, and $0<\Om<\Om^*:=2$ is fixed. Assume $0<a_1,a_2<a^*$,
$0<\B<\B^*:=a^*+\sqrt{(a^*-a_1)(a^*-a_2)}$, and let $\big(v_{1\B},v_{2\B}\big)$ be a solution of the following elliptic system
\begin{equation}\label{v1Bv2B.system}
\left\{\begin{array}{lll}
\begin{split}
  &\quad -\Delta v_{1\B}(x)+i\,\eps_\B^2\,\Om\big[x^\perp\cdot\nabla v_{1\B}(x)\big]
         +\eps_\B^4V(x)v_{1\B}(x)\\
&=-v_{1\B}(x)+\ds\frac{a_1}{a^*}|v_{1\B}|^2v_{1\B}(x)
  +\frac{\B}{a^*}|v_{2\B}|^2v_{1\B}(x)
\quad \hbox{in}\  \ \R^2,\\[3mm]
  &\quad -\Delta v_{2\B}(x)+i\,\eps_\B^2\,\Om\big[x^\perp\cdot\nabla v_{2\B}(x)\big]
         +\eps_\B^4V(x)v_{2\B}(x)\\
&=-v_{2\B}(x)+\ds\frac{a_2}{a^*}|v_{2\B}|^2v_{2\B}(x)
  +\frac{\B}{a^*}|v_{1\B}|^2v_{2\B}(x)
\quad\hbox{in}\  \ \R^2,
\end{split}
\end{array}\right.
\end{equation}
satisfying
\begin{equation}\label{vjB.xinfty}
\lim_{|x|\to\infty}|v_{j\B}(x)|=0, \ \ j=1,2,
\end{equation}
where $\eps_\B>0$ is small. Then we have
\begin{equation}\label{pohozaev v1B plus v2B}
\inte\frac{\partial V_\Om(x)}{\partial x_j}\big(|v_{1\B}|^2+|v_{2\B}|^2\big)dx=0,
\ \ j=1,2,
\end{equation}
where $V_\Om(x)=V(x)-\frac{\Om^2}{4}|x|^2\geq 0$.
\end{lem}

{\noindent \bf Proof.}
Rewrite \eqref{v1Bv2B.system} as
\begin{equation}\label{rewrite v1Bv2B.equation}
\left\{\begin{array}{lll}
\begin{split}
  &\quad -\Big(\nabla-i\frac{\Om}{2}\eps_\B^2 x^\perp\Big)^2 v_{1\B}(x)
         +\eps_\B^4V_\Om(x)v_{1\B}(x) \\
&=-v_{1\B}(x)+\ds\frac{a_1}{a^*}|v_{1\B}|^2v_{1\B}(x)
  +\frac{\B}{a^*}|v_{2\B}|^2v_{1\B}(x)
\quad \hbox{in}\  \ \R^2,\\[3mm]
  &\quad -\Big(\nabla-i\frac{\Om}{2}\eps_\B^2 x^\perp\Big)^2 v_{2\B}(x)
         +\eps_\B^4V_\Om(x)v_{2\B}(x)\\
&=-v_{2\B}(x)+\ds\frac{a_2}{a^*}|v_{2\B}|^2v_{2\B}(x)
  +\frac{\B}{a^*}|v_{1\B}|^2v_{2\B}(x)
\quad\hbox{in}\  \ \R^2.
\end{split}
\end{array}\right.
\end{equation}
Multiplying the first equation of \eqref{rewrite v1Bv2B.equation} by
$\ds\Big[\frac{\partial\bar{v}_{1\B}}{\partial x_j}
+\frac{i\eps_\B^2\Om\big(-\delta_{2j}x_1+\delta_{1j}x_2\big)\bar{v}_{1\B}}{2}\Big]$,
where $\delta_{1j}$ and $\delta_{2j}$ are $\delta-$functions,
integrating over $\R^2$ and taking its real part, we obtain that for $j=1,2$,
\begin{equation*}
\begin{split}
&\quad Re\Big(\inte\Big[-\Big(\nabla-i\frac{\Om}{2}\eps_\B^2 x^\perp\Big)^2 v_{1\B}
                   +\eps_\B^4V_\Om(x)v_{1\B}\Big]
\Big[\frac{\partial\bar{v}_{1\B}}{\partial x_j}
+\frac{i\eps_\B^2\Om\big(-\delta_{2j}x_1+\delta_{1j}x_2\big)\bar{v}_{1\B}}{2}\Big]\Big)\\
&=Re\Big(\inte\Big[-v_{1\B}+\ds\frac{a_1}{a^*}|v_{1\B}|^2v_{1\B}
              +\frac{\B}{a^*}|v_{2\B}|^2v_{1\B}\Big]
\Big[\frac{\partial\bar{v}_{1\B}}{\partial x_j}
    +\frac{i\eps_\B^2\Om\big(-\delta_{2j}x_1+\delta_{1j}x_2\big)\bar{v}_{1\B}}{2}\Big]\Big),
\end{split}
\end{equation*}
where and below $\bar{f}$ denotes the conjugate of $f$. By the comparison principle, one can derive from \eqref{v1Bv2B.system} and \eqref{vjB.xinfty} that
\begin{equation}\label{vjB.decay}
|v_{j\B}(x)|\le  Ce^{-\frac{2}{3}|x|} \ \ \hbox{as}\ \ |x|\to\infty, \ \ j=1,2.
\end{equation}
Following \eqref{vjB.decay}, we calculate that for $j=1,2$,
\begin{equation}\label{A1}
\begin{split}
&\quad Re\Big(\inte-\Big(\nabla-i\frac{\Om}{2}\eps_\B^2 x^\perp\Big)^2 v_{1\B}
\Big[\frac{\partial\bar{v}_{1\B}}{\partial x_j}
    +\frac{i\eps_\B^2\Om\big(-\delta_{2j}x_1+\delta_{1j}x_2\big)\bar{v}_{1\B}}{2}\Big]\Big)\\
&=Re\Big(\inte\Big(\nabla-i\frac{\Om}{2}\eps_\B^2 x^\perp\Big)v_{1\B}
\Big[\frac{\partial}{\partial x_j}
\Big(\nabla\bar{v}_{1\B}+i\frac{\Om}{2}\eps_\B^2 x^\perp\bar{v}_{1\B}\Big)\\
&\qquad\qquad
-\frac{i\eps_\B^2\Om(-\delta_{2j},\delta_{1j})\bar{v}_{1\B}}{2}
+\Big(\nabla+i\frac{\Om}{2}\eps_\B^2 x^\perp\Big)
\frac{i\eps_\B^2\Om\big(-\delta_{2j}x_1+\delta_{1j}x_2\big)\bar{v}_{1\B}}{2}\Big]\Big)\\
&=Re\Big(\inte\frac{1}{2}\frac{\partial}{\partial x_j}
\Big|\Big(\nabla-i\frac{\Om}{2}\eps_\B^2 x^\perp\Big)v_{1\B}\Big|^2
+\Big(\nabla-i\frac{\Om}{2}\eps_\B^2 x^\perp\Big)v_{1\B}
\Big[-\frac{i\eps_\B^2\Om(-\delta_{2j},\delta_{1j})\bar{v}_{1\B}}{2}\\
&\qquad\qquad+\Big(\nabla+i\frac{\Om}{2}\eps_\B^2 x^\perp\Big)
\frac{i\eps_\B^2\Om\big(-\delta_{2j}x_1+\delta_{1j}x_2\big)\bar{v}_{1\B}}{2}\Big]\Big)\\
&=Re\Big(\inte
\Big(\nabla-i\frac{\Om}{2}\eps_\B^2 x^\perp\Big)v_{1\B}
\Big[-\frac{i\eps_\B^2\Om(-\delta_{2j},\delta_{1j})\bar{v}_{1\B}}{2}\\
&\qquad\qquad+\Big(\nabla+i\frac{\Om}{2}\eps_\B^2 x^\perp\Big)
\frac{i\eps_\B^2\Om\big(-\delta_{2j}x_1+\delta_{1j}x_2\big)\bar{v}_{1\B}}{2}\Big]\Big)\\
&=Re\Big(\inte
\frac{i\eps_\B^2\Om\big(-\delta_{2j}x_1+\delta_{1j}x_2\big)}{2}
\Big|\Big(\nabla-i\frac{\Om}{2}\eps_\B^2 x^\perp\Big)v_{1\B}\Big|^2\Big)=0.
\end{split}
\end{equation}
Similarly, we have for $j=1,2$,
\begin{equation}\label{A2}
\begin{split}
Re\Big(\inte \eps_\B^4V_\Om(x)v_{1\B}
\Big[\frac{\partial\bar{v}_{1\B}}{\partial x_j}
    +\frac{i\eps_\B^2\Om\big(-\delta_{2j}x_1+\delta_{1j}x_2\big)\bar{v}_{1\B}}{2}\Big]\Big)
=\frac{\eps_\B^4}{2}\inte V_\Om(x)\frac{\partial|v_{1\B}|^2}{\partial x_j},
\end{split}
\end{equation}
\begin{equation}\label{A3}
\begin{split}
&\quad Re\Big(\inte\Big[-v_{1\B}+\ds\frac{a_1}{a^*}|v_{1\B}|^2v_{1\B}\Big]
\Big[\frac{\partial\bar{v}_{1\B}}{\partial x_j}
    +\frac{i\eps_\B^2\Om\big(-\delta_{2j}x_1+\delta_{1j}x_2\big)\bar{v}_{1\B}}{2}\Big]\Big)\\
&=-\frac{1}{2}\inte \frac{\partial |v_{1\B}|^2}{\partial x_j}
  +\frac{a_1}{4a^*}\inte \frac{\partial |v_{1\B}|^4}{\partial x_j},
\end{split}
\end{equation}
and
\begin{equation}\label{A4}
Re\Big(\inte\frac{\B}{a^*}|v_{2\B}|^2v_{1\B}
\Big[\frac{\partial\bar{v}_{1\B}}{\partial x_j}
+\frac{i\eps_\B^2\Om\big(-\delta_{2j}x_1+\delta_{1j}x_2\big)\bar{v}_{1\B}}{2}\Big]\Big)
=\frac{\B}{2a^*}\inte |v_{2\B}|^2\frac{\partial |v_{1\B}|^2}{\partial x_j}.
\end{equation}

We now obtain from \eqref{A1}--\eqref{A4} that for $j=1,2$,
\begin{equation}\label{pohozaev v1B}
\frac{\eps_\B^4}{2}\inte V_\Om(x)\frac{\partial|v_{1\B}|^2}{\partial x_j}
=-\frac{1}{2}\inte \frac{\partial |v_{1\B}|^2}{\partial x_j}
 +\frac{a_1}{4a^*}\inte \frac{\partial |v_{1\B}|^4}{\partial x_j}
 +\frac{\B}{2a^*}\inte |v_{2\B}|^2\frac{\partial |v_{1\B}|^2}{\partial x_j}.
\end{equation}
Similarly, 
we deduce from the second equation of \eqref{rewrite v1Bv2B.equation} that for $j=1,2$,
\begin{equation}\label{pohozaev v2B}
\frac{\eps_\B^4}{2}\inte V_\Om(x)\frac{\partial |v_{2\B}|^2}{\partial x_j}
=-\frac{1}{2}\inte \frac{\partial |v_{2\B}|^2}{\partial x_j}
 +\frac{a_2}{4a^*}\inte \frac{\partial |v_{2\B}|^4}{\partial x_j}
 +\frac{\B}{2a^*}\inte |v_{1\B}|^2\frac{\partial |v_{2\B}|^2}{\partial x_j}.
\end{equation}
Using \eqref{vjB.decay}, we conclude from \eqref{pohozaev v1B}--\eqref{pohozaev v2B} that for $j=1,2,$
\begin{equation*}
\begin{split}
&\quad\eps_\B^4\inte\frac{\partial V_\Om(x)}{\partial x_j}\big(|v_{1\B}|^2+|v_{2\B}|^2\big)\\
&=-\eps_\B^4\inte V_\Om(x)\Big(\frac{\partial |v_{1\B}|^2}{\partial x_j}
+\frac{\partial |v_{2\B}|^2}{\partial x_j}\Big)\\
&=\lim_{R\to\infty}\int_{B_R(0)}\Big[\frac{\partial |v_{1\B}|^2}{\partial x_j}
 -\frac{a_1}{2a^*}\frac{\partial |v_{1\B}|^4}{\partial x_j}
+\frac{\partial |v_{2\B}|^2}{\partial x_j}
-\frac{a_2}{2a^*}\frac{\partial |v_{2\B}|^4}{\partial x_j}
-\frac{\B}{a^*}\frac{\partial( |v_{1\B}|^2 |v_{2\B}|^2)}{\partial x_j}\Big]\\
&=\lim_{R\to\infty}\int_{\partial B_R(0)}\Big[|v_{1\B}|^2\nu_j
 -\frac{a_1}{2a^*}|v_{1\B}|^4\nu_j
+|v_{2\B}|^2\nu_j
 -\frac{a_2}{2a^*}|v_{2\B}|^4\nu_j
 -\frac{\B}{a^*}
 |v_{1\B}|^2 |v_{2\B}|^2\nu_j\Big]=0,
\end{split}
\end{equation*}
where $\nu=(\nu_1,\nu_2)$ denotes the outward unit  normal of $\partial B_R(0)$.
This implies that \eqref{pohozaev v1B plus v2B} holds true, and we are done.
\qed

\subsection{The proofs of \eqref{4u19.bar.ujBvjB.Linfty} and Lemma \ref{lem 5}}
This appendix is focused on the proofs of \eqref{4u19.bar.ujBvjB.Linfty} and Lemma \ref{lem 5}. We begin to prove \eqref{4u19.bar.ujBvjB.Linfty} as follows.
\vskip 0.05truein

{\noindent \bf Proof of \eqref{4u19.bar.ujBvjB.Linfty}.}
We first prove the second inequality of \eqref{4u19.bar.ujBvjB.Linfty}. On the contrary, assume
\begin{equation}\label{4u20}
 \liminf_{\B\nearrow\B^*}\frac{\|\hat u_{2,1\B}-\hat u_{1,1\B}\|_{L^\infty(\R^2)}}
                              {\|\hat u_{2,2\B}-\hat u_{1,2\B}\|_{L^\infty(\R^2)}}
 =\infty.
\end{equation}
Set
\begin{equation}\label{4u21.eta.define}
  \eta_{\B}(x):=\frac{\hat u_{2,1\B}(x)-\hat u_{1,1\B}(x)}
                     {\|\hat u_{2,1\B}-\hat u_{1,1\B}\|_{L^\infty(\R^2)}}
  =R_{\eta_{\B}}(x)+iI_{\eta_{\B}}(x),
\end{equation}
where $R_{\eta_{\B}}(x)$ and $I_{\eta_{\B}}(x)$ denote the real and imaginary parts of $\eta_{\B}(x)$, respectively. We then derive from \eqref{4u14.bar.ujBvjB.equation} that $\eta_{\B}(x)$ satisfies
\begin{equation}\label{4u22.eta.equation}
\begin{split}
\mathcal{L}_\B\eta_{\B}
&=\alpha^2_\B\mu_{2\B}\eta_{\B}
+\frac{\alpha^2_\B(\mu_{2\B}-\mu_{1\B})}
      {\|\hat u_{2,1\B}-\hat u_{1,1\B}\|_{L^\infty(\R^2)}}\hat u_{1,1\B}\\
&\quad+\frac{a_1}{a^*}|\hat u_{2,1\B}|^2\eta_\B
      +\frac{a_1}{a^*}\Big[R_{\eta_{\B}}\big(R_{2,1\B}+R_{1,1\B}\big)
                      +I_{\eta_{\B}}\big(I_{2,1\B}+I_{1,1\B}\big)\Big]\hat u_{1,1\B}\\
&\quad+\frac{\B}{a^*}|\hat u_{1,2\B}|^2\eta_{\B}
      +\frac{\B}{a^*}\frac{|\hat u_{2,2\B}|^2-|\hat u_{1,2\B}|^2}
                    {\|\hat u_{2,1\B}-\hat u_{1,1\B}\|_{L^\infty(\R^2)}}\hat u_{2,1\B}
\quad \hbox{in}\  \ \R^2,
\end{split}
\end{equation}
where the operator $\mathcal{L}_\B$ is as in \eqref{LB}.

We begin to argue that there exists a constant $C>0$, independent of $\B$, such that
\begin{equation}\label{4u23}
\frac{\alpha^2_\B|\mu_{2\B}-\mu_{1\B}|}
     {\|\hat u_{2,1\B}-\hat u_{1,1\B}\|_{L^\infty(\R^2)}}\leq C.
\end{equation}
Indeed,
multiplying the first equation of \eqref{4u14.bar.ujBvjB.equation} by $\overline{\hat{u}_{j,1\B}}$ and integrating over $\R^2$, we have for $j=1,2$,
\begin{equation}\label{4u15.eps2BmujB.-1}
\inte\mathcal{L}_\B\hat{u}_{j,1\B}\cdot\overline{\hat{u}_{j,1\B}}dx
=\alpha^2_\B\mu_{j\B}\inte\big|\hat u_{j,1\B}\big|^2dx
 +\frac{a_1}{a^*}\inte\big|\hat u_{j,1\B}\big|^4dx
 +\frac{\B}{a^*}\inte\big|\hat u_{j,2\B}\big|^2\big|\hat u_{j,1\B}\big|^2dx,
\end{equation}
where and below $\bar{f}$ denotes the conjugate of $f$.
Similarly,
we derive that for $j=1,2$,
\begin{equation}\label{4u15.eps2BmujB.-2}
\inte\mathcal{L}_\B\hat{u}_{j,2\B}\cdot\overline{\hat{u}_{j,2\B}}dx
=\alpha^2_\B\mu_{j\B}\inte\big|\hat u_{j,2\B}\big|^2dx
 +\frac{a_2}{a^*}\inte\big|\hat u_{j,2\B}\big|^4dx
 +\frac{\B}{a^*}\inte\big|\hat u_{j,1\B}\big|^2\big|\hat u_{j,2\B}\big|^2dx.
\end{equation}
Noting from \eqref{4u11.bar.ujBvjB.define} that
$$\inte\Big(|\hat u_{j,1\B}|^2+|\hat u_{j,2\B}|^2\Big)dx=a^*,\ \ j=1,2,$$
we then derive from \eqref{4u15.eps2BmujB.-1} and \eqref{4u15.eps2BmujB.-2} that
\begin{equation}\label{4u23-1}
\begin{split}
\frac{\alpha^2_\B(\mu_{2\B}-\mu_{1\B})a^*}
  {\|\hat u_{2,1\B}-\hat u_{1,1\B}\|_{L^\infty(\R^2)}}
  &=\frac{A_2-A_1}{\|\hat u_{2,1\B}-\hat u_{1,1\B}\|_{L^\infty(\R^2)}}
   +\frac{B_2-B_1}{\|\hat u_{2,1\B}-\hat u_{1,1\B}\|_{L^\infty(\R^2)}}\\
  &\quad+\frac{C_2-C_1}{\|\hat u_{2,1\B}-\hat u_{1,1\B}\|_{L^\infty(\R^2)}},
\end{split}
\end{equation}
where
\begin{equation*}\label{4u23-2}
A_j:=\inte\Big(\mathcal{L}_\B\hat{u}_{j,1\B}\cdot\overline{\hat{u}_{j,1\B}}
     +\mathcal{L}_\B\hat{u}_{j,2\B}\cdot\overline{\hat{u}_{j,2\B}}\Big)dx,\ \ j=1,2,
\end{equation*}
\begin{equation*}\label{4u23-3}
B_j:=-\frac{a_1}{a^*}\inte\big|\hat u_{j,1\B}\big|^4dx
     -\frac{a_2}{a^*}\inte\big|\hat u_{j,2\B}\big|^4dx,\ \ j=1,2,
\end{equation*}
and
\begin{equation*}\label{4u23-4}
C_j:=-\frac{2\B}{a^*}\inte \big|\hat u_{j,1\B}\big|^2\big|\hat u_{j,2\B}\big|^2dx,
\ \ j=1,2.
\end{equation*}
Direct calculations give from \eqref{4u14.bar.ujBvjB.equation} that
\begin{equation}\label{4u23-5}
\begin{split}
  &\quad\frac{A_2-A_1}{\|\hat u_{2,1\B}-\hat u_{1,1\B}\|_{L^\infty(\R^2)}}\\
&=\inte\frac{\mathcal{L}_\B\hat{u}_{2,1\B}\cdot\overline{\hat{u}_{2,1\B}}
             -\mathcal{L}_\B\hat{u}_{1,1\B}\cdot\overline{\hat{u}_{1,1\B}}}
  {\|\hat u_{2,1\B}-\hat u_{1,1\B}\|_{L^\infty(\R^2)}}dx
  +\inte\frac{\mathcal{L}_\B\hat{u}_{2,2\B}\cdot\overline{\hat{u}_{2,2\B}}
             -\mathcal{L}_\B\hat{u}_{1,2\B}\cdot\overline{\hat{u}_{1,2\B}}}
  {\|\hat u_{2,1\B}-\hat u_{1,1\B}\|_{L^\infty(\R^2)}}dx\\
&=\inte\Big(\mathcal{L}_\B\hat{u}_{2,1\B}\cdot\bar{\eta}_\B
            +\eta_\B\cdot\overline{\mathcal{L}_\B\hat{u}_{1,1\B}}\Big)dx\\
  &\quad+\inte\Big(\mathcal{L}_\B\hat{u}_{2,2\B}
  \cdot\frac{\overline{\hat{u}_{2,2\B}-\hat{u}_{1,2\B}}}
            {\|\hat u_{2,1\B}-\hat u_{1,1\B}\|_{L^\infty(\R^2)}}
  +\frac{\hat{u}_{2,2\B}-\hat{u}_{1,2\B}}
        {\|\hat u_{2,1\B}-\hat u_{1,1\B}\|_{L^\infty(\R^2)}}
  \cdot\overline{\mathcal{L}_\B\hat{u}_{1,2\B}}\Big)dx\\
&=\inte\Big(\alpha^2_\B\mu_{2\B}\hat u_{2,1\B}
      +\frac{a_1}{a^*}|\hat u_{2,1\B}|^2\hat u_{2,1\B}
      +\frac{\B}{a^*}|\hat u_{2,2\B}|^2\hat u_{2,1\B}\Big)\cdot\bar{\eta}_\B dx\\
&\quad+\inte\Big(\alpha^2_\B\mu_{1\B}\overline{\hat u_{1,1\B}}
      +\frac{a_1}{a^*}|\hat u_{1,1\B}|^2\overline{\hat u_{1,1\B}}
      +\frac{\B}{a^*}|\hat u_{1,2\B}|^2\overline{\hat u_{1,1\B}}\Big)\cdot\eta_\B dx\\
&\quad+\inte\Big(\mathcal{L}_\B\hat{u}_{2,2\B}\cdot
       \frac{\overline{\hat{u}_{2,2\B}-\hat{u}_{1,2\B}}}
            {\|\hat u_{2,1\B}-\hat u_{1,1\B}\|_{L^\infty(\R^2)}}
      +\frac{\hat{u}_{2,2\B}-\hat{u}_{1,2\B}}
            {\|\hat u_{2,1\B}-\hat u_{1,1\B}\|_{L^\infty(\R^2)}}
  \cdot\overline{\mathcal{L}_\B\hat{u}_{1,2\B}}\Big)dx,
\end{split}
\end{equation}
\begin{align}\label{4u23-6}
  &\quad\frac{B_2-B_1}{\|\hat u_{2,1\B}-\hat u_{1,1\B}\|_{L^\infty(\R^2)}}\notag\\
&=-\frac{a_1}{a^*}\inte\frac{\big|\hat u_{2,1\B}\big|^4-\big|\hat u_{1,1\B}\big|^4}
                            {\|\hat u_{2,1\B}-\hat u_{1,1\B}\|_{L^\infty(\R^2)}}dx
-\frac{a_2}{a^*}\inte\frac{\big|\hat u_{2,2\B}\big|^4-\big|\hat u_{1,2\B}\big|^4}
                      {\|\hat u_{2,1\B}-\hat u_{1,1\B}\|_{L^\infty(\R^2)}}dx\notag\\
&=-\frac{a_1}{a^*}
     \inte\Big(\big|\hat u_{2,1\B}\big|^2+\big|\hat u_{1,1\B}\big|^2\Big)
     \cdot\Big(\overline{\hat u_{2,1\B}}\eta_\B+\hat u_{1,1\B}\bar{\eta}_\B\Big)dx\\
&\quad-\frac{a_2}{a^*}
     \inte\Big(\big|\hat u_{2,2\B}\big|^2+\big|\hat u_{1,2\B}\big|^2\Big)
     \cdot\Big(\overline{\hat u_{2,2\B}}\frac{\hat{u}_{2,2\B}-\hat{u}_{1,2\B}}
                    {\|\hat u_{2,1\B}-\hat u_{1,1\B}\|_{L^\infty(\R^2)}}
+\hat u_{1,2\B}\frac{\overline{\hat{u}_{2,2\B}-\hat{u}_{1,2\B}}}
                    {\|\hat u_{2,1\B}-\hat u_{1,1\B}\|_{L^\infty(\R^2)}}\Big)dx,\notag
\end{align}
and
\begin{equation}\label{4u23-7}
\begin{split}
  &\quad\frac{C_2-C_1}{\|\hat u_{2,1\B}-\hat u_{1,1\B}\|_{L^\infty(\R^2)}}\\
&=-\frac{2\B}{a^*}\inte\frac{\big|\hat u_{2,1\B}\big|^2\big|\hat u_{2,2\B}\big|^2
-\big|\hat u_{1,1\B}\big|^2\big|\hat u_{1,2\B}\big|^2}
  {\|\hat u_{2,1\B}-\hat u_{1,1\B}\|_{L^\infty(\R^2)}}dx\\
&=-\frac{2\B}{a^*}
     \inte\Big(\overline{\hat u_{2,1\B}}\eta_\B+\hat u_{1,1\B}\bar{\eta}_\B\Big)
           \big|\hat u_{2,2\B}\big|^2dx\\
&\quad-\frac{2\B}{a^*}
\inte\Big(\overline{\hat u_{2,2\B}}\frac{\hat{u}_{2,2\B}-\hat{u}_{1,2\B}}
                             {\|\hat u_{2,1\B}-\hat u_{1,1\B}\|_{L^\infty(\R^2)}}
                    +\hat u_{1,2\B}\frac{\overline{\hat{u}_{2,2\B}-\hat{u}_{1,2\B}}}
{\|\hat u_{2,1\B}-\hat u_{1,1\B}\|_{L^\infty(\R^2)}}\Big)\big|\hat u_{1,1\B}\big|^2dx.
\end{split}
\end{equation}
Applying \eqref{Sec 5 eps2mu-1}, \eqref{4u13.bar.ujBvjB.lim} and \eqref{4u20}--\eqref{4u21.eta.define}, we derive from \eqref{4u23-1}--\eqref{4u23-7} that \eqref{4u23} holds true.

In addition, we obtain from \eqref{4u13.bar.ujBvjB.lim} and \eqref{4u21.eta.define} that there exists a constant $C>0$, independent of $0<\B<\B^*$, such that
\begin{equation}\label{4u24}
\Big\|R_{\eta_{\B}}(R_{2,1\B}+R_{1,1\B})
+I_{\eta_{\B}}(I_{2,1\B}+I_{1,1\B})\Big\|_{L^\infty(\R^2)}\leq C.
\end{equation}
Similar to \cite [Lemma 3.2] {GLP}, together with \eqref{4u6.baruBvB.decay}, one can then deduce from \eqref{Sec 5 eps2mu-1}, \eqref{4u20}, \eqref{4u22.eta.equation}, \eqref{4u23} and \eqref{4u24} that there exists a constant $C>0$, independent of $0<\B<\B^*$, such that
\begin{equation}\label{4u25.eta.decay}
      |\eta_{\B}(x)|\leq Ce^{-\frac{2}{3}|x|} \quad  \text{and} \quad
      |\nabla\eta_{\B}(x)|\leq Ce^{-\frac{1}{2}|x|} \  \ \text{in} \  \  \R^2\  \  \text{as}\  \ \B\nearrow\B^*.
     \end{equation}

We obtain from \eqref{4u25.eta.decay} that $(x^\perp\cdot\nabla\eta_{\B})$ is bounded uniformly and decays exponentially for sufficiently large $|x|$ as $\B\nearrow\B^*$. The standard elliptic regularity theory then implies from \eqref{Sec 5 eps2mu-1}, \eqref{4u6.baruBvB.decay}, \eqref{4u20}, \eqref{4u22.eta.equation}, \eqref{4u23} and \eqref{4u24} that $\eta_{\B}\in C^{1,\alpha}_{loc}(\R^2,\C)$ and $\|\eta_{\B}\|_{C^{1,\alpha}_{loc}(\R^2,\C)}\leq C$ uniformly as $\B\nearrow\B^*$ for some $\alpha\in(0,1)$. Therefore, there exists a subsequence, still denoted by $\{\eta_{\B}\}$, of $\{\eta_{\B}\}$ such that
 \begin{equation}\label{4u23-8}
   \eta_{\B}=R_{\eta_{\B}}+iI_{\eta_{\B}}\to\eta_{0}=R_{\eta_{0}}+iI_{\eta_{0}}
   \ \ \hbox{uniformly in}\ \ C^{1}_{loc}(\R^2,\C) \ \ \hbox{as} \ \ \B\nearrow\B^*.
 \end{equation}
Applying \eqref{Sec 5 eps2mu-1}--\eqref{4u6.baruBvB.decay}, \eqref{4u20}, \eqref{4u25.eta.decay} and \eqref{4u23-8}, we then derive from \eqref{4u23-1}--\eqref{4u23-7} that
\begin{align}\label{4u23-4}
\lim_{\B\nearrow\B^*}\frac{\alpha^2_\B(\mu_{2\B}-\mu_{1\B})}
{\|\hat u_{2,1\B}-\hat u_{1,1\B}\|_{L^\infty(\R^2)}}
=&-\frac{2a_1\gamma_1^{\frac{3}{2}}}{(a^*)^2}\inte w^3R_{\eta_{0}}dx
  -\frac{2\B^*\gamma_2\sqrt{\gamma_1}}{(a^*)^2}\inte w^3R_{\eta_{0}}dx.
\end{align}
Using \eqref{Sec 5 eps2mu-1}, \eqref{4u13.bar.ujBvjB.lim}, \eqref{4u20} and \eqref{4u25.eta.decay}--\eqref{4u23-4}, we derive from  \eqref{4u22.eta.equation} that $\eta _0$ satisfies
\begin{equation*}
\begin{split}
&\quad-\Delta\eta_{0}+\eta_{0}
-\frac{2a_1\gamma_1}{a^*}w^2R_{\eta_{0}}-\frac{a_1\gamma_1}{a^*}w^2\eta_{0}
-\frac{\B^*\gamma_2}{a^*}w^2\eta_{0}\\
&=-\frac{2a_1\gamma_1^2}{(a^*)^2}w\inte w^3R_{\eta_{0}}dx
  -\frac{2\B^*\gamma_1\gamma_2}{(a^*)^2}w\inte w^3R_{\eta_{0}}dx
\quad \hbox{in}\  \ \R^2.
\end{split}
\end{equation*}
This implies that $(R_{\eta_{0}},I_{\eta_{0}})$ satisfies the following system
\begin{equation*}
\left\{\begin{array}{lll}
-\ds\Delta R_{\eta_{0}}+R_{\eta_{0}}
-\Big[\frac{3a_1\gamma_1}{a^*}+\frac{\B^*\gamma_2}{a^*}\Big]w^2R_{\eta_{0}}
=-\frac{2\gamma_1}{a^*}\Big[\frac{a_1\gamma_1}{a^*}
 +\frac{\B^*\gamma_2}{a^*}\Big]w\inte w^3R_{\eta_{0}}dx
\  \ \hbox{in}\  \ \R^2,\\[3mm]
-\ds\Delta I_{\eta_{0}}+I_{\eta_{0}}-\Big[\frac{a_1\gamma_1}{a^*}
+\frac{\B^*\gamma_2}{a^*}\Big]w^2I_{\eta_{0}}=0
\quad \hbox{in}\  \ \R^2,
   \end{array}\right.
   \end{equation*}
which can be further simplified as
\begin{equation}\label{4u27.eta1.eta2.equation}
\left\{\begin{array}{lll}
-\ds\Delta R_{\eta_{0}}+R_{\eta_{0}}
-\Big[1+\frac{2a_1\gamma_1}{a^*}\Big]w^2R_{\eta_{0}}
=-\frac{2\gamma_1}{a^*}w\inte w^3R_{\eta_{0}}dx
\quad \hbox{in}\  \ \R^2,\\[3mm]
-\Delta I_{\eta_{0}}+I_{\eta_{0}}-w^2I_{\eta_{0}}=0
\quad \hbox{in}\  \ \R^2,
   \end{array}\right.
   \end{equation}
due to the fact that $a_1\gamma_1+\B^*\gamma_2=a^*$.

Multiplying the first equation of \eqref{4u27.eta1.eta2.equation} by $w$ and integrating over $\R^2$, we have
\begin{equation}\label{lem4.4-1}
\ds\inte\nabla w \nabla R_{\eta_{0}}dx+\inte w R_{\eta_{0}}dx
=\Big[1+2\gamma_1\Big(\frac{a_1}{a^*}-1\Big)\Big]
 \inte w^3R_{\eta_{0}}dx,
   \end{equation}
and it follows from \eqref{1.8equation:w} that
\begin{equation}\label{lem4.4-2}
\inte\nabla w \nabla R_{\eta_{0}}dx+\inte w R_{\eta_{0}}dx
=\inte w^3R_{\eta_{0}}dx.
   \end{equation}
Since $a_1\in(0,a^*)$ and $\gamma_1\in(0,1)$, the combination of  \eqref{lem4.4-1} and \eqref{lem4.4-2} yields that $\inte w^3R_{\eta_{0}}dx=0$. The first equation of \eqref{4u27.eta1.eta2.equation} is then reduced to
\begin{equation}\label{lem4.4-3}
-\ds\Delta R_{\eta_{0}}+R_{\eta_{0}}
-\Big[1+\frac{2a_1\gamma_1}{a^*}\Big]w^2R_{\eta_{0}}=0
\quad \hbox{in}\  \ \R^2,\quad \hbox{where}\ \ 1+\frac{2a_1\gamma_1}{a^*}\in(1,3).
   \end{equation}
By \cite [Lemma 2.2] {DW}, we thus conclude from \eqref{lem4.4-3} that $R_{\eta_{0}}\equiv0$. Note from \eqref{4u4} and \eqref{4u21.eta.define} that $\inte w(x)I_{\eta_{\B}}(x)\equiv0$, which further yields that $\inte w(x)I_{\eta_{0}}(x)\equiv0$. We then deduce from \eqref{G5} and \eqref{4u27.eta1.eta2.equation} that $I_{\eta_{0}}\equiv0$, and hence $\eta_0=R_{\eta_{0}}+i I_{\eta_{0}}\equiv0$.

On the other hand, since $\|\eta_\B\|_{L^\infty(\R^2)}=1$, we can derive from \eqref{4u25.eta.decay} and \eqref{4u23-8} that $\eta_\B\to\eta_0\not\equiv0$ uniformly in $\R^2$ as $\B\nearrow\B^*$, which however contradicts to the above conclusion that $\eta_0\equiv0$ in $\R^2$. This implies that \eqref{4u20} is false, and hence the second inequality of \eqref{4u19.bar.ujBvjB.Linfty} holds true.

Similarly, one can get the first inequality of \eqref{4u19.bar.ujBvjB.Linfty}, which therefore completes the proof of \eqref{4u19.bar.ujBvjB.Linfty}.
\qed

\vskip 0.05truein

{\noindent \bf Proof of Lemma \ref{lem 5}.}
Following \eqref{4u14.bar.ujBvjB.equation}, we obtain that the real part $(R_{j,1\B},R_{j,2\B})$ of
$(\hat u_{j,1\B},\hat u_{j,2\B})$ satisfies
\begin{equation}\label{Rj1B Rj2B.equation}
\left\{\begin{array}{lll}
  \quad-\Delta R_{j,1\B}(x)-\alpha^2_\B\Om\big(x^\perp\cdot\nabla I_{j,1\B}\big)
       +\alpha_\B^4V(x)R_{j,1\B}(x)\\[5mm]
=\ds\alpha^2_\B\mu_{j\B}R_{j,1\B}(x)+\frac{a_1}{a^*}|\hat u_{j,1\B}|^2R_{j,1\B}(x)
                                    +\frac{\B}{a^*}|\hat u_{j,2\B}|^2R_{j,1\B}(x)
\quad\hbox{in}\  \ \R^2,\\[5mm]
  \quad-\Delta R_{j,2\B}(x)-\alpha^2_\B\Om\big(x^\perp\cdot\nabla I_{j,2\B}\big)
       +\alpha_\B^4V(x)R_{j,2\B}(x)\\[5mm]
=\ds\alpha^2_\B\mu_{j\B}R_{j,2\B}(x)+\frac{a_2}{a^*}|\hat u_{j,2\B}|^2R_{j,2\B}(x)
                                    +\frac{\B}{a^*}|\hat u_{j,1\B}|^2R_{j,2\B}(x)
\quad\hbox{in}\  \ \R^2,
\end{array}\right.
\end{equation}
and the imaginary part $(I_{j,1\B},I_{j,2\B})$ of
$(\hat u_{j,1\B},\hat u_{j,2\B})$ satisfies
\begin{equation}\label{Ij1B Ij2B.equation}
\left\{\begin{array}{lll}
  \quad-\Delta I_{j,1\B}(x)+\alpha^2_\B\Om\big(x^\perp\cdot\nabla R_{j,1\B}\big)
       +\alpha_\B^4V(x)I_{j,1\B}(x)\\[5mm]
=\ds\alpha^2_\B\mu_{j\B}I_{j,1\B}(x)+\frac{a_1}{a^*}|\hat u_{j,1\B}|^2I_{j,1\B}(x)
                                    +\frac{\B}{a^*}|\hat u_{j,2\B}|^2I_{j,1\B}(x)
\quad\hbox{in}\  \ \R^2,\\[5mm]
  \quad-\Delta I_{j,2\B}(x)+\alpha^2_\B\Om\big(x^\perp\cdot\nabla R_{j,2\B}\big)
                           +\alpha_\B^4V(x)I_{j,2\B}(x)\\[5mm]
=\ds\alpha^2_\B\mu_{j\B}I_{j,2\B}(x)+\frac{a_2}{a^*}|\hat u_{j,2\B}|^2I_{j,2\B}(x)
                                    +\frac{\B}{a^*}|\hat u_{j,1\B}|^2I_{j,2\B}(x)
\quad\hbox{in}\  \ \R^2,
\end{array}\right.
\end{equation}
where $j=1, 2$. We then deduce from \eqref{4u11.bar.ujBvjB.define},
\eqref{Rj1B Rj2B.equation} and \eqref{Ij1B Ij2B.equation} that for $j=1,2$,
\begin{equation}\label{alpha_B2 mu_jB}
\begin{split}
\alpha_\B^2\mu_{j\B}a^*
&=\alpha_\B^2\mu_{j\B}\inte\Big(R_{j,1\B}^2+R_{j,2\B}^2+I_{j,1\B}^2+I_{j,2\B}^2\Big)\\
&=\inte\big(|\nabla R_{j,1\B}|^2+|\nabla R_{j,2\B}|^2\big)
      +\inte\big(|\nabla I_{j,1\B}|^2+|\nabla I_{j,2\B}|^2\big)\\
&\quad-\inte\Big(\alpha_\B^2\Om\big(x^\perp\cdot\nabla I_{j,1\B}\big)R_{j,1\B}
+\alpha_\B^2\Om\big(x^\perp\cdot\nabla I_{j,2\B}\big)R_{j,2\B}
\Big)\\
&\quad+\inte\Big(\alpha_\B^2\Om\big(x^\perp\cdot\nabla R_{j,1\B}\big)I_{j,1\B}
+\alpha_\B^2\Om\big(x^\perp\cdot\nabla R_{j,2\B}\big)I_{j,2\B}
\Big)\\
&\quad+\inte\alpha_\B^4V(x)\Big(R_{j,1\B}^2+R_{j,2\B}^2+I_{j,1\B}^2+I_{j,2\B}^2\Big)\\
&\quad-\ds\inte
\Big(\frac{a_1}{a^*}|\hat u_{j,1\B}|^2\big(R_{j,1\B}^2+I_{j,1\B}^2\big)
+\frac{a_2}{a^*}|\hat u_{j,2\B}|^2\big(R_{j,2\B}^2+I_{j,2\B}^2\big)\Big)\\
&\quad-\ds\inte \frac{\B}{a^*}
\Big(|\hat u_{j,2\B}|^2\big(R_{j,1\B}^2+I_{j,1\B}^2\big)
+|\hat u_{j,1\B}|^2\big(R_{j,2\B}^2+I_{j,2\B}^2\big)\Big).\\
\end{split}
\end{equation}

By Theorem \ref{thm-expan solution}, we obtain from \eqref{4u11.bar.ujBvjB.define} that
\begin{equation}\label{A.alphaB2 muB expan}
\alpha_\B^2\mu_{j\B}=-1+C(\lambda_0,a_1,a_2,\B^*)\alpha_\B^4+O(\alpha_\B^8)\ \ \hbox{as}\ \ \B\nearrow\B^*,
\ \ j=1,2,
\end{equation}
and
\begin{equation}\label{A.expan Rj1B Rj2B Ij1B Ij2B}
R_{j,l\B}=\rho_{l\B}w+O(\alpha_\B^{4}),\ \  I_{j,l\B}=O(\alpha_\B^{6})\ \ \hbox{in}\ \ \R^2
\ \ \hbox{as} \ \ \B\nearrow\B^*,\ \ j,l=1,2,
   \end{equation}
where $\rho_{l\B}>0$ is as in \eqref{rho jB define}. Here the constant $C(\lambda_0,a_1,a_2,\B^*)$ is independent of $\Om$ and satisfies
\begin{equation*}
C(\lambda_0,a_1,a_2,\B^*):=\frac{3\langle\widetilde{\mathcal{L}}\psi_0,\psi_0\rangle}{2\lambda_0}
                          +\frac{\lambda_0\big(4\gamma_1\gamma_2-1\big)}
                                {\big(2\B^*-a_1-a_2\big)8\gamma_1^2\gamma_2^2},
\end{equation*}
where $0<\gamma_1,\gamma_2<1$ are as in \eqref{1.13gamma}, $\lambda_0>0$ is given by \eqref{1.14lam0}, and $\widetilde{\mathcal{L}}$ is defined by \eqref{line oper L}.

Applying \eqref{4u31.xijB.estimates}, \eqref{A.alphaB2 muB expan} and \eqref{A.expan Rj1B Rj2B Ij1B Ij2B}, we derive from
\eqref{Rj1B Rj2B.equation}--\eqref{alpha_B2 mu_jB} that as $\B\nearrow\B^*$,
\begin{equation*}
\begin{split}
&\quad\frac{\alpha^2_\B(\mu_{2\B}-\mu_{1\B})a^*}
{\|\hat u_{2,1\B}-\hat u_{1,1\B}\|^{\frac{1}{2}}_{L^\infty(\R^2)}
  \|\hat u_{2,2\B}-\hat u_{1,2\B}\|^{\frac{1}{2}}_{L^\infty(\R^2)}}\\
  &=\inte\Big[\big(\nabla R_{2,1\B}+\nabla R_{1,1\B}\big)\nabla R_{\xi_{1\B}}
     -\Big(\alpha_\B^2\Om(x^\perp\cdot\nabla I_{2,1\B})
     +\alpha_\B^2\Om(x^\perp\cdot\nabla I_{1,1\B})\Big)R_{\xi_{1\B}}\Big]\\
  &\quad+\ds\inte\alpha_\B^4V(x)\big(R_{2,1\B}+R_{1,1\B}\big)R_{\xi_{1\B}}
        -\frac{a_1}{a^*}\inte\big(R_{2,1\B}^2+R_{1,1\B}^2\big)
                             \big(R_{2,1\B}+R_{1,1\B}\big)R_{\xi_{1\B}}\\
&\quad-\ds\frac{\B}{a^*}\inte\Big[R_{2,2\B}^2\big(R_{2,1\B}+R_{1,1\B}\big)R_{\xi_{1\B}}
                            +\big(R_{2,2\B}+R_{1,2\B}\big)R_{1,1\B}^2R_{\xi_{2\B}}\Big]\\
&\quad+\inte\Big[\big(\nabla R_{2,2\B}+\nabla R_{1,2\B}\big)\nabla R_{\xi_{2\B}}
      -\Big(\alpha_\B^2\Om(x^\perp\cdot\nabla I_{2,2\B})
           +\alpha_\B^2\Om(x^\perp\cdot\nabla I_{1,2\B})\Big)R_{\xi_{2\B}}\Big]\\
  &\quad+\ds\inte\alpha_\B^4V(x)\big(R_{2,2\B}+R_{1,2\B}\big)R_{\xi_{2\B}}
        -\frac{a_2}{a^*}\inte\big(R_{2,2\B}^2+R_{1,2\B}^2\big)
                             \big(R_{2,2\B}+R_{1,2\B}\big)R_{\xi_{2\B}}\\
  &\quad-\ds\frac{\B}{a^*}\inte\Big[R_{2,2\B}^2\big(R_{2,1\B}+R_{1,1\B}\big)R_{\xi_{1\B}}
                            +\big(R_{2,2\B}+R_{1,2\B}\big)R_{1,1\B}^2R_{\xi_{2\B}}\Big]\\
&\quad+\inte\Big[\big(\nabla I_{2,1\B}+\nabla I_{1,1\B}\big)\nabla I_{\xi_{1\B}}
      +\Big(\alpha_\B^2\Om(x^\perp\cdot\nabla R_{2,1\B})
           +\alpha_\B^2\Om(x^\perp\cdot\nabla R_{1,1\B})\Big)I_{\xi_{1\B}}\Big]\\
  &\quad+\inte\alpha_\B^4V(x)\big(I_{2,1\B}+I_{1,1\B}\big)I_{\xi_{1\B}}
        +\inte\alpha_\B^4V(x)\big(I_{2,2\B}+I_{1,2\B}\big)I_{\xi_{2\B}}\\
&\quad+\inte\Big[\big(\nabla I_{2,2\B}+\nabla I_{1,2\B}\big)\nabla I_{\xi_{2\B}}
      +\Big(\alpha_\B^2\Om(x^\perp\cdot\nabla R_{2,2\B})
           +\alpha_\B^2\Om(x^\perp\cdot\nabla R_{1,2\B})\Big)I_{\xi_{2\B}}\Big]
  +o(\alpha_\B^4)\\
&=\alpha_\B^2\mu_{1\B}\inte\Big[\big(R_{2,1\B}+R_{1,1\B}\big)R_{\xi_{1\B}}
                               +\big(R_{2,2\B}+R_{1,2\B}\big)R_{\xi_{2\B}}\Big]\\
&\quad+\alpha_\B^2(\mu_{2\B}-\mu_{1\B})
       \inte\big(R_{2,1\B}R_{\xi_{1\B}}+R_{2,2\B}R_{\xi_{2\B}}\big)\\
&\quad-\ds\frac{a_1}{a^*}
       \inte R_{2,1\B}R_{1,1\B}\big(R_{2,1\B}+R_{1,1\B}\big)R_{\xi_{1\B}}
      -\frac{a_2}{a^*}
      \inte R_{2,2\B}R_{1,2\B}\big(R_{2,2\B}+R_{1,2\B}\big)R_{\xi_{2\B}}\\
&\quad+\ds\frac{\B}{a^*}\inte\Big[R_{2,2\B}^2R_{2,1\B}+R_{1,2\B}^2R_{1,1\B}
                   -2R_{2,2\B}^2\big(R_{2,1\B}+R_{1,1\B}\big)\Big]R_{\xi_{1\B}}\\
&\quad+\ds\frac{\B}{a^*}\inte\Big[R_{2,1\B}^2R_{2,2\B}+R_{1,1\B}^2R_{1,2\B}
                   -2R_{1,1\B}^2\big(R_{2,2\B}+R_{1,2\B}\big)\Big]R_{\xi_{2\B}}
                   +o(\alpha_\B^4)\\
&=-\ds\frac{a_1}{a^*}
       \inte R_{2,1\B}R_{1,1\B}\big(R_{2,1\B}+R_{1,1\B}\big)R_{\xi_{1\B}}
      -\frac{a_2}{a^*}
      \inte R_{2,2\B}R_{1,2\B}\big(R_{2,2\B}+R_{1,2\B}\big)R_{\xi_{2\B}}\\
&\quad+\ds\frac{\B}{a^*}\inte\Big[R_{2,2\B}^2R_{2,1\B}+R_{1,2\B}^2R_{1,1\B}
                   -2R_{2,2\B}^2\big(R_{2,1\B}+R_{1,1\B}\big)\Big]R_{\xi_{1\B}}\\
&\quad+\ds\frac{\B}{a^*}\inte\Big[R_{2,1\B}^2R_{2,2\B}+R_{1,1\B}^2R_{1,2\B}
                   -2R_{1,1\B}^2\big(R_{2,2\B}+R_{1,2\B}\big)\Big]R_{\xi_{2\B}}
                   +o(\alpha_\B^4),
\end{split}
\end{equation*}
where we have used the fact that
\begin{equation*}
\begin{split}
&\quad\inte\big(R_{2,l\B}+R_{1,l\B}\big)R_{\xi_{l\B}}\\
&=\frac{\inte\big(R_{2,l\B}^2-R_{1,l\B}^2\big)}
{\|\hat u_{2,1\B}-\hat u_{1,1\B}\|^{\frac{1}{2}}_{L^\infty(\R^2)}
   \|\hat u_{2,2\B}-\hat u_{1,2\B}\|^{\frac{1}{2}}_{L^\infty(\R^2)}}\\
&=\frac{a^*-\inte I_{2,l\B}^2-\big(a^*-\inte I_{1,l\B}^2\big)}
{\|\hat u_{2,1\B}-\hat u_{1,1\B}\|^{\frac{1}{2}}_{L^\infty(\R^2)}
   \|\hat u_{2,2\B}-\hat u_{1,2\B}\|^{\frac{1}{2}}_{L^\infty(\R^2)}}\\
&=-\inte\big(I_{2,l\B}+I_{1,l\B}\big)I_{\xi_{l\B}}
=o(\alpha_\B^4)\ \ \hbox{as}\ \ \B\nearrow\B^*,\ \ l=1,2.
\end{split}
\end{equation*}
Using \eqref{4u31.xijB.estimates} and \eqref{A.expan Rj1B Rj2B Ij1B Ij2B}, we then deduce from above that as $\B\nearrow\B^*$,
\begin{equation*}
\begin{split}
&\quad\frac{\alpha^2_\B(\mu_{2\B}-\mu_{1\B})}
{\|\hat u_{2,1\B}-\hat u_{1,1\B}\|^{\frac{1}{2}}_{L^\infty(\R^2)}
  \|\hat u_{2,2\B}-\hat u_{1,2\B}\|^{\frac{1}{2}}_{L^\infty(\R^2)}}\\
&\quad+\frac{a_1}{2(a^*)^2}\inte\big(R_{2,1\B}^2+R_{1,1\B}^2\big)
                                \big(R_{2,1\B}+R_{1,1\B}\big)R_{\xi_{1\B}}\\
&\quad+\frac{a_2}{2(a^*)^2}\inte\big(R_{2,2\B}^2+R_{1,2\B}^2\big)
                                \big(R_{2,2\B}+R_{1,2\B}\big)R_{\xi_{2\B}}\\
&\quad+\ds\frac{\B}{(a^*)^2}\inte
                         \Big[\big(R_{2,1\B}+R_{1,1\B}\big)R_{2,2\B}^2R_{\xi_{1\B}}
                             +\big(R_{2,2\B}+R_{1,2\B}\big)R_{1,1\B}^2R_{\xi_{2\B}}\Big]\\
&=-\frac{a_1}{(a^*)^2}\inte R_{2,1\B}R_{1,1\B}\big(R_{2,1\B}+R_{1,1\B}\big)
                            R_{\xi_{1\B}}\\
&\quad-\frac{a_2}{(a^*)^2}\inte R_{2,2\B}R_{1,2\B}\big(R_{2,2\B}+R_{1,2\B}\big)
                            R_{\xi_{2\B}}\\
&\quad+\frac{\B}{(a^*)^2}\inte\Big[R_{2,2\B}^2R_{2,1\B}+R_{1,2\B}^2R_{1,1\B}
                   -2R_{2,2\B}^2\big(R_{2,1\B}+R_{1,1\B}\big)\Big]R_{\xi_{1\B}}\\
&\quad+\frac{\B}{(a^*)^2}\inte\Big[R_{2,1\B}^2R_{2,2\B}+R_{1,1\B}^2R_{1,2\B}
                   -2R_{1,1\B}^2\big(R_{2,2\B}+R_{1,2\B}\big)\Big]R_{\xi_{2\B}}\\
&\quad+\frac{a_1}{2(a^*)^2}\inte\big(R_{2,1\B}^2+R_{1,1\B}^2\big)
                                \big(R_{2,1\B}+R_{1,1\B}\big)R_{\xi_{1\B}}\\
&\quad+\frac{a_2}{2(a^*)^2}\inte\big(R_{2,2\B}^2+R_{1,2\B}^2\big)
                                \big(R_{2,2\B}+R_{1,2\B}\big)R_{\xi_{2\B}}\\
&\quad+\ds\frac{\B}{(a^*)^2}\inte
                         \Big[\big(R_{2,1\B}+R_{1,1\B}\big)R_{2,2\B}^2R_{\xi_{1\B}}
                             +\big(R_{2,2\B}+R_{1,2\B}\big)R_{1,1\B}^2R_{\xi_{2\B}}\Big]+o(\alpha_\B^4)\\
&=\frac{a_1}{2(a^*)^2}\inte\big(R_{2,1\B}-R_{1,1\B}\big)^2
                  \big(R_{2,1\B}+R_{1,1\B}\big)R_{\xi_{1\B}}\\
&\quad+\frac{a_2}{2(a^*)^2}\inte\big(R_{2,2\B}-R_{1,2\B}\big)^2
                  \big(R_{2,2\B}+R_{1,2\B}\big)R_{\xi_{2\B}}\\
&\quad+\frac{\B}{(a^*)^2}\inte\Big[\big(R_{1,2\B}^2-R_{2,2\B}^2\big)R_{1,1\B}R_{\xi_{1\B}}
             +\big(R_{2,1\B}^2-R_{1,1\B}^2\big)R_{2,2\B}R_{\xi_{2\B}}\Big]+o(\alpha_\B^4)\\
&=o(\alpha_\B^4),
\end{split}
\end{equation*}
which therefore implies that \eqref{alpha_B2 mu_jB expan} holds true, and we are done.
\qed




\begin{thebibliography}{99}
\bibitem{A} A. Aftalion,  Vortices in Bose-Einstein condensates, Progress in Nonlinear Differential Equations and their Applications, 67. Birkh$\ddot{a}$user Boston, Inc., Boston, MA,  2006.

\bibitem{ABN}  A. Aftalion, X. Blanc and F. Nier, {\em Lowest Landau level functional and Bargmann spaces for Bose-Einstein condensates}, J. Funct. Anal. {\bf 241} (2006), 661--702.

\bibitem{AJ} A. Aftalion, R. L. Jerrard and J. Royo-Letelier, {\em Non-existence of vortices in the small density region of a condensate}, J. Funct. Anal. {\bf 260} (2011), 2387--2406.


\bibitem{AMW} A. Aftalion, P. Mason and J. C. Wei, {\em Vortex-peak interaction and lattice shape in rotating two-component Bose-Einstein condensates},  Phys. Rev. A {\bf 85} (2012), 033614.



\bibitem{ANS}  A. Aftalion, B. Noris and C. Sourdis, {\em Thomas-Fermi approximation for coexisting two component Bose-Einstein condensates and nonexistence of vortices for small rotation},  Comm. Math. Phys. {\bf 336} (2015), no. 2, 509--579.

\bibitem{AS} A. Aftalion and E. Sandier, {\em Vortex patterns and sheets in segregated two component Bose-Einstein condensates}, Calc. Var. Partial Differ. Eqns. {\bf59} (2020), No. 19, 38 pp.

\bibitem{ANS1} J. Arbunich, I. Nenciu and C. Sparber, {\em Stability and instability properties of rotating Bose-Einstein condensates},  Lett. Math. Phys. {\bf 109} (2019), 1415--1432.

\bibitem{BC} W. Z. Bao and Y. Y. Cai, {\em Mathematical theory and numerical methods for Bose-Einstein condensation}, Kinet. Relat. Models {\bf6} (2013), 1--135.





\bibitem{C} T. Cazenave, Semilinear Schr\"{o}dinger Equations, Courant Lecture Notes in Mathematics  Vol. 10, Courant Institute of Mathematical Science/AMS, New York, 2003.

\bibitem{CR} M. Correggi and N. Rougerie, {\em Boundary behavior of the Ginzburg-Landau order parameter in the surface superconductivity regime}, Arch. Ration. Mech. Anal. {\bf 219} (2016), 553--606.


\bibitem{DW} E. N. Dancer and J. C. Wei, {\em Spike solutions in coupled nonlinear Schr\"{o}dinger equations with attractive interaction}, Trans. Amer. Math. Soc. {\bf361} (2009),  1189--1208.


\bibitem{EL} M. J. Esteban and P. L. Lions, {\em Stationary solutions of nonlinear Schr\"{o}dinger equations with an external magnetic field}, Partial differential equations and the calculus of variations, Vol. I, 401--449, Progr. Nonlinear Differential Eqns. Appl. 1, Birkhuser Boston, Boston, MA, 1989.

\bibitem{GRPG}  J. J. Garc{\'{\i}}a-Ripoll, V. M. P\'{e}rez-Garc{\'{\i}}a and F. Sols, {\em Split vortices in optically coupled Bose-Einstein condensates},  Phys. Rev. A {\bf66} (2002), 021602.

\bibitem{GNN} B. Gidas, W. M. Ni and L. Nirenberg, {\em Symmetry of positive solutions of nonlinear elliptic equations in $\R^{n}$}, Mathematical analysis and applications Part A, Adv. Math. Suppl. Stud. Vol. 7, Academic Press, New York (1981), 369--402.

\bibitem{GT} D. Gilbarg and N. S. Trudinger,  Elliptic Partial Differential Equations of Second Order, Springer, Berlin 1997.

\bibitem{G} M. Grossi, {\em On the number of single-peak solutions of the nonlinear Schr\"{o}dinger equations}, Ann. Inst. H. Poincar\'{e} Anal. Non Lin\'{e}aire {\bf19} (2002), 261--280.

\bibitem{GX} Q. Guo and H. F. Xie, {\em Existence and local uniqueness of normalized solutions for two-component Bose-Einstein condensates},  Z. Angew. Math. Phys. {\bf72} (2021), 189.

\bibitem{Guo} Y. J. Guo, {\em The nonexistence of vortices for rotating Bose-Einstein condenstates in non-radially symmetric traps}, J. Math. Pure. Appl. (2022), to appear, 31 pages.

\bibitem{GLWZ1} Y. J. Guo, S. Li, J. C. Wei and X. Y. Zeng, {\em Ground states of two-component attractive Bose-Einstein condenstates I: Existence and uniqueness}, J. Funct. Anal. {\bf276} (2019), 183--230.

\bibitem{GLWZ2} Y. J. Guo, S. Li, J. C. Wei and X. Y. Zeng, {\em Ground states of two-component attractive Bose-Einstein condenstates II: Semi-trivial limit behavior}, Trans. Amer. Math. Soc. {\bf371} (2019), 6903--6948.

\bibitem{GLW} Y. J. Guo, C. S. Lin and J. C. Wei, {\em Local uniqueness and refined spike profiles of ground states for two-dimensional attractive Bose-Einstein condensates}, SIAM J. Math. Anal. {\bf49} (2017), 3671--3715.

\bibitem{GLP} Y. J. Guo, Y. Luo and S. J. Peng, {\em Local uniqueness of ground states for rotating Bose-Einstein condenstates with attractive interactions}, Calc. Var. Partial Differ. Eqns. {\bf60} (2021), 237.

\bibitem{GLY} Y. J. Guo, Y. Luo and W. Yang, {\em The nonexistence of vortices for rotating Bose-Einstein condenstates with attractive interactions},  Arch. Ration. Mech. Anal. {\bf238} (2020), 1231--1281.

\bibitem{GS} Y. J. Guo and R. Seiringer, {\em On the mass concentration for Bose-Einstein condensates with attractive interactions}, Lett. Math. Phys. {\bf104} (2014), 141--156.



\bibitem{GZZ2} Y. J. Guo, X. Y. Zeng and H. S. Zhou, {\em Blow-up solutions for two coupled Gross-Pitaevskii equations with attractive interactions}, Discrete Contin. Dyn. Syst. Ser. A {\bf37} (2017), 3749--3786.

\bibitem{HMEWC} D. S. Hall, M. R. Matthews, J. R. Ensher, C. E. Wieman and E. A. Cornell, {\em Dynamics of component separation in a binary mixture of Bose-Einstein condensates},
    Phys. Rev. Lett. {\bf81} (1998), 1539--1542.


\bibitem{HL} Q. Han and F. H. Lin, Elliptic Partial Differential Equations, Courant Lecture Notes in Mathematics Vol. 1, Courant Institute of Mathematical Science/AMS, New York, 2011.

\bibitem{HS} T.-L. Ho and V. B. Shenoy, {\em Binary mixtures of Bose condensates of alkali atoms}, Phys. Rev. Lett. {\bf77} (1996), 3276--3279.





\bibitem{IM-1}  R. Ignat and V. Millot, {\em The critical velocity for vortex existence in a two-dimensional rotating Bose-Einstein condensate},  J. Funct. Anal. {\bf 233} (2006),  260--306.




\bibitem{KTU1} K. Kasamatsu, M. Tsubota and M. Ueda, {\em Vortex phase diagram in rotating two-component Bose-Einstein condensates}, Phys. Rev. Lett. {\bf91} (2003), 150406.

\bibitem{KTU2} K. Kasamatsu, M. Tsubota and M. Ueda, {\em Vortices in multicomponent Bose-Einstein condensates}, Internat. J. Modern Phys. B {\bf19} (2005), 1835--1904.


\bibitem{K} M. K. Kwong, {\em Uniqueness of positive solutions of $\Delta u-u+u^{p}=0$ in $\R^{N}$},  Arch. Ration. Mech. Anal. {\bf105} (1989), 243--266.

\bibitem{LNR} M. Lewin, P. T. Nam and N. Rougerie, {\em Blow-up profile of rotating 2D focusing Bose gases}, Macroscopic Limits of Quantum Systems, a conference in honor of Herbert Spohn's 70th birthday, Springer Verlag, 2018. 

\bibitem{LL} E. H. Lieb and M. Loss,  Analysis, Graduate Studies in Mathematics Vol. 14,  Amer. Math. Soc., Providence, RI, second edition, 2001.

\bibitem{LS} E. H. Lieb and R. Seiringer, {\em Derivation of the Gross-Pitaevskii equation for rotating Bose gases}, Comm. Math. Phys. {\bf264} (2006), 505--537.

\bibitem{LS1} E. H. Lieb and J. P. Solovej, {\em Ground state energy of the two-component charged Bose gas}, Comm. Math. Phys. {\bf252} (2004), 485--534.




\bibitem{MA} P. Mason and A. Aftalion, {\em Classification of the ground states and topological defects in a rotating two-component Bose-Einstein condensate},
   Phys. Rev. A {\bf84} (2011), 033611.

\bibitem{MH} E. J. Mueller and T.-L. Ho, {\em Two-component Bose-Einstein condensates with a large number of vortices}, Phys. Rev. Lett. {\bf88} (2002), 180403.



\bibitem{RS} M. Reed and B. Simon, Methods of Modern Mathematical Physics. IV. Analysis of Operators. Academic Press, New York/London, 1978.

\bibitem{Roug} N. Rougerie, {\em Scaling limits of bosonic ground states, from many-body to nonlinear Schr\"{o}dinger}, EMS Surveys in Math. Sciences  {\bf 7}  (2020), 253--408.


\bibitem{SSbook} E. Sandier and S. Serfaty,  Vortices in the Magnetic Ginzburg-Landau Model, Progress in Nonlinear Differential Equations and their Applications {\bf 70}, Basel: Birkh\'auser, 2007.


\bibitem{Wei96} J. C. Wei, {\em On the construction of single-peaked solutions to a singularly perturbed semilinear Dirichlet problem},  J. Diff. Eqns. {\bf 129} (1996),  315--333.

\bibitem{WY} J. C. Wei and W. Yao, {\em Uniqueness of positive solutions to some coupled nonlinear Schr\"{o}dinger equations}, Commun. Pure Appl. Anal. {\bf11} (2012), 1003--1011.

\bibitem{W} M. I. Weinstein, {\em Nonlinear Schr\"{o}dinger equations and sharp interpolations estimates}, Commun. Math. Phys. {\bf87} (1983), 567--576.

\bibitem{ZBL} Y. Z. Zhang, W. Z. Bao and H. L. Li, {\em Dynamics of rotating two-component Bose-Einstein condensates and its efficient computation}, Phys. D {\bf234} (2007), 49--69.

\end{thebibliography}
\end{document}